\documentclass[useAMS,usenatbib,usegraphicx]{mn2e}

\usepackage{times}
\usepackage{totpages}
\usepackage{amsmath}
\usepackage{amssymb}
\usepackage{xcolor}
\usepackage{lscape}

\newcommand{\myemail}{chales@aoc.nrao.edu}

\newcommand{\farcsecd}{\mbox{\ensuremath{.\!\!^{\prime\prime}}}}

\newcommand{\degree}{\ensuremath{^\circ}}
\newcommand{\ms}{\scriptscriptstyle}
\newcommand{\tnm}{\textnormal}
\newcommand{\trm}{\textrm}
\renewcommand{\S}{Section}

\title[ATLAS 1.4~GHZ DR2: OBSERVATIONS AND METHODS]{ATLAS 1.4~GHz Data Release 2 -- I.
Observations of the CDF-S and ELAIS-S1 fields and methods for constructing differential number counts}
\author[HALES ET Al.]{
\parbox{\textwidth}{
C.~A. Hales,$^{1,2}$\thanks{E-mail: \myemail}\thanks{Current address:
National Radio Astronomy Observatory, P.O. Box 0, Socorro, NM 87801,
USA; Jansky Fellow of the National Radio Astronomy Observatory.}
R.~P. Norris,$^{2,3}$
B.~M. Gaensler,$^{1,3}$
E. Middelberg,$^{4}$
K.~E. Chow,$^{2}$
A.~M. Hopkins,$^{5,3}$
M.~T. Huynh,$^{6}$
E. Lenc,$^{1,3}$
and M.~Y. Mao$^{7}$
}\vspace{4mm}\\
$^{1}$Sydney Institute for Astronomy, School of Physics,
The University of Sydney, NSW 2006, Australia\\
$^{2}$Australia Telescope National Facility, CSIRO Astronomy and Space Science,
P.O. Box 76, Epping, NSW 1710, Australia\\
$^{3}$ARC Centre of Excellence for All-sky Astrophysics (CAASTRO)\\
$^{4}$Astronomisches Institut, Ruhr-Universit\"{a}t, Universit\"{a}tsstr. 150,
44801 Bochum, Germany\\
$^{5}$Australian Astronomical Observatory, P.O. Box 915, North Ryde, NSW 1670, Australia\\
$^{6}$International Centre for Radio Astronomy Research, University of Western Australia,
Crawley, WA 6009, Australia\\
$^{7}$National Radio Astronomy Observatory, P.O. Box 0, Socorro, NM 87801, USA
}
\begin{document}

\date{Draft version \today}

\pagerange{\pageref{firstpage}--\ref{TotPages}} \pubyear{\the\year}

\maketitle

\label{firstpage}

\begin{abstract}
This is the first of two papers describing the second data release (DR2) of the Australia
Telescope Large Area Survey (ATLAS) at 1.4~GHz, which comprises deep wide-field
observations in total intensity, linear polarization, and circular polarization over the
{\it Chandra} Deep Field-South and European Large Area {\it Infrared Space Observatory}
Survey-South 1 regions. DR2 improves upon the first data release by maintaining consistent
data reductions across the two regions, including polarization analysis, and including
differential number counts in total intensity
and linear polarization. Typical DR2 sensitivities across the mosaicked multi-pointing images
are $30$~$\mu$Jy~beam$^{-1}$ at approximately $12\arcsec\times6\arcsec$ resolution over a
combined area of 6.4~deg$^2$. In this paper we present detailed descriptions of our data
reduction and analysis procedures, including corrections for instrumental effects such as
positional variations in image sensitivity, bandwidth smearing with a non-circular beam,
and polarization leakage, and application of the {\tt BLOBCAT} source extractor. We present
the DR2 images and catalogues of components (discrete regions of radio emission) and sources
(groups of physically associated radio components). We describe new analytic methods to
account for resolution bias and Eddington bias when constructing differential number counts
of radio components.
\end{abstract}


\section{Introduction}\label{ch4:SecIntr}

Radio surveys are a cornerstone of modern astronomy. Counts of extragalactic radio sources
per steradian per unit flux density provide fundamental constraints on galaxy evolution,
as they implicitly encapsulate both the underlying redshift and luminosity
distributions of source populations \citep[e.g.][]{1966MNRAS.133..421L}.

In total intensity, the 1.4~GHz source counts are observed to flatten below 1~mJy, though
the extent of this flattening is controversial because the results from deep surveys exhibit
a large degree of scatter. To illustrate, see the compilation of surveys in Fig.~3 from
\citet{2013PASA...30...20N} where there is a factor of 2 variation in the counts
below 1~mJy. Some studies have attributed the large scatter in the faint counts to
cosmic variance, namely to intrinsic differences between survey fields caused
by source clustering \citep[e.g.][]{2004MNRAS.352..131S}. However, significant
differences in the counts for fields observed by separate studies, such as the
Lockman Hole \citep{2009MNRAS.397..281I}, indicate that calibration and data
processing errors may be largely responsible for the scatter. Issues to
consider include corrections for bandwidth smearing \citep[e.g.][]{2009MNRAS.397..281I},
Eddington bias \citep[e.g.][]{2006MNRAS.372..741S}, resolution bias
\citep[e.g.][]{2008ApJ...681.1129B}, and non-instrumental factors such
as source clustering in the field \citep[e.g.][]{2013MNRAS.432.2625H}.
The present conclusion in the literature is that the scatter
in the sub-millijansky counts is likely to be significantly affected by data processing
differences between surveys \citep{2006MNRAS.371..963B,2007ASPC..380..189C,
2009MNRAS.397..281I,2010A&ARv..18....1D,2012ApJ...758...23C,2013MNRAS.432.2625H}.
This conclusion motivates the need for studies that describe data reduction and
analysis procedures in detail, so as to facilitate robust comparisons with other
works and encourage future improvements.

To date, very few surveys dedicated to extragalactic polarized radio sources have
been conducted, primarily because of correlator limitations that have
required polarization capabilities to be sacrificed for spectral resolution.
Polarization surveys at 1.4~GHz include the NRAO VLA Sky Survey
\citep[NVSS;][]{1998AJ....115.1693C} observed with the Very Large Array,
which encompasses 82\% of the sky at resolution full-width at
half-maximum (FWHM) 45\arcsec\ to a root mean square (rms) sensitivity
in polarization of 0.29~mJy~beam$^{-1}$, surveys of the European Large Area {\it Infrared
Space Observatory} Survey-North 1 (ELAIS-N1) region observed using the Dominion Radio
Astrophysical Observatory (DRAO) Synthesis Telescope by \citet{2007ApJ...666..201T}
over 7.43~deg$^2$ with resolution FWHM $\sim50\arcsec$ to 78~$\mu$Jy~beam$^{-1}$
and in a deeper follow-up study with the same facility by \citet{2010ApJ...714.1689G}
over 15.16~deg$^2$ to 45~$\mu$Jy~beam$^{-1}$, and the Australia Telescope
Low-Brightness Survey \citep[ATLBS;][]{2010MNRAS.402.2792S} which encompasses
two fields observed with the Australia Telescope Compact Array (ATCA) over
a total of 8.42~deg$^2$ with resolution FWHM $\sim50\arcsec$ to
$\sim80\mu$Jy~beam$^{-1}$. A summary of polarization surveys at other radio
wavelengths is presented by \citet{2012AdAst2012E..52T}. By cross-matching
polarized 1.4~GHz sources with mid-infrared counterparts, \citet{2007ApJ...666..201T}
identified the population of polarized millijansky sources as being extragalactic
radio sources powered by AGNs. \citet{2010ApJ...714.1689G} found that the polarized
emission from these sources was likely to originate in extended radio lobes.
\citet{2002A&A...396..463M} and \citet{2004MNRAS.349.1267T} found an
anti-correlation between the fractional linear polarization and total intensity
flux densities of NVSS sources; faint sources were more highly polarized. This finding
was supported for ELAIS-N1 sources by \citet{2007ApJ...666..201T} and
\citet{2010ApJ...714.1689G}, and for ATLBS sources by \citet{2010MNRAS.402.2792S}.
\citet{2004MNRAS.349.1267T}, \citet{2007ApJ...666..201T}, and \citet{2010ApJ...714.1689G}
found that the Euclidean-normalised differential number-counts of polarized sources
flattened at linearly polarized flux densities $L$~{\footnotesize $\lesssim$}~1~mJy
to levels greater than those predicted by \citet{2004NewAR..48.1289B}; the latter
predicted polarized source counts to $\mu$Jy levels by convolving total
intensity source counts with a fractional polarization distribution modelled on NVSS data.
\citet{2008evn..confE.107O} were unable to reproduce the observed flattening in a
population modelling study. The observed flattening suggests the emergence of systematic
changes in polarized source properties with decreasing flux density, such as higher
ordering of magnetic fields in fainter sources, or perhaps the emergence of an
unexpected faint population. To examine the emerging fractional polarization
anti-correlation and source count flattening trends in more detail, deeper and higher
angular resolution observations of the 1.4~GHz polarized sky are required.

In this work we present reprocessed and new 1.4~GHz observations
of the {\it Chandra} Deep Field-South \citep[CDF-S; Galactic coordinates $l\approx224\degree$,
$b\approx-55\degree$;][]{2006AJ....132.2409N} and ELAIS-South 1 \citep[ELAIS-S1; $l\approx314\degree$,
$b\approx-73\degree$;][]{2008AJ....135.1276M} regions, obtained as part
of the Australia Telescope Large Area Survey (ATLAS) project with the ATCA. We
collectively refer to these previous ATLAS papers as Data Release 1 (DR1)
and denote the present work Data Release 2 (DR2). Given that DR1 did not
include polarization analysis of the ATLAS data, we have chosen to
reprocess the original observations to ensure consistent and improved
data reduction and analysis between both the total intensity and
polarization data and the two independent ATLAS regions. In preparation
for ATLAS DR2, we have developed new tools to ensure accurate calculation
of the statistical significance of flux density measurements in linear
polarization \citep{2012MNRAS.424.2160H} and to ensure accurate measurement
of these flux densities using the {\tt BLOBCAT} source extractor \citep{2012MNRAS.425..979H}.

The motivations for ATLAS DR2 are to (i) present a detailed description of
our data reduction and analysis procedures to inform future deep surveys
such as  those being developed for SKA Pathfinder facilities around the world
(see summary of facilities described by \citealt{2012A&A...543A.113B} and
\citealt{2013PASA...30...20N}), (ii) compute differential number counts for total
intensity and linearly polarized objects (total intensity counts were not included in DR1),
and (iii) investigate the nature of faint polarized sources and consider possible
explanations for the fractional polarization trend seen in previous studies.
Clearly, biases introduced at an early stage of data reduction have the
potential to propagate through to the final data in a non-linear fashion,
affecting the ability for that data to be used for unplanned and novel
experiments in the future \citep[e.g.][]{2009ApJ...692..887C}. In this
paper (Paper I) we focus on point (i) from above, regarding data reduction and the
development of new techniques to produce high fidelity data suitable for
investigating points (ii) and (iii). Results and discussion regarding
points (ii) and (iii) will be presented in Paper II \citep{halesPII}.

This paper is organised as follows. In \S~\ref{ch4:SecObs} we describe our
ATLAS radio data and ancillary mid-infrared and optical data. In \S~\ref{ch4:SecDat}
we outline our radio data reduction and post-processing procedures to
obtain mosaicked images of total intensity, linear polarization (using
rotation measure synthesis), and additionally circular polarization
for the two ATLAS regions. In \S~\ref{ch4:SecInst} we describe instrumental
effects of time-average smearing, bandwidth smearing, and polarization
leakage, our methods to account for them in our ATLAS data, and the effective
survey area boundaries. In \S~\ref{ch4:SecExt} we detail how radio components were
detected and extracted in total intensity, linear polarization, and circular
polarization, and how their flux densities were corrected to account for subtle
noise-induced systematics. In \S~\ref{ch4:SecClass} we describe our implementation
of two cross-identification and classification schemes: the first to group
components into sources, to associate these sources with infrared
sources, and to classify them according to their multiwavelength properties;
and the second to associate linearly polarized components or polarization
upper limits with total intensity counterparts and to classify these associations
based on their polarized morphologies. In \S~\ref{ch4:SecCNC} we describe in detail
corrections required to calculate total intensity and linear polarization
differential number-counts, including a new fully analytic method to account
for resolution bias. In \S~\ref{ch5:SecRes} we present the ATLAS DR2 total intensity
and linear polarization images, and the radio component and source catalogues.
We conclude in \S~\ref{ch4:SecConcl}. For reference, a selection of important
symbols used in this work is presented in Table~\ref{tbl:symbols}.
\begin{table*}
\centering
\caption{Selection of important symbols used in this work.\label{tbl:symbols}}
\begin{tabular}{@{}cll@{}}
\hline
Symbol & Description & Defined \\
\hline
$I_{\trm{\tiny MFS}}$ & total intensity mosaic produced using multi-frequency synthesis approach & \S~\ref{ch4:SecDatSubMFS} \\
$V_{\trm{\tiny MFS}}$ & circular polarization mosaic produced using multi-frequency synthesis approach & \S~\ref{ch4:SecDatSubMFS} \\
$I_{\ms i}$, $Q_{\ms i}$, $U_{\ms i}$ & $i$'th frequency channel mosaic in total intensity, Stokes $Q$, or Stokes $U$ & \S~\ref{ch4:SecDatSubPCI} \\
$\sigma_{\ms Q,i},\sigma_{\ms U,i}$ & rms noise map of $i$'th frequency channel mosaic in Stokes $Q$ or Stokes $U$ & \S~\ref{ch4:SecDatSubRM} \\
$\sigma_{\ms Q,U,i}$ & map of combined rms noise for Stokes $Q$ and Stokes $U$ in $i$'th frequency channel & \S~\ref{ch4:SecDatSubRM} \\
$I_{\trm{\tiny CA}}$ & total intensity mosaic produced using channel average approach & \S~\ref{ch4:SecDatSubRM} \\
$L_{\trm{\tiny RM}}$ & linear polarization mosaic produced using rotation measure synthesis & \S~\ref{ch4:SecDatSubRM} \\
$\sigma_{\trm{\tiny RM}}(x,y)$ & rms noise map for $L_{\trm{\tiny RM}}$ & \S~\ref{ch4:SecDatSubRM} \\
$\varpi$ & bandwidth smearing ratio (observed divided by true surface brightness) & \S~\ref{ch4:SecInstSubBS} \\
$K_{\trm{\tiny LEAK}}$ & total intensity to linear polarization leakage mosaic & \S~\ref{ch4:SecInstSubLeak} \\
$L_{\trm{\tiny RM}}^{\trm{\tiny CORR}}$ & $L_{\trm{\tiny RM}}$ corrected for polarization leakage & \S~\ref{ch4:SecInstSubLeak} \\
$F^{\trm{\tiny AREA}}$ & survey area & \S~\ref{ch4:SecInstSubArea} \\
$S_{\trm{\tiny peak}}$,$S_{\trm{\tiny int}}$ & peak or integrated surface brightness (more generally, $S$ denotes flux density) & \S~\ref{ch4:SecExtSubFlood} \\
$A_{\trm{\tiny S}}$ & detection signal-to-noise ratio & \S~\ref{ch4:SecExtSubFlood} \\
$V^{\trm{\tiny AREA}}$ & visibility area for detection & \S~\ref{ch4:SecExtSubFlood} \\
$\theta,B$ & observed or beam full-width at half-maximum & \S~\ref{ch4:SecExtSubDeconv} \\
$\Theta$ & deconvolved angular size & \S~\ref{ch4:SecExtSubDeconv} \\
$\gamma$ & slope of differential number-counts, $dN/dS \propto S^{-\gamma}$ & \S~\ref{ch4:SecExtSubDB} \\
$S_{\trm{\tiny ML}}$ & deboosted flux density using maximum-likelihood scheme & \S~\ref{ch4:SecExtSubDB} \\
$dN_{\trm{\tiny H03}}/dS$ & differential number-count fit from \citet{2003AJ....125..465H} & \S~\ref{ch4:SecExtSubDB} \\
$dN_{\trm{\tiny H03M}}/dS$ & modified version of $dN_{\trm{\tiny H03}}/dS$ & \S~\ref{ch4:SecExtSubDB} \\
$f_{\ms \Pi}$ & distribution of fractional linear polarization ($\Pi \equiv L/I$) & \S~\ref{ch4:SecExtSubDB} \\
$L_{\trm{\tiny UL}}$ & linear polarization upper limit & \S~\ref{ch4:SecClassSubPACI} \\
$r,e$ & resolution or Eddington bias corrections & \S~\ref{ch4:SecCNCSubRB} \\
$\Theta_\trm{\scriptsize max}$ & maximum intrinsic angular size for detectable component & \S~\ref{ch4:SecCNCSubRB1} \\
$\widetilde{\sigma}$ & local rms noise divided by local bandwidth smearing ratio & \S~\ref{ch4:SecCNCSubRB1} \\
$f_{\widetilde{\sigma}}$ & probability distribution for $\widetilde{\sigma}$ & \S~\ref{ch4:SecCNCSubRB1} \\
$h$ & integral angular size distribution & \S~\ref{ch4:SecCNCSubRB1} \\
$\eta$ & angular filling factor for linearly polarized emission ($\Theta_\tnm{\scriptsize L}/\Theta_\tnm{\scriptsize I}$) & \S~\ref{ch4:SecCNCSubRB1} \\
$dN_{\tnm{\scriptsize detectable}}/dS$ & differential number-counts that are observable & \S~\ref{ch4:SecCNCSubRB1} \\
$\Theta_\tnm{\tiny med}$ & median largest angular size & \S~\ref{ch4:SecCNCSubRB1} \\
$\Theta_\trm{\scriptsize min}$ & minimum intrinsic angular size for detected component to be classified as resolved & \S~\ref{ch4:SecCNCSubRB2} \\
$dN_{\tnm{\scriptsize resolved}}/dS$ & resolved detectable number-counts & \S~\ref{ch4:SecCNCSubRB2} \\
$dN_{\tnm{\scriptsize unresolved}}/dS$ & unresolved detectable number-counts, assuming ideal case without measurement bias & \S~\ref{ch4:SecCNCSubRB2} \\
$dN_{\tnm{\scriptsize unresolved-obs}}/dS$ & unresolved detectable number-counts, accounting for measurement bias & \S~\ref{ch4:SecCNCSubRB2} \\
\hline
\end{tabular}
\end{table*}

\section{Observational Data}\label{ch4:SecObs}

\subsection{Radio Data}\label{ch4:SecObsSubRadio}

ATLAS observations of the CDF-S and ELAIS-S1 fields were obtained with the ATCA \citep{1992JEEEA..12..103F},
a synthesis telescope consisting of six 22~m alt-az antennas on an east-west baseline. Each antenna is
equipped with linearly polarized feeds used to measure all four polarization products ($XX,YY,XY,YX$), from
which all four Stokes parameters ($I,Q,U,V$) can be derived. Noise diodes in the feed horns of each
antenna replace the need to observe a polarization position angle calibrator to derive absolute
$XY$ phase.

The two ATLAS fields were observed in mosaic mode using ATCA's standard continuum correlator setup,
{\tt FULL\_128\_2}. This correlator configuration enabled observation of 2$\,\times\,$128 MHz bandwidth
windows centred on 1344 and 1432 MHz, with each 128 MHz window divided into 32$\,\times\,$4 MHz
non-independent channels. A correlator cycle time of 10 seconds was used. The full width at half maximum
(FWHM) of the primary beam at these frequencies is $\sim$35\arcmin. The standard ATCA primary
flux density calibrator PKS~B1934$-$638 \citep{reynolds} was used for both ATLAS fields. The
secondary calibrators\footnote{http://www.narrabri.atnf.csiro.au/calibrators/} for the
CDF-S and ELAIS-S1 fields were PKS~B0237$-$233 and PKS~B0022$-$423, respectively.

\begin{figure*}
 \centering
 \includegraphics[trim = 0mm 0mm 0mm 120mm, height=65mm]{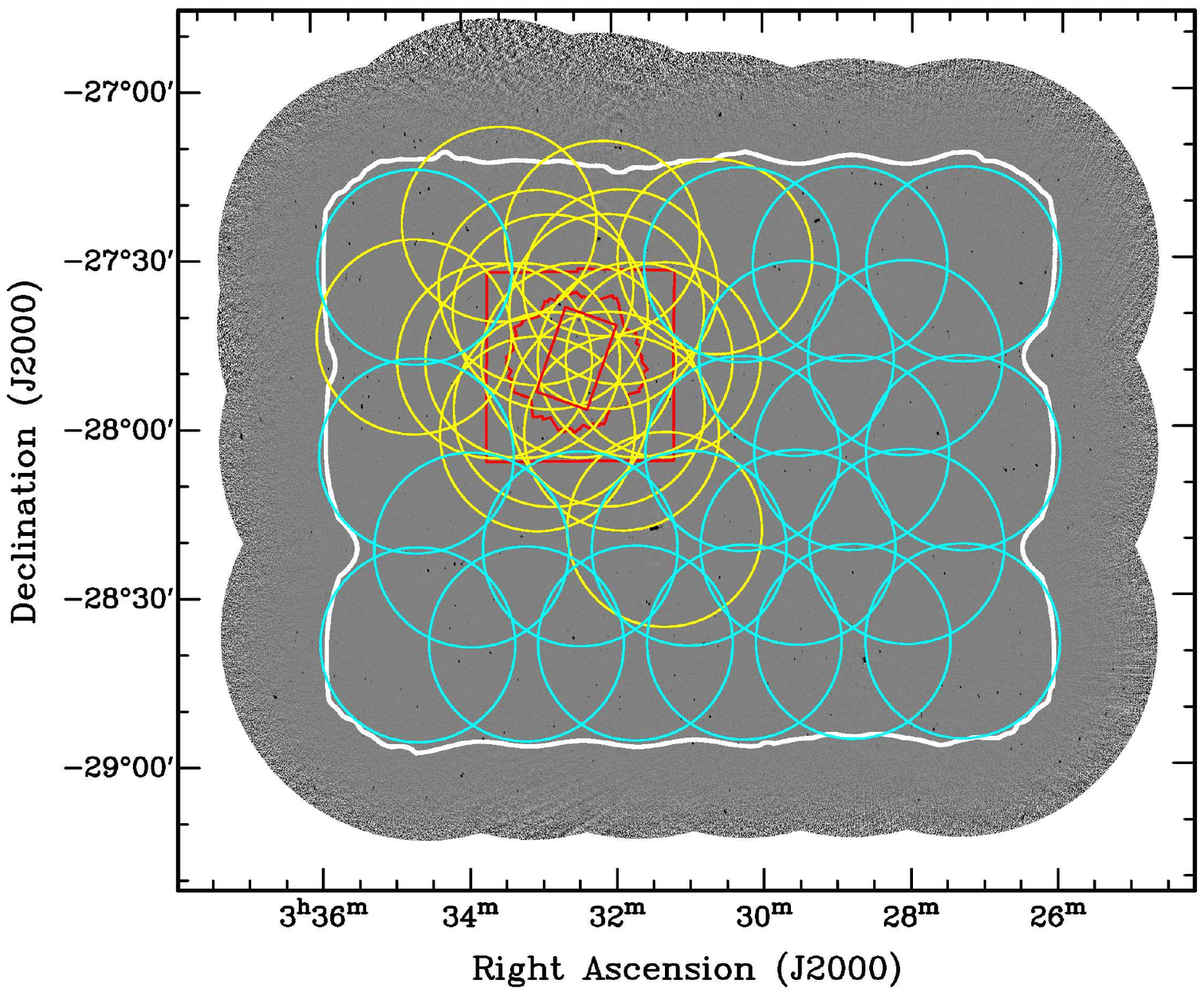}
 \hspace{0.2pc}
 \includegraphics[trim = 0mm 0mm 0mm 100mm, height=65mm]{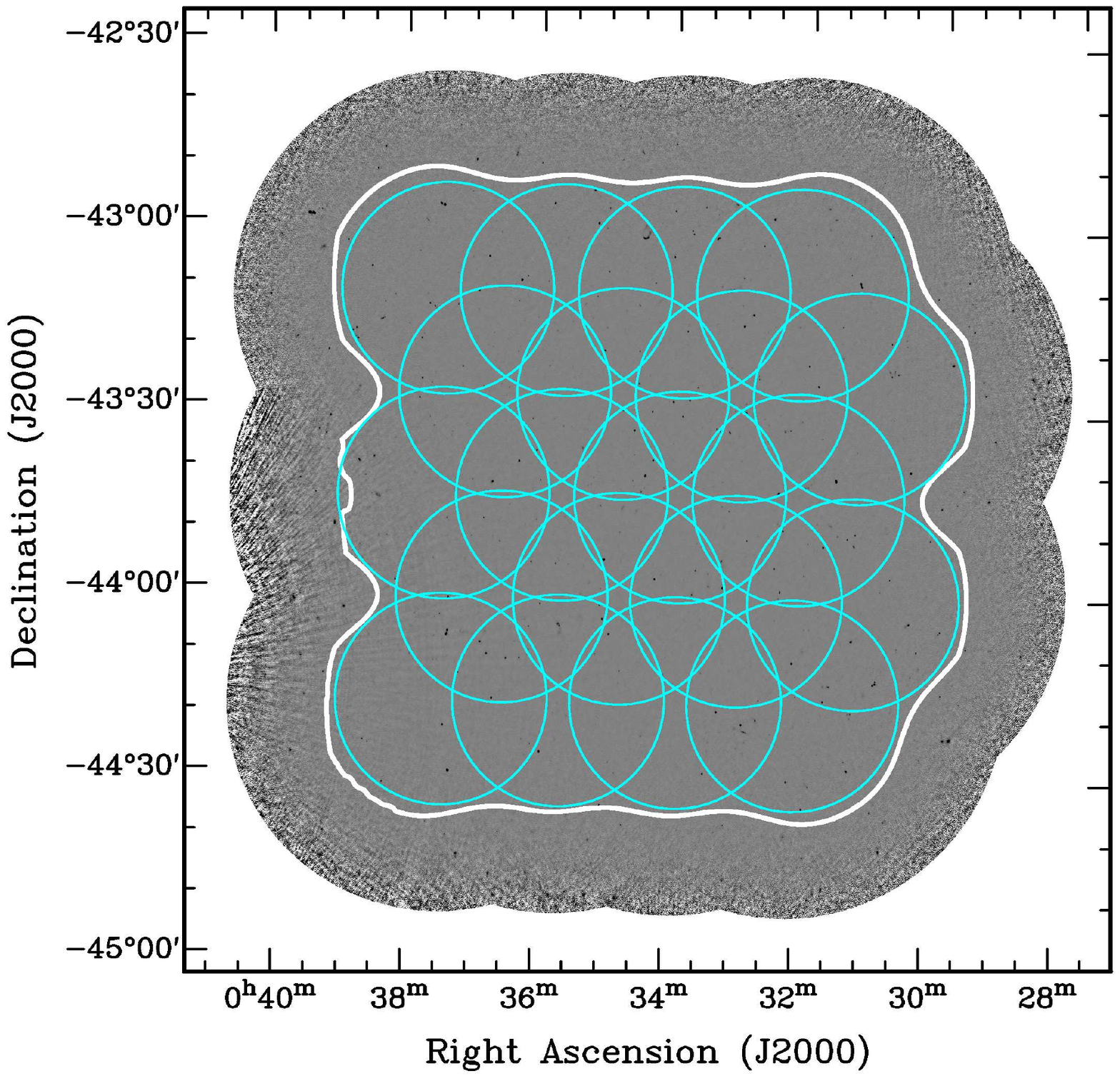}
 \begin{minipage}{140mm}
 \caption{Overview of mosaicked observations of the CDF-S (left) and ELAIS-S1 (right) ATLAS fields;
	  background images are of continuum total intensity (see \S~\ref{ch4:SecDatSubMFS}) with
	  equally-scaled shading levels. Circles indicate the locations and 35\arcmin\ half-power
	  primary beam widths of the pointings. The CDF-S field consists of 39 pointings: 18 of these
	  were observed by \citet{2003NewAR..47..391K} (yellow circles), while the remaining 21
	  pointings were observed solely by ATLAS (cyan circles). The rotated red rectangle indicates
	  the GOODS-South field \citep{2004ApJ...600L..93G}. The red irregular polygon and outer red
	  square indicate the 2MS CDF-S \citep{2008ApJS..179...19L} and extended-CDF-S
	  \citep{2005ApJS..161...21L} fields, respectively. All 20 pointings within the ELAIS-S1
	  field were observed solely by ATLAS. The thick outer contour (white) in each field
	  indicates the survey area boundary (see \S~\ref{ch4:SecInstSubArea}). SWIRE observations
	  \citep{2003PASP..115..897L} encompass each ATLAS field.}
 \label{ch4:fig:cdfspoint}
\end{minipage}
\end{figure*}

\begin{figure*}
 \centering
 \includegraphics[trim = 0mm 10mm 0mm 220mm, height=62mm]{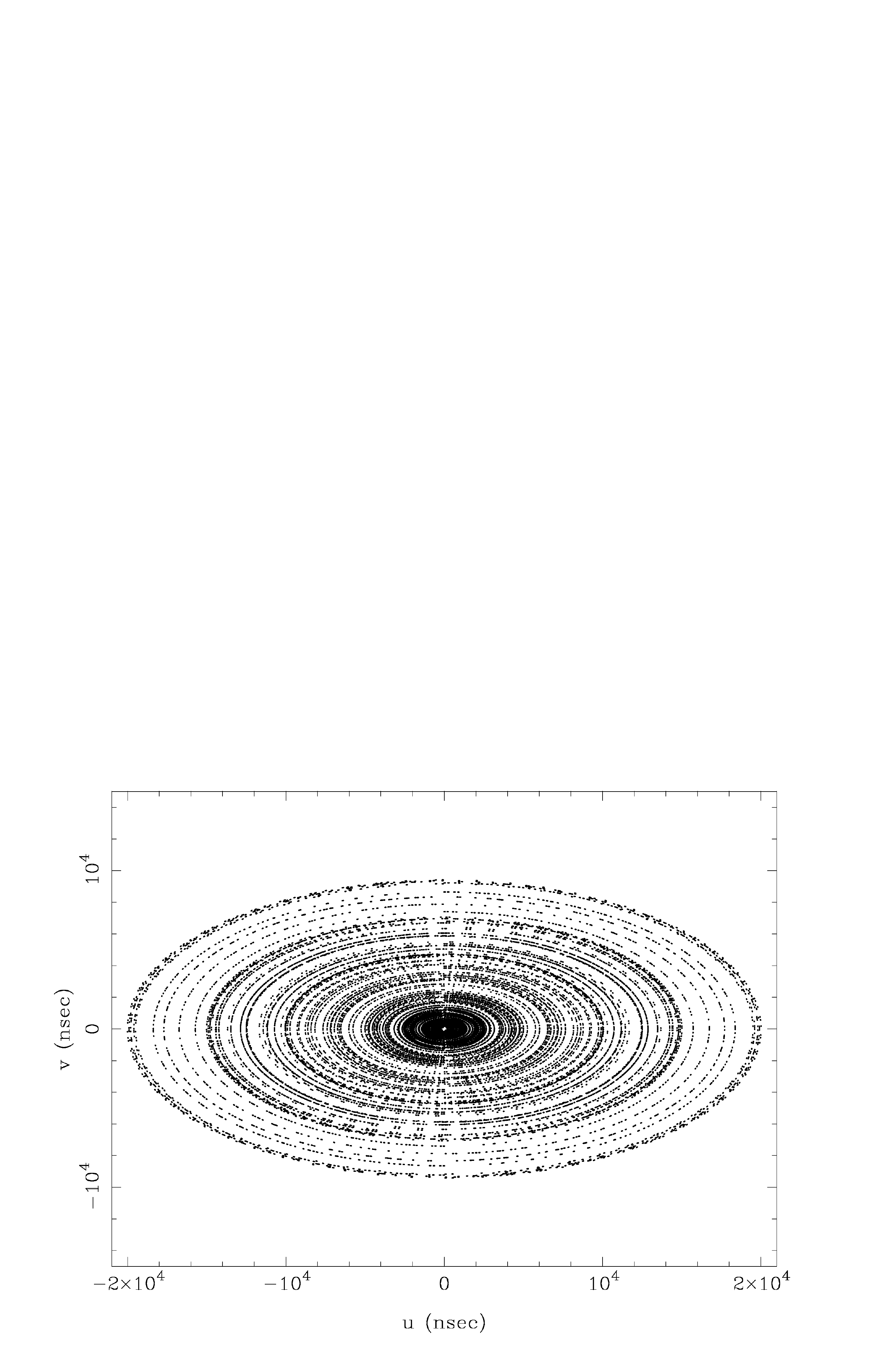}
 \hspace{0.2pc}
 \includegraphics[trim = 0mm 10mm 0mm 220mm, height=62mm]{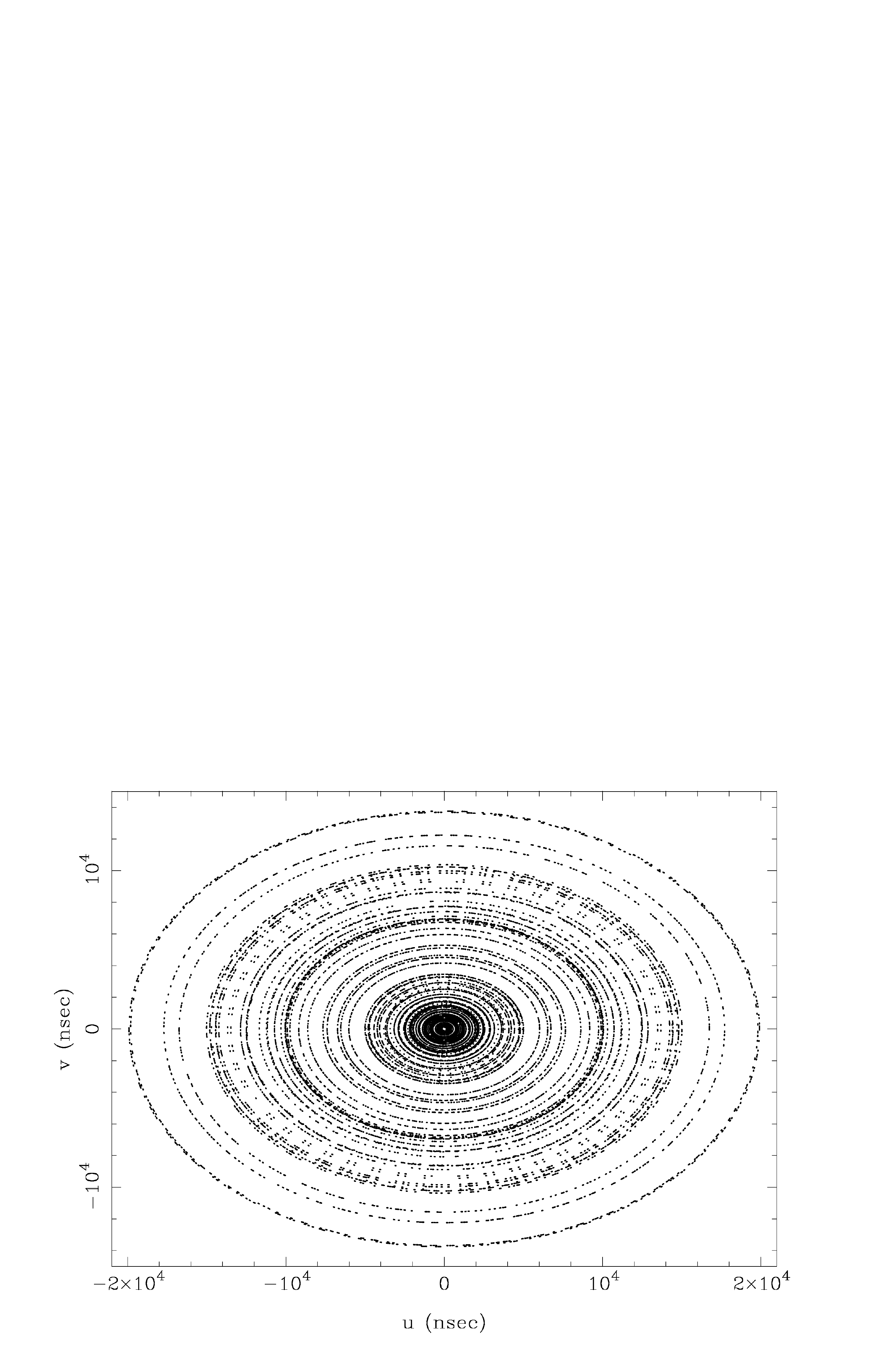}\\
 \includegraphics[trim = 0mm 10mm 0mm 220mm, height=62mm]{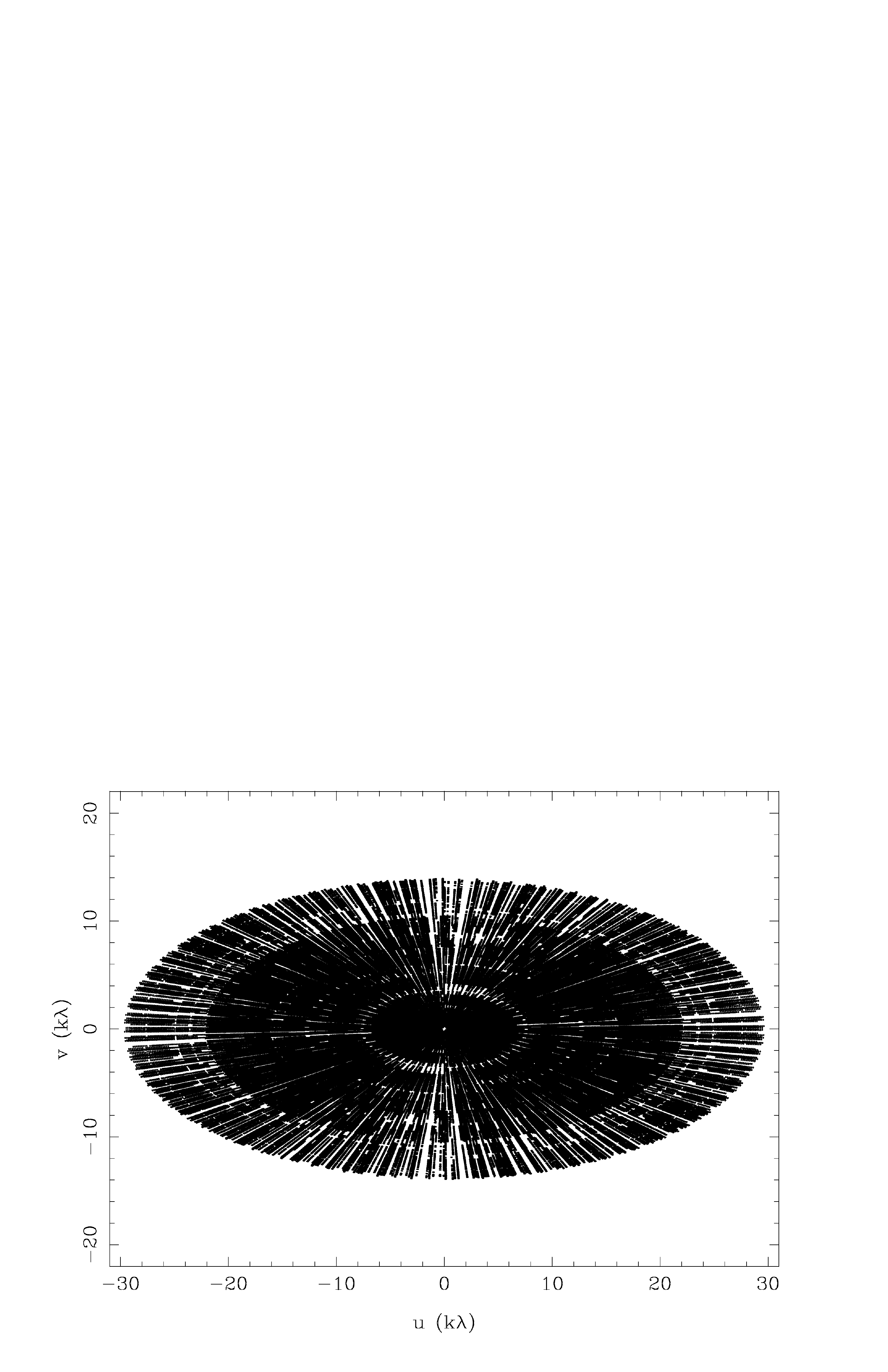}
 \hspace{0.2pc}
 \includegraphics[trim = 0mm 10mm 0mm 220mm, height=62mm]{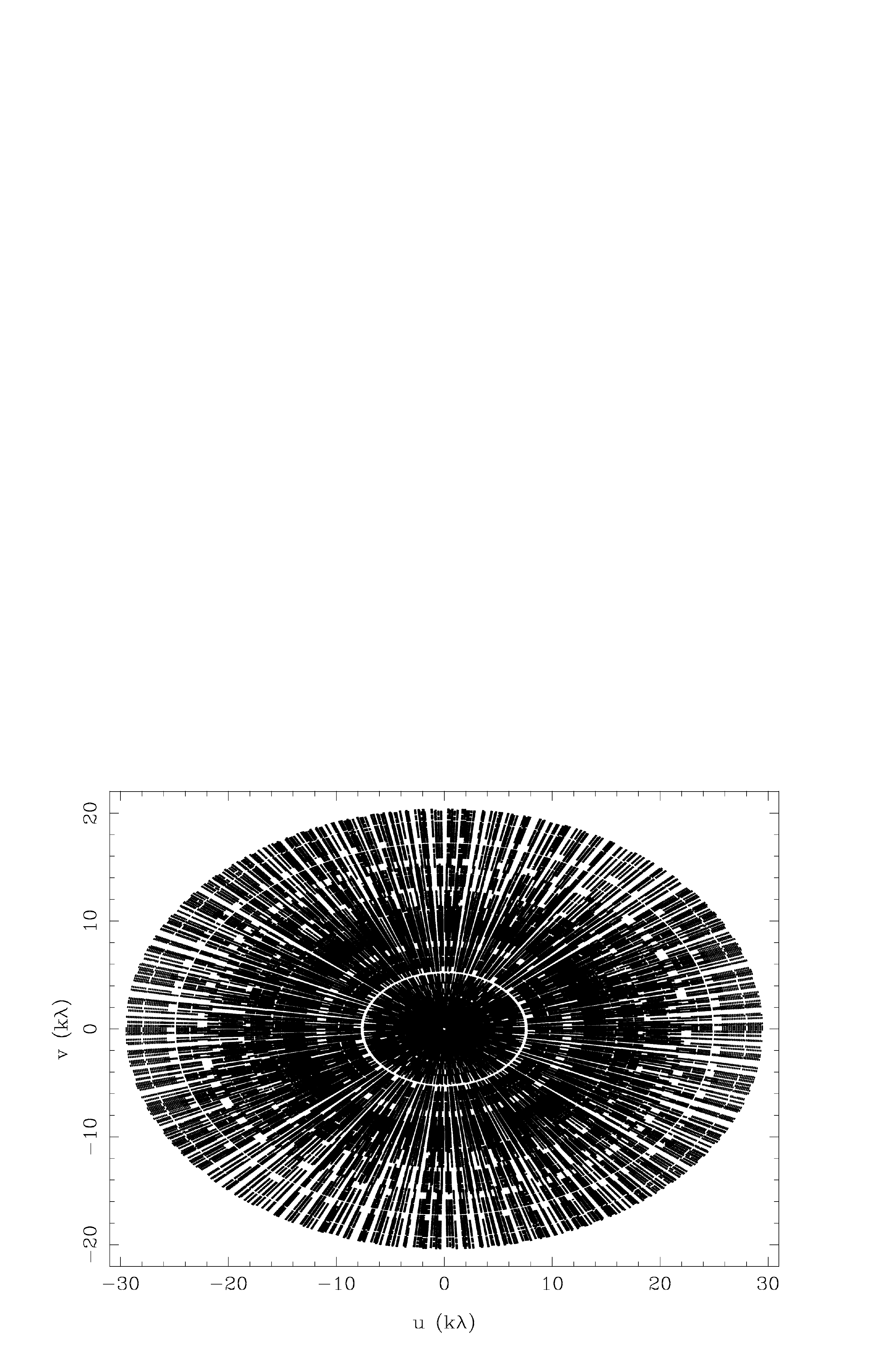}
 \begin{minipage}{140mm}
 \caption{Typical $uv$-plane coverage for a pointing in each of the CDF-S (left column)
          and ELAIS-S1 (right column) fields in units of nano-seconds (upper row; independent of
	  frequency) and kilo-lambda (lower row; indicates multi-channel coverage).}
 \label{ch4:fig:uvcov}
\end{minipage}
\end{figure*}

Both ATLAS fields consist of multiple pointings, as shown in Fig.~\ref{ch4:fig:cdfspoint}.
Each pointing was observed using a number of complementary ATCA array configurations to
maximise $uv$-plane coverage; we show the typical $uv$-plane coverage for ATLAS pointings
in Fig.~\ref{ch4:fig:uvcov}.
Some pointings are more sensitive than others due to non-uniform time allocation. The
original DR1 CDF-S observations combined 7 pointings from \citet{2003NewAR..47..391K}
(ATCA Project ID C1035) with 21 ATLAS pointings (ATCA Project ID C1241). To boost
sensitivity in the CDF-S field, in DR2 we have included 11 additional pointings from
\citet{2003NewAR..47..391K}, making use of all 18 suitable pointings from their data.
We have also included new CDF-S observations of the 21 ATLAS pointings, obtained in
the period 2005 January to 2006 March. Observing dates, array configurations, and net
integration times of the CDF-S data used in this work are shown in
Table~\ref{ch4:tbl:CDFSobsdates}; pointing centres are given in
Table~\ref{ch4:tbl:CDFSobspnts}.
\begin{table}
\centering
\caption{	Observing dates, array configurations, and net integration times on
		source for the ATLAS DR2 CDF-S field\label{ch4:tbl:CDFSobsdates}.
	}
\begin{tabular}{@{}llcc@{}}
\hline
Project ID & Date & Array & Net Integration\\
 & & & Time (h) \\
\hline
C1035 \ldots &2002 Apr 4$-$7, 10, 12$-$13 &6A &72.9 \\
&2002 Aug 23$-$24, 27$-$29 &6C &29.6 \\
C1241 \ldots &2004 Jan 7$-$8, 12 &6A &23.9 \\
&2004 Feb 3$-$5 &6B &24.7 \\
&2004 Jun 6, 8$-$12 &750D &37.4 \\
&2004 Nov 24$-$30 &6D &50.4 \\
&2004 Dec 28$-$30 &1.5D &22.6 \\
&2005 Jan 7$-$8, 18$-$19, 23 &750B &31.9 \\
&2005 Apr 9$-$10 &6A &18.5 \\
&2005 Apr 14 &1.5A &8.9 \\
&2005 Apr 22; 2005 May 2 &750A &15.0 \\
&2005 Jun 1, 10 &EW367 &11.7 \\
&2005 Jun 25$-$26 &6B &18.1 \\
&2005 Dec 6 &6A &8.7 \\
&2006 Mar 23$-$24, 27 &6C &23.0 \\
\hline
\end{tabular}
\end{table}
\begin{table}
\begin{minipage}{\columnwidth}
\centering
\caption{	Coordinates of ATLAS DR2 CDF-S calibrators and pointing
		centres\label{ch4:tbl:CDFSobspnts}.}
\begin{tabular}{@{}lrr@{}}
\hline
Source/Pointing\footnote{The prefix K indicates a pointing from
\citet{2003NewAR..47..391K} (Project ID C1035).} & R.A. (J2000) & Decl. (J2000) \\
\hline
B1934$-$638 &19:39:25.026 &-63:42:45.63 \\
B0237$-$233 &2:40:08.175 &-23:09:15.73 \\
K2 &3:31:42.777 &-27:48:30.00 \\
K3 &3:32:05.390 &-27:57:09.62 \\
K4 &3:32:50.610 &-27:57:09.62 \\
K5 &3:33:13.223 &-27:48:30.00 \\
K6 &3:32:50.610 &-27:39:50.37 \\
K7 &3:32:05.390 &-27:39:50.37 \\
K8 &3:31:19.682 &-28:18:41.18 \\
K9 &3:34:40.330 &-27:44:03.49 \\
K10 &3:33:30.909 &-27:24:10.98 \\
K11 &3:32:08.381 &-27:26:51.93 \\
K12 &3:30:38.126 &-27:30:14.68 \\
1 &3:28:47.330 &-28:38:37.98 \\
1a &3:27:18.362 &-28:38:31.14 \\
2 &3:28:03.890 &-28:21:46.74 \\
3 &3:28:48.482 &-28:05:05.58 \\
3a &3:27:18.362 &-28:05:05.58 \\
4 &3:28:05.258 &-27:48:14.34 \\
5 &3:28:49.610 &-27:31:32.82 \\
5a &3:27:18.362 &-27:31:32.82 \\
10 &3:30:16.970 &-27:31:40.02 \\
11 &3:29:32.834 &-27:48:22.98 \\
12 &3:30:16.298 &-28:05:12.42 \\
13 &3:29:31.922 &-28:21:55.74 \\
14 &3:30:15.602 &-28:38:44.82 \\
15 &3:31:43.874 &-28:38:48.42 \\
16 &3:30:59.954 &-28:22:00.78 \\
27 &3:32:27.986 &-28:22:02.58 \\
28 &3:33:12.122 &-28:38:48.42 \\
29 &3:34:40.394 &-28:38:44.82 \\
30 &3:33:56.018 &-28:22:00.78 \\
31 &3:34:39.698 &-28:05:12.42 \\
33 &3:34:39.026 &-27:31:40.02 \\
41 &3:32:28.000 &-27:48:30.00 \\
42 &3:31:20.166 &-27:48:30.00 \\
43 &3:31:54.083 &-28:01:29.43 \\
44 &3:33:01.917 &-28:01:29.43 \\
45 &3:33:35.834 &-27:48:30.00 \\
46 &3:33:01.917 &-27:35:30.56 \\
47 &3:31:54.083 &-27:35:30.56 \\
\hline
\end{tabular}
\end{minipage}
\end{table}
For the ELAIS-S1 field with 20 pointings, we have reprocessed the same raw DR1 data as
outlined by \citet{2008AJ....135.1276M}. The baselines measured for the CDF-S and ELAIS-S1
fields cover the range 31$-$6000~m and 46$-$5969~m, respectively.

\subsection{Ancillary Data}\label{ch4:SecObsSubAD}

We supplemented our 1.4~GHz radio observations with data at infrared
and optical wavelengths, as described below, to enable source classifications
using multiwavelength cross-identifications.

\subsubsection{Infrared Data}\label{ch4:SecObsSubSubIR}

{\it Spitzer Space Telescope} \citep{2004ApJS..154....1W} observations
encompassing the CDF-S and ELAIS-S1 ATLAS fields were carried out
as part of the {\it Spitzer} Wide-Area Infrared Extragalactic Survey
\citep[SWIRE;][]{2003PASP..115..897L} Legacy Project. We obtained flux densities
for SWIRE sources from a pre-release version of the SWIRE Public Data
Release\footnote{See http://irsa.ipac.caltech.edu/data/SPITZER/docs/spitzermission/\linebreak
observingprograms/legacy/swire/~.}~3
catalogue (SDR3; Fall 2005) in five wavelength bands: 3.6, 4.5, 5.8, and 8.0~$\mu$m observed
with the Infrared Array Camera \citep[IRAC;][]{2004ApJS..154...10F}, and 24.0~$\mu$m observed
with the Multiband Imaging Photometer for {\it Spitzer} \citep[MIPS;][]{2004ApJS..154...25R}.
The flux density limits for the pre-release SDR3 catalogue in each
band were approximately 4, 5, 43, 38, and 230~$\mu$Jy, respectively. These limits
are less conservative than those applied to the general release SDR3 catalogue
\citep{2005AAS...207.6301S}\footnote{Also see http://swire.ipac.caltech.edu/swire/astronomers/publications/\linebreak
SWIRE2\_doc\_ 083105.pdf~.}. SWIRE data is available over 97\% and 100\% of the
ATLAS DR2 CDF-S and ELAIS-S1 survey areas, respectively.

\subsubsection{Optical Data}\label{ch4:SecObsSubSubOpt}

In the optical band, \citet{2012MNRAS.426.3334M} obtained spectroscopic observations
of SWIRE sources associated with ATLAS DR1 radio sources.
Spectra were obtained for optical counterparts to 160 SWIRE sources in the CDF-S field,
and 306 SWIRE sources in the ELAIS-S1 field, to limiting magnitudes of typically
$R\approx20$, extending to $R\approx23$ for the faintest sources.

\section{Data Reduction and Post-Processing}\label{ch4:SecDat}

We developed a semi-automated analysis pipeline to edit, calibrate, image, and post-process the
ATLAS radio data using a combination of the {\tt MIRIAD} package \citep{1995ASPC...77..433S} and custom
software, as described in the following sections.

\subsection{Flagging and Calibration}\label{ch4:SecDatSubFlag}

We used the {\tt MIRIAD} task {\tt ATLOD} to re-weight the spectrum for each visibility in the lag
domain in order to reduce the Gibbs phenomenon, which can cause ghost images to be reflected about the
phase centre for strong sources. The FWHM of the effective spectral
resolution resulting from this process was 2.11 channels \citep{killeenMemo1}. {\tt ATLOD} was
also used to discard a number of channels including those centred on harmonics of 128 MHz that
suffered from self-interference, those located near the edges of each frequency window, and every
second channel (which does not result in sensitivity loss because the 4 MHz channels are not
independent). The net result is a total of $23\times8$ MHz channels collectively spanning
1292$-$1484 MHz, with a gap about the 11th harmonic at 1404$-$1412 MHz. Given the small
amount of correlation between these 8 MHz channels (effective channel widths are 8.44~MHz),
in this work we have generally assumed that our channels are statistically independent,
with one exception as described in \S~\ref{ch4:SecInstSubBS}.

For each observational epoch we manually inspected and flagged the primary calibrator data for
radio frequency interference (RFI). We then bandpass-calibrated the secondary calibrator and field
data in preparation for automated RFI removal with {\tt PIEFLAG} \citep{2006PASA...23...64M},
which uses baseline-based statistics derived from a reference channel that is checked to be
minimally affected by RFI. We carried out rms-based flagging on the
calibrator data, and both amplitude and rms-based flagging on the field data.
Shadowed antennas were flagged, limiting projected baselines to $>30$~m.
We then manually inspected the data, removing any residual RFI. On average, $\sim$8\% and
$\sim$20\% of the data were flagged in the 1344 and 1432 MHz frequency windows, respectively.
The resulting net integration times for the CDF-S and ELAIS-S1 fields were 397 and 245
hours, respectively. For completeness, we note that both the Moon and the Sun were separated
by $>42^\circ$ from either ATLAS field throughout their observation; separations from calibrators were
$>36^\circ$ for B1934$-$638, $>20^\circ$ for B0237$-$233, and $>37^\circ$ for B0022$-$423.
Given these large angular separations, we assumed that any influences on our data from
the Moon \citep[e.g. as relevant to polarization:][]{2007R&QE...50..542V,2012RAA....12.1297Z}
or the Sun were negligible.

{\tt MIRIAD} was used to derive and apply the bandpass, complex gain, complex leakage, and
flux density calibrations. Optimised circular polarization calibration was not pursued,
limiting on-axis circular polarization leakage to no better than $V/I\sim0.1\%$ \citep{rayner}.
The ATCA's absolute flux density scale is accurate to within 2\% \citep{reynolds}.
Parallactic angle coverage was sufficient within all epochs
to ensure that leakage solutions could be accurately determined for each antenna. The
misalignment\footnote{To good approximation, the real part of a leakage term corresponds to feed
misalignment (in which the Y feed signal leaks into the X feed), whereas the imaginary part corresponds
to feed ellipticity (in which the Y feed has a finite response to the X feed, seen with a phase lag of
90\degree) \citep{saultMemo1}.} and ellipticity were found to be small (magnitude
{\footnotesize $\lesssim$~}$8\times10^{-3}$) and stable (rms
{\footnotesize $\lesssim$~}$10^{-3}$) for each antenna over the course of 4 years
of ATLAS observing. The data were then split into individual pointings in preparation
for both multi-frequency synthesis and per-channel imaging.

\subsection{Multi-Frequency Synthesis Imaging}\label{ch4:SecDatSubMFS}

In this section we describe the production of mosaics for the CDF-S and ELAIS-S1 ATLAS DR2 fields
in total intensity (Stokes $I$) and circular polarization (Stokes $V$), whereby continuum
images for each individual pointing were created using multi-frequency synthesis (MFS),
deconvolved using {\tt MIRIAD}'s multi-frequency cleaning routine {\tt MFCLEAN},
primary beam corrected, and then linearly mosaicked. We term this the MFS approach in order to
differentiate it from the per-channel approach described in \S~\ref{ch4:SecDatSubPCI}. We note that
the volume of ATLAS data prevented joint deconvolution of all pointings simultaneously.

We set a common pixel size of 1\arcsec\ for all ATLAS DR2 images. This size was
limited by the computational capability of {\tt MFCLEAN} to respond to strong sources significantly
beyond the primary beam of some pointings. We explored a range of weighting schemes, balancing the
trade-off between beam characteristics and sensitivity, selecting superuniform and uniform
weighting for the CDF-S and ELAIS-S1 pointings, respectively.

For each pointing we first lightly cleaned the Stokes $I$ image with {\tt MFCLEAN} to extract model
components with surface brightness {\footnotesize $\gtrsim$}~2 mJy~beam$^{-1}$ (higher for pointings
containing strong sources). We then used these components to correct for residual phase errors
by applying one iteration of phase self-calibration to the data in each of the 1344 and 1432 MHz
observing windows, assuming frequency-independent corrections within each window.
The self-calibration solution interval was selected to be 3~minutes, allowing for sufficient
time to accumulate statistics in relatively faint pointings.
Typical rms values for the variations in the resulting phase corrections were found to be $9.5^\circ$ and
$7.0^\circ$ for pointings in the CDF-S and ELAIS-S1 fields, respectively. Using the corrected phases,
the Stokes $I$ data for each pointing were re-imaged and re-cleaned. The Stokes $V$ data were then
imaged using the corrected phases; no cleaning was required.

To efficiently clean each Stokes $I$ image we tracked the maximum residual surface brightness,
$r_{max}$, against number of clean iterations, $k$, which roughly displayed a power law decline.
We empirically determined that a robust way to halt the cleaning process (specifically for our
ATLAS data) was to stop when the slope flattened off to
$\Delta log_{10}(r_{max})/\Delta k$~{\footnotesize $\gtrsim$}~$-10^{-4}$~Jy~beam$^{-1}$~iteration$^{-1}$.
Using this approach in an automated manner (and checking the results manually), we were able to clean deeply
enough in each pointing to ensure that sidelobes from strong sources were below the thermal noise, but
shallow enough to prevent the cleaning of noise and the development of significant clean bias. Approximately
2500$-$5000 iterations were performed per Stokes $I$ image, dependent on how many bright sources
were visible. We checked the resultant images for clean bias, finding no significant surface brightness
attenuation, as discussed further in \S~\ref{ch4:SecDatSubCB}.

For each pointing, the clean components were convolved with a Gaussian fit to the dirty beam
(i.e. the `native' pointing resolution), as calculated by the {\tt MIRIAD} task {\tt RESTOR}, and added to the residuals
to produce an image. We did not set a common FWHM for all pointings in the {\tt RESTOR} step because that would
have decoupled the resolution of the cleaned sources from the resolution of the noise, rendering any
subsequent image analysis statistically compromised.
Each Stokes $I$ and $V$ image was then convolved with the task {\tt CONVOL} to a common resolution, chosen
to be no better than the worst resolution of all the pointings within each ATLAS field (see
also discussion in \S~\ref{ch4:SecDatSubPCI}). The final resolutions of all MFS CDF-S and ELAIS-S1
pointing images were 13\farcsecd0$\times$6\farcsecd0 and 9\farcsecd6$\times$7\farcsecd6,
respectively, each with position angle $0^\circ$ (North).

A spatial map of rms noise was produced for each pointing image, as described in \S~\ref{ch4:SecDatSubNoise},
from which an average observational rms noise value for each pointing was obtained. The pointings for
each respective ATLAS field were then primary beam corrected and linearly mosaicked, weighting
each pointing by the inverse of its average observational noise variance. The use of observational noise
values, as opposed to predicted theoretical values, enabled us to take into account the decreased
sensitivity in pointings containing difficult-to-clean strong sources beyond the primary beam,
as well as variations in the degree of data flagging, in order to produce optimally sensitive
mosaics. The resulting Stokes $I$ and $V$ mosaics were then regridded from the ATCA's native north-celestial-pole
(NCP) projection into a zenithal equal-area (ZEA) projection \citep{2002A&A...395.1077C} in
preparation for source extraction. For clarity, we denote these ZEA mosaics $I_{\trm{\tiny MFS}}$
and $V_{\trm{\tiny MFS}}$, respectively.

Noise properties of $I_{\trm{\tiny MFS}}$ and $V_{\trm{\tiny MFS}}$ are described in \S~\ref{ch4:SecDatSubNoise}. The use
of {\tt MFCLEAN} and 1\arcsec\ pixels significantly improved image fidelity in the CDF-S
field in comparison with DR1, particularly in reducing sidelobes about a strong $\sim$1~Jy
source (PKS~B0326$-$288) in the south-west of the field.

\subsection{Per-Channel Imaging}\label{ch4:SecDatSubPCI}

In this section we discuss the production of Stokes $I$, $Q$, and $U$ mosaics in each
of the 23 frequency channels for the two ATLAS fields (20 pointings in ELAIS-S1 and 39
pointings in CDF-S), obtained by imaging, primary beam correcting, and linearly
mosaicking a total of $3\times23\times(20+39)=4071$ individual images. We term this
the per-channel (PC) approach. As with the MFS approach, the volume of ATLAS data
prevented joint deconvolution of all pointings simultaneously in each frequency channel.

The frequency-independent gain solutions from the MFS 1344 and 1432 MHz self-calibration
process were applied to the channel data in each respective frequency window for each pointing. We explored a range of
suitable weighting schemes, checking that the central core of the dirty beam could be appropriately
modelled with a Gaussian\footnote{Natural weighting tended to produce beams with central plateaus that were
non-Gaussian in appearance, due to the prevalence of short $uv$-spacing data. To approximate such
beams with a Gaussian would have been inappropriate, and would have rendered overly complex any
subsequent attempts to clean and eventually measure flux densities from the images.}. We selected
near-natural weighting with a robustness parameter of $-0.25$ to optimise the dirty beam,
applying this weighting scheme to all pointings in both ATLAS fields.

We cleaned the Stokes $I$, $Q$, and $U$ images for each pointing with {\tt CLEAN}\footnote{{\tt MIRIAD}'s
implementation of {\tt CLEAN} takes into account sources with negative surface brightness.}, following the
procedure outlined in \S~\ref{ch4:SecDatSubMFS}. We note that our cleaning approach avoided
the need to set a clean cutoff related to the theoretical noise in each pointing; the theoretical
noise could have easily decoupled from the true noise in those pointings in which strong sources were
present, or for which significant data-flagging had been carried out. Approximately 1000
iterations were performed per Stokes $I$ image. Approximately 400 iterations were performed
per Stokes $Q$ or $U$ image. We checked the resultant images for the effects of clean bias,
finding no significant impact, as discussed further in \S~\ref{ch4:SecDatSubCB}.

For the PC images in each pointing, the clean components were convolved to the pointing's
native resolution and added to the residuals. Each image was then convolved to the worst
resolution of any other image at any frequency within each respective ATLAS field. These two convolution
steps ensured that the final images of all pointings in all channels had the same resolution,
taking into account both the differing wavelength and $uv$-plane coverage (due to RFI
flagging) in each channel. The final resolutions of all PC CDF-S and ELAIS-S1 images were
14\farcsecd6$\times$5\farcsecd4 and 10\farcsecd6$\times$6\farcsecd2, respectively, both
with position angle $0^\circ$ (North). Mosaics of Stokes $I$, $Q$, and $U$ in ZEA projection
were then produced for each frequency channel for the two ATLAS fields, incorporating the
same procedure to weight each constituent image by the inverse of their average observational
noise variance as described earlier in \S~\ref{ch4:SecDatSubMFS}. For each of these resultant PC
mosaics, which we denote $I_{\ms i}(x,y)$, $Q_{\ms i}(x,y)$ and $U_{\ms i}(x,y)$ for
the $i$'th channel over spatial pixels $(x,y)$, we computed a spatial rms noise map,
as described in \S~\ref{ch4:SecDatSubNoise}. In subsequent discussion we will typically
drop the pixel $(x,y)$ notation, unless required for clarity.

To illustrate the importance of the second convolution step described above, we note that
the ratio between native beam volumes for images at either end of the observed frequency
range in the CDF-S was 1.7. Attempting to combine such native images
for subsequent analysis (e.g. as required of channel mosaics in \S~\ref{ch4:SecDatSubRM})
would bias all measurements of integrated surface brightness.
The two convolution steps were therefore critical for maintaining statistical control over
the final mosaics.

\subsection{Clean Bias}\label{ch4:SecDatSubCB}

Clean bias is a deconvolution effect that redistributes surface brightness from real sources
to noise peaks, systematically reducing the observed surface brightness of sources independent
of their signal-to-noise ratio (SNR) \citep{1998AJ....115.1693C}. The effect is worse for
observations with poor $uv$-coverage due to increased sidelobe levels. Despite our good
$uv$-coverage, we have checked for clean bias in our MFS and PC data processing to ensure
that ATLAS sources have errors dominated by noise and not bias.

We injected 190 point sources with
SNRs ranging from 5$\sigma$ to 100$\sigma$ at random positions into the $uv$-data for a representative
sample of ATLAS pointings. The data were then imaged and cleaned following both the MFS and PC approaches.
For each injected source we compared the peak surface brightness with the input flux density,
repeating multiple times to accumulate statistics. We found that our implemented
cleaning strategy produced no discernible clean bias; we measured differences between input
and output peak surface brightness of $0\pm 5~\mu$Jy~beam$^{-1}$ in the MFS Stokes $I$ approach (image
rms $\approx 30 \mu$Jy~beam$^{-1}$) and $ 0 \pm 35~\mu$Jy~beam$^{-1}$
in the PC Stokes $I$ approach (image rms per channel $\approx 160 \mu$Jy~beam$^{-1}$). We found that the number
of clean cycles would need to increase by a factor of $\sim$10 to induce a clean
bias of {\footnotesize $\gtrsim$}5\% for a 5$\sigma$ source \citep[e.g. see results of DR1
clean bias calculations in Fig.~3 of][]{2008AJ....135.1276M}.

To examine the potential effects of clean bias on polarization position angles
\citep[e.g. see][]{2008MNRAS.385..274B}, we injected 40 sources with a range of SNRs into a
representative sample of PC Stokes $Q$ and $U$ $uv$-data. For each injected source we specified
a linearly polarized flux density and a random position angle. The Stokes $Q$ and $U$ data were then imaged and
cleaned per the PC approach, combined in quadrature, and corrected for Ricean bias using the
first-order scheme described by \citet{vla161}. We then compared both the output peak
polarized surface brightness and position angle with the input values for each source,
repeating the entire test multiple times. We found that neither statistic displayed
significant clean bias; we measured differences between input and output peak linearly
polarized surface brightness of $0 \pm 30~\mu$Jy~beam$^{-1}$ (image rms $\approx 120
\mu$Jy~beam$^{-1}$) and found no discernible tendency for position angles to be oriented toward multiples
of 45\degree. We found that in order to induce discernible clean bias in linear
polarization, approximately 50 times more clean cycles than originally implemented were required;
an additional factor of 50 times more cycles were required to produce a clear position angle bias.

\subsection{Rotation Measure Synthesis}\label{ch4:SecDatSubRM}

We used rotation measure (RM) synthesis \citep{2005A&A...441.1217B} and RM clean \citep{2009A&A...503..409H}
to produce a map of linearly polarized emission for each ATLAS field, processing the Stokes
$Q_{\ms i}(x,y)$ and $U_{\ms i}(x,y)$ mosaics and their associated rms noise maps,
$\sigma_{\ms Q,i}(x,y)$ and $\sigma_{\ms U,i}(x,y)$, for all 23 spectral channels.
For each spatial pixel, we weighted the spectral data by their combined variance,
$\sigma_{\ms Q,U,i}^{\ms 2}(x,y)$, which we calculated according to
Equations~(A2)$-$(A3) from \citet{2012MNRAS.424.2160H}; our data are always consistent
with $0.8 < \sigma_{\ms Q,i}(x,y)/\sigma_{\ms U,i}(x,y) < 1.2$.

\begin{figure}
 \centering
 \includegraphics[trim=45mm 35mm 120mm 100mm,width=75mm]{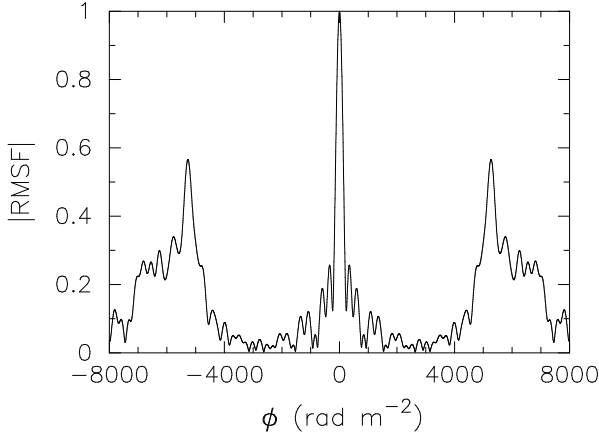}
 \caption{Absolute value of the RMSF versus Faraday depth for a typical spatial pixel in ATLAS,
 	  corresponding to the weighted spectral coverage of observations along that sightline.
	  The observed FWHM of the main peak is 293~rad~m$^{\ms -2}$.}
 \label{ch4:fig:rmsf}
\end{figure}

In implementing RM synthesis we sampled the Faraday dispersion function at each spatial pixel,
$F(x,y,\phi)$, in steps of 5 rad~m$^{\ms -2}$ between Faraday depths $-4000<\phi<4000$
rad~m$^{\ms -2}$. This range was selected to ensure sensitivity up to the maximum scale
afforded by our spectral resolution;
$\phi_{\trm{\scriptsize max}} \approx \sqrt{3}/\!\min [ \delta(\lambda^{\ms 2}_{\ms i}) ] = 3900$~rad~m$^{\ms -2}$,
where $\delta(\lambda^{\ms 2}_{\ms i})$ is channel width in wavelength-squared
space, $\lambda^{\ms 2}$, for the $i$'th channel [see Equation~(63)
from \citealt{2005A&A...441.1217B}]. A typical rotation measure sampling function (RMSF)
for our data is shown in Fig.~\ref{ch4:fig:rmsf}.
The main peak has a measured FWHM of $\delta\phi=293$~rad~m$^{\ms -2}$ with sidelobes
of order 25\% and grating lobes of order 55\%. Measured FWHM's for all spatial pixels
are $293\pm0.2$~rad~m$^{\ms -2}$; the spread reflects the slightly different spectral
weighting used to process each spatial pixel. For comparison, the
theoretical value of $\delta\phi$ obtained by assuming uniform spectral weighting is
265~rad~m$^{\ms -2}$ [see Equation~(61) from \citealt{2005A&A...441.1217B}]. Strong
grating lobes are present beyond $\pm5000$~rad~m$^{\ms -2}$. Given our spectral
coverage, our data are insensitive to Faraday thicknesses greater than
$\sim$76~rad~m$^{\ms -2}$
[$\tnm{max-thickness}\approx\pi/\lambda_{\trm{\scriptsize min}}^{\ms 2}$, where
$\lambda_{\trm{\scriptsize min}}$ is the shortest wavelength observed; see
Equation~(62) from \citealt{2005A&A...441.1217B}]. Therefore, our Faraday
spectra are only sensitive to unresolved RM components.

We did not correct our data for ionospheric Faraday rotation. We note that the ionosphere
will typically produce a RM that varies between approximately $+$0.2 and $+$1.0
rad~m$^{\ms -2}$ for ATCA observations at zenith and at the array's elevation limit (12\degree),
respectively \citep{2008AdSpR..42..599B,2010GeoJI.183.1216F}. RM fluctuations about these mean
values due to ionospheric density variations are typically $\sim$0.5~rad~m$^{\ms -2}$. Given
the FWHM of our RMSF, and to some extent the phase self-calibration applied to the data, we
assume that the influence of ionospheric Faraday rotation on both measured RMs and potential
depolarization is negligible.

For each ATLAS field we constructed a map of linearly polarized emission, which we
denote $L_{\trm{\tiny RM}}(x,y)$, by applying a 3-point parabolic (3PP) fit to extract the
fitted peak polarized surface brightness from within the cleaned Faraday dispersion
spectrum for each spatial pixel, $F^{\trm{\scriptsize cleaned}}(x,y,\phi)$, namely
\begin{equation}\label{ch4:eqn:LrmDEF}
	L_{\trm{\tiny RM}}(x,y)\equiv\tnm{3PP-fit-max}\left[\,\left|F^{\trm{\scriptsize cleaned}}(x,y,\phi)\right|\,\right] \,.
\end{equation}
RM cleaning \citep{2009A&A...503..409H} was performed down to a level of $4.4\sigma_{\trm{\tiny RM}}(x,y)$
[Gaussian equivalent SNR of $4\sigma$; see Equation~(15) from \citealt{2012MNRAS.424.2160H}],
where $\sigma_{\trm{\tiny RM}}(x,y)$ is the rms noise at each spatial pixel in $L_{\trm{\tiny RM}}(x,y)$.
$\sigma_{\trm{\tiny RM}}(x,y)$ was calculated by combining $\sigma_{\ms Q,i}(x,y)$ and
$\sigma_{\ms U,i}(x,y)$ from each spectral channel according to
Equations~(20)$-$(23) from \citet{2012MNRAS.424.2160H}. Properties of $\sigma_{\trm{\tiny RM}}(x,y)$
for each ATLAS field are presented in \S~\ref{ch4:SecDatSubNoise}.
We note that for each pixel, $L_{\trm{\tiny RM}}(x,y)$ was sampled from
$M\equiv2\phi_{\trm{\scriptsize max}}/\delta\phi\approx 28$ independent measurements.
The non-Gaussian statistics exhibited by $L_{\trm{\tiny RM}}(x,y)$, taking into account
the value of $M$, are discussed by \citet{2012MNRAS.424.2160H}.

We chose to represent the polarized emission at each pixel by Equation~(\ref{ch4:eqn:LrmDEF})
for two key reasons. First, our data are insensitive to resolved sources in Faraday depth space,
enabling us to represent integrated measurements of surface brightness per unit $\phi$ by peak
measurements of surface brightness per unit $\phi$ for any Faraday component. Second, we do not
expect to find many polarized sources with multiple Faraday components that are separated
in Faraday space by more than the FWHM of our RMSF. For example, \citet{2011AJ....141..191F}
found that less than\footnote{For this comparison we neglect components from \citet{2011AJ....141..191F}
with SNR~$<$~6 because the statistical significance of such polarization detections drop
below the Gaussian equivalent of 5$\sigma$; see \citet{2012MNRAS.424.2160H} with $M\approx70$
as relevant to their data.}
5\% of polarized 1.4~GHz sources consisted of multiple RM components separated by more than $\sim$280
rad~m$^{\ms -2}$, when observed with high resolution in Faraday depth space. We therefore assumed that,
even in cases where multiple Faraday components may be present within the width of one RMSF, the total
polarized emission in Faraday space for our data could be approximated by the dominant peak.
Examination of $F^{\trm{\scriptsize cleaned}}(x,y,\phi)$ for our data revealed this to be a suitable
approximation (no lines of sight with multiple RM components were detected), though we note that we
did not attempt to compare the widths of Faraday components with the width of the RMSF.

\begin{figure}
 \centering
 \includegraphics[trim = 20mm 20mm 20mm 50mm, width=82mm]{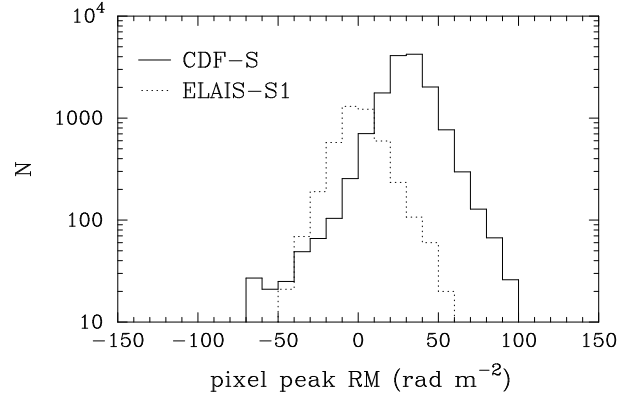}
 \caption{Distribution of RMs for all pixels with $L_{\trm{\tiny RM}}(x,y)/\sigma_{\trm{\tiny RM}}(x,y)\ge7$.
	  This ratio (which is not used elsewhere in this work) is chosen to be higher than the source
	  detection threshold defined in \S~\ref{ch4:SecExtSubFlood} so as to avoid most contamination
	  by spurious high-RM pixels associated with obvious image artefacts. The higher pixel count
	  and increased width of the CDF-S distribution relative to the ELAIS-S1 distribution is due
	  to the increased presence of strong artefacts in the CDF-S field, such as those about the
	  1~Jy source PKS~B0326$-$288. Note that the y-axis is logarithmic.}
 \label{ch4:fig:rm}
\end{figure}

Typical RMs for all significant pixels in the CDF-S and ELAIS-S1 polarization images were found to be
{\footnotesize $\lesssim$}~$70$~rad~m$^{\ms -2}$ in magnitude, as indicated in Fig.~\ref{ch4:fig:rm}.
Detailed analysis of the RM properties of ATLAS sources is beyond the scope of this work
and will be presented in a future ATLAS data release.

We examined the data for RM clean bias (i.e. the Faraday space analogue of clean bias described
in \S~\ref{ch4:SecDatSubCB}) by manually inspecting the locations of RM clean components.
Qualitatively, we did not find any misplaced components for the vast majority of spatial pixels,
indicating negligible RM clean bias. We note that clean bias due to grating lobes could only
be produced by sources with $|\trm{RM}|>1000$~rad~m$^{\ms -2}$; no such sources were found
in our data.

In parallel with RM synthesis, we also assembled a channel-averaged Stokes $I$ mosaic,
denoted by $I_{\trm{\tiny CA}}$, to be used for correcting the $L_{\trm{\tiny RM}}$
mosaics for spurious instrumental polarized emission (see \S~\ref{ch4:SecExtSubFlood}).
The $I_{\trm{\tiny CA}}$ mosaic was formed by stacking the PC Stokes $I$ mosaics,
$I_{\ms i}(x,y)$, with weighting factors identical to those used to form
$L_{\trm{\tiny RM}}(x,y)$, namely
\begin{equation}\label{ch4:eqn:PCIstack}
	I_{\trm{\tiny CA}}(x,y) = \left[ \sum_{\ms i=1}^{\ms 23}
			  I_{\ms i}(x,y)/\sigma_{\ms Q,U,i}^{\ms 2}(x,y) \right]
			  \left[ \sum_{\ms i=1}^{\ms 23} 1/\sigma_{\ms Q,U,i}^{\ms 2}(x,y)
			  \right]^{\ms \!-1} .
\end{equation}
Noise properties of the $I_{\trm{\tiny CA}}$ mosaic are described in \S~\ref{ch4:SecDatSubNoise}.
We do not use the $I_{\trm{\tiny CA}}$ mosaics for radio component extraction,
even though they contain regions with rms noise levels less than those
in the $I_{\trm{\tiny MFS}}$ mosaics, because they contain disruptive sidelobes from
strong sources interspaced between the optimal low noise regions.

\subsection{Noise Distribution in Images}\label{ch4:SecDatSubNoise}

We used the SExtractor package \citep[v.~2.5.0;][]{1996A&AS..117..393B,2005astro.ph.12139H}
to map spatial variations in rms noise across all channel, pointing, and mosaicked
images of Stokes $I$, $Q$, $U$, and $V$.
As outlined in \S~\ref{ch4:SecDatSubRM}, maps of $\sigma_{\trm{\tiny RM}}$ for our $L_{\trm{\tiny RM}}$
mosaics were produced by combining $\sigma_{\ms Q,i}$ and $\sigma_{\ms U,i}$ for each
spectral channel following the equations presented by \citet{2012MNRAS.424.2160H}.

SExtractor calculates the rms noise at each spatial pixel in an image by analysing the
distribution of pixel values within a local background mesh, taking into account not
only local variations in image sensitivity, but also the possible presence of DC offsets
due to artefacts (e.g. sidelobes). Following Equation~(3) from \citet{2012MNRAS.425..979H},
we set the mesh size for each image analysed to the area enclosed by $N_{\ms b}=150$ independent
resolution elements. Uncertainties in our estimates of local rms noise are therefore
$\{[1+0.75/(N_{\ms b}-1)]^{\ms 2}[1-1/N_{\ms b}]-1\}^{\ms 0.5}=6\%$
\citep[using an approximation to the variance of the standard error estimator,
suitable for $N_{\ms b}>10$; p.~63,][]{johnson}.

\begin{figure*}
 \centering
 \includegraphics[trim = 10mm 50mm 10mm 40mm, height=62mm, angle=-90]{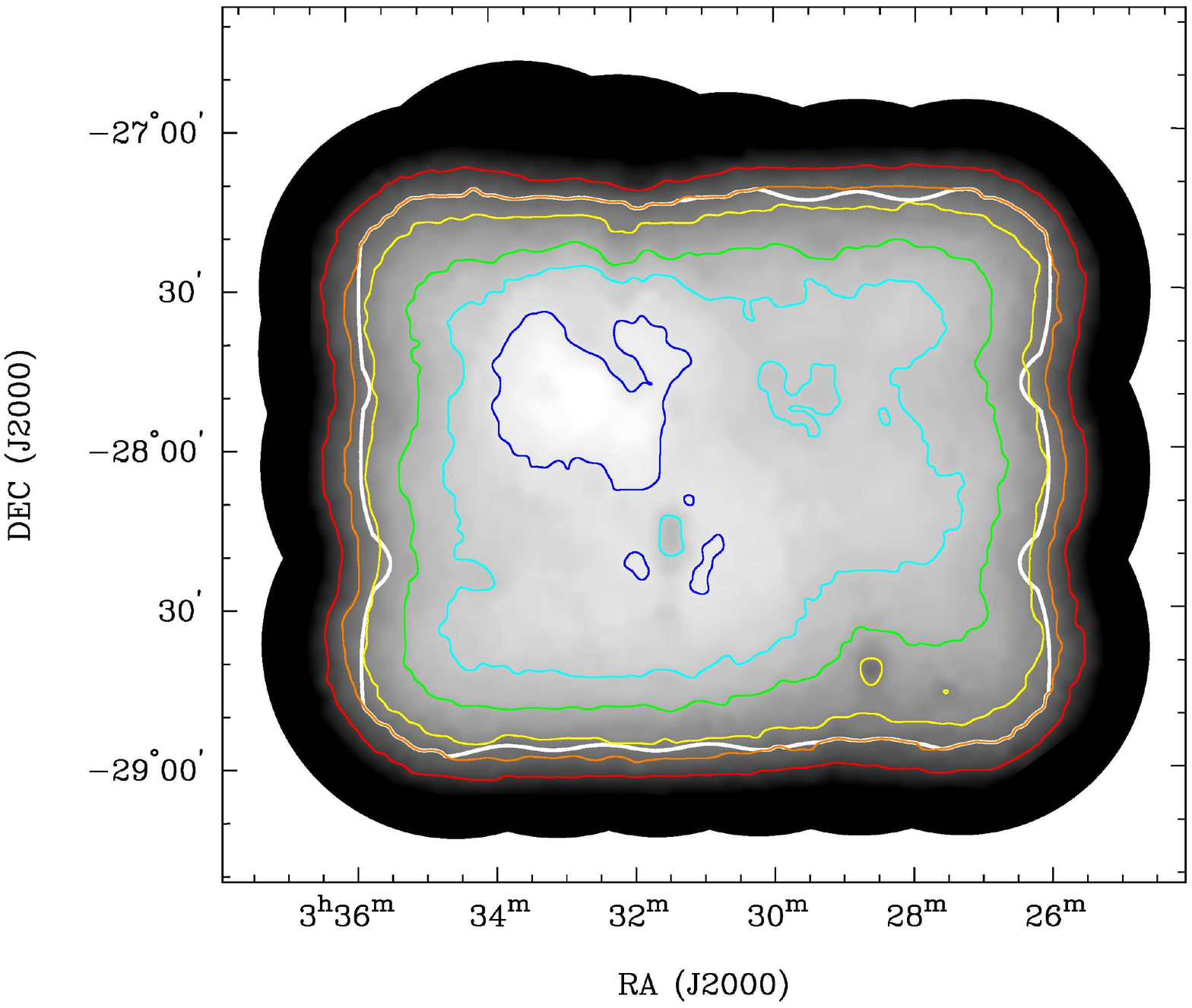}
 \hspace{5pc}
 \includegraphics[trim = 10mm 50mm 10mm 40mm, height=62mm, angle=-90]{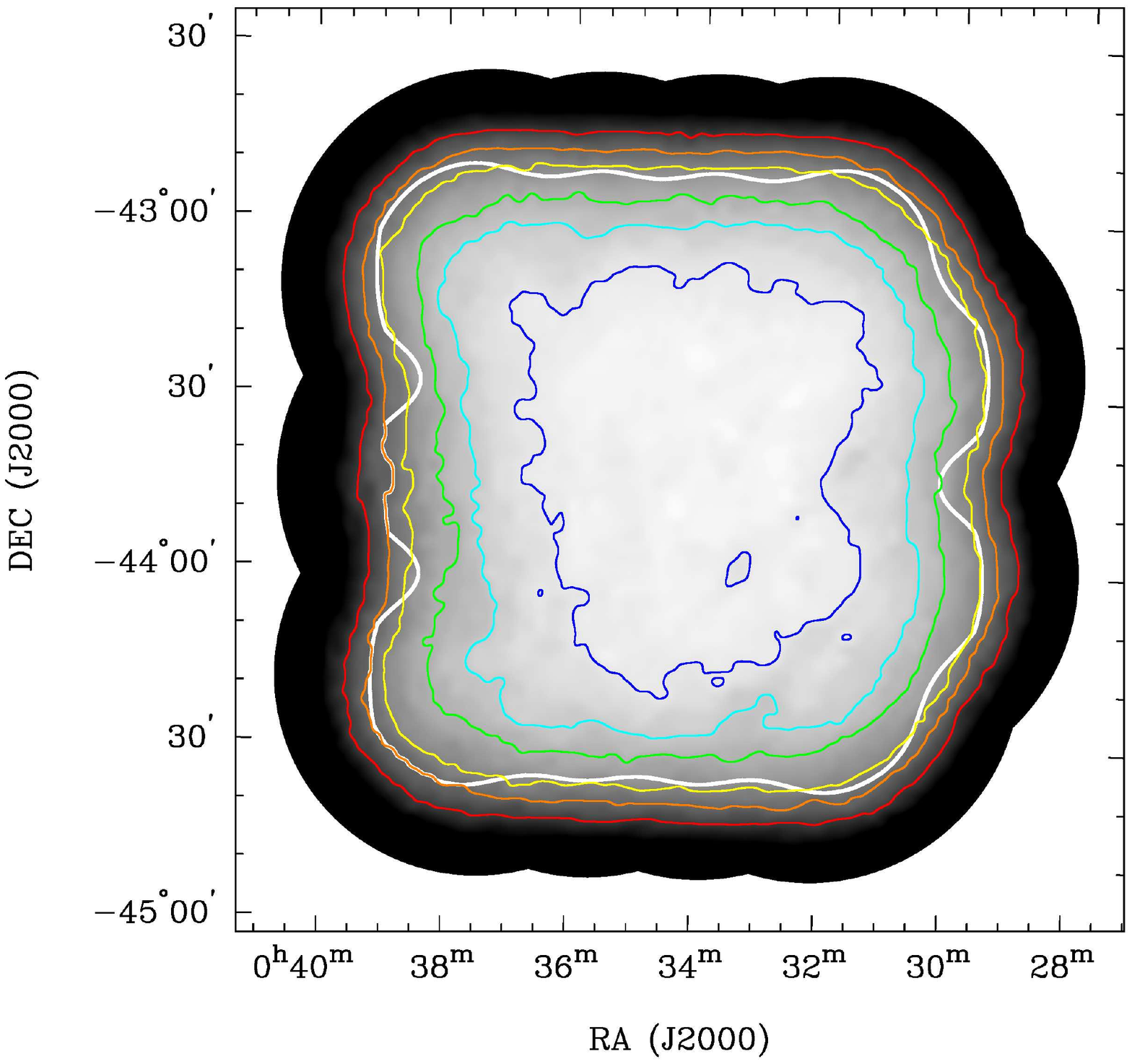}\vspace{0.5pc}\\
 \includegraphics[trim = 10mm 50mm 10mm 40mm, height=62mm, angle=-90]{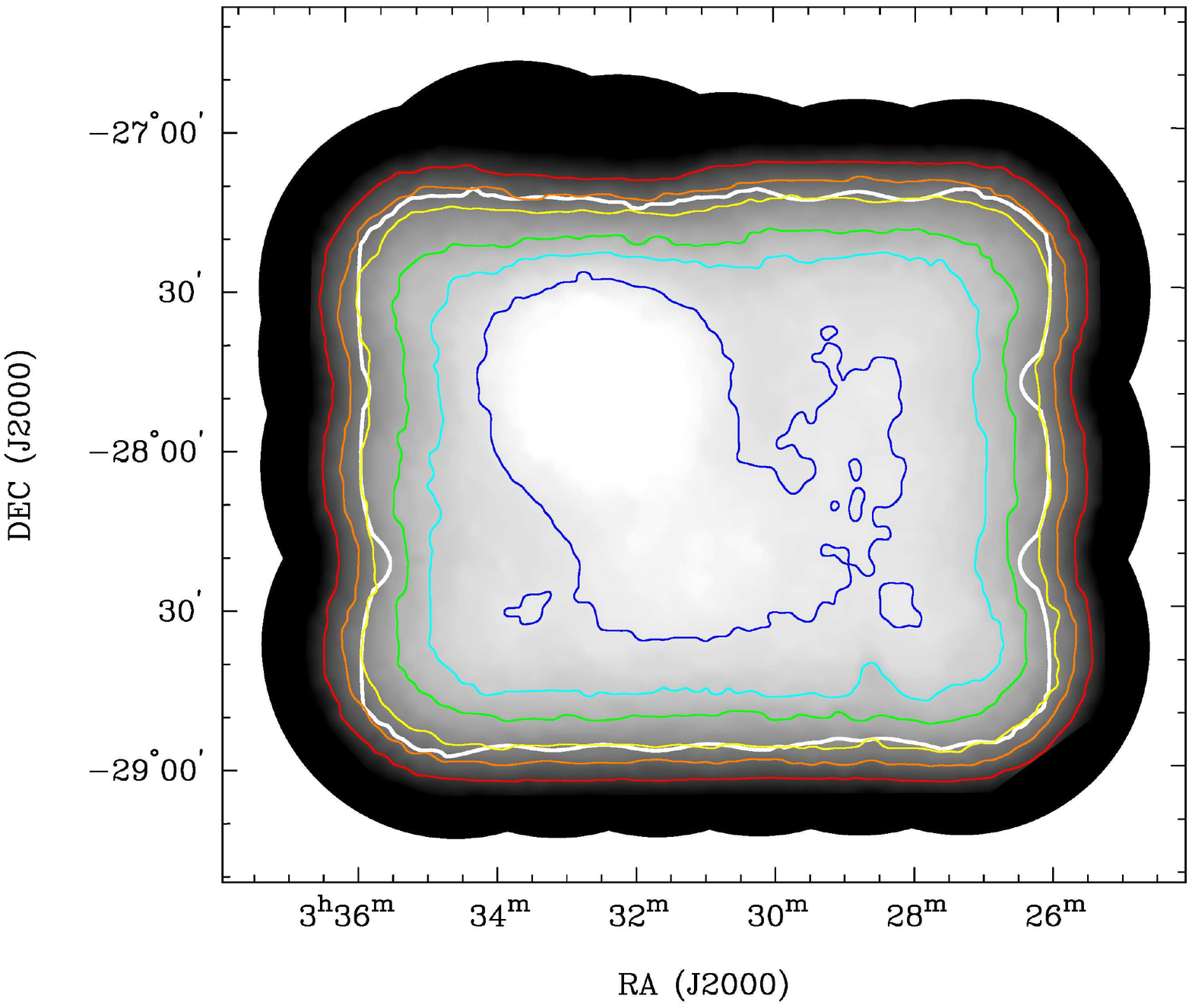}
 \hspace{5pc}
 \includegraphics[trim = 10mm 50mm 10mm 40mm, height=62mm, angle=-90]{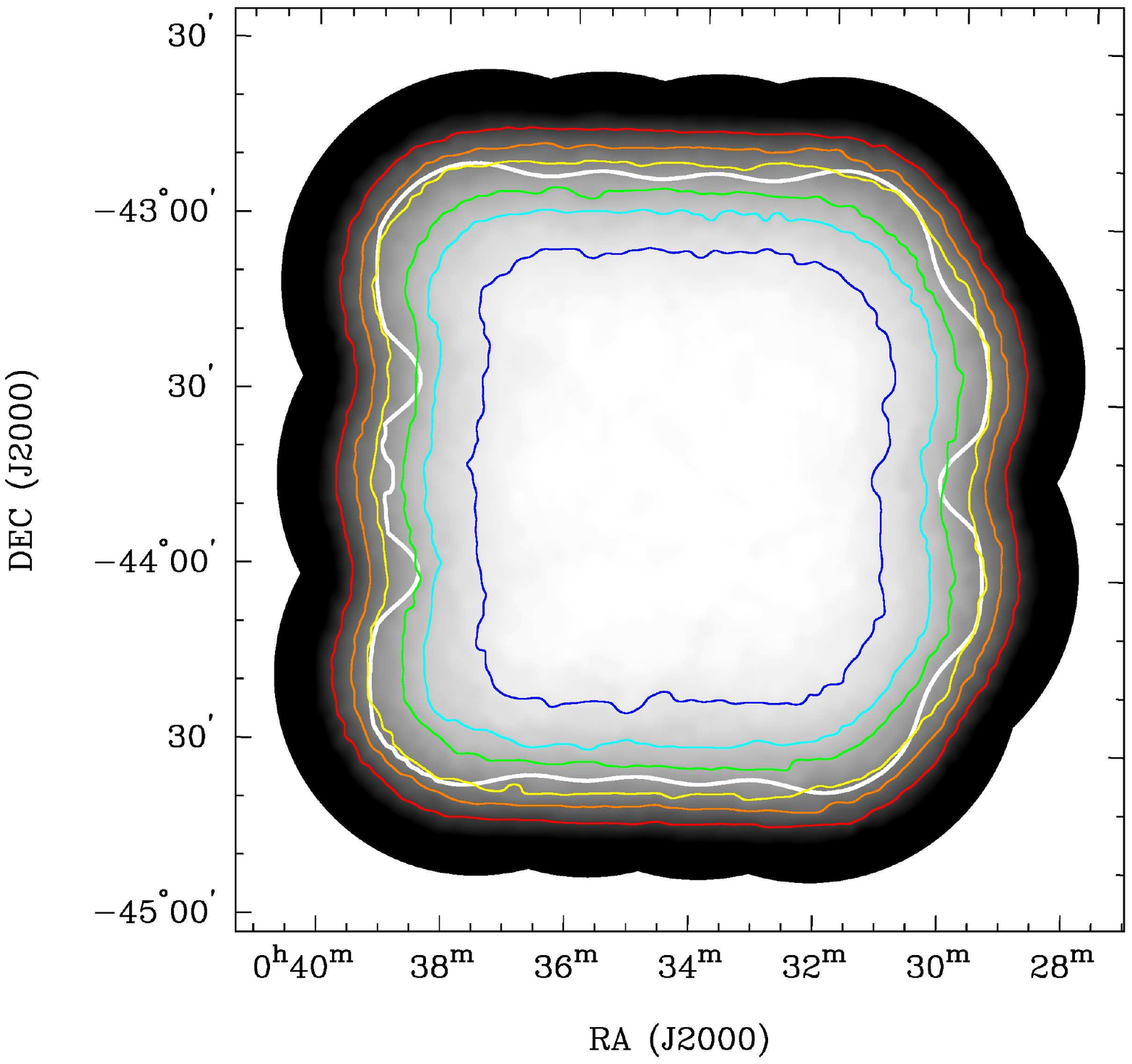}\vspace{0.5pc}\\
 \includegraphics[trim = 10mm 50mm 10mm 40mm, height=62mm, angle=-90]{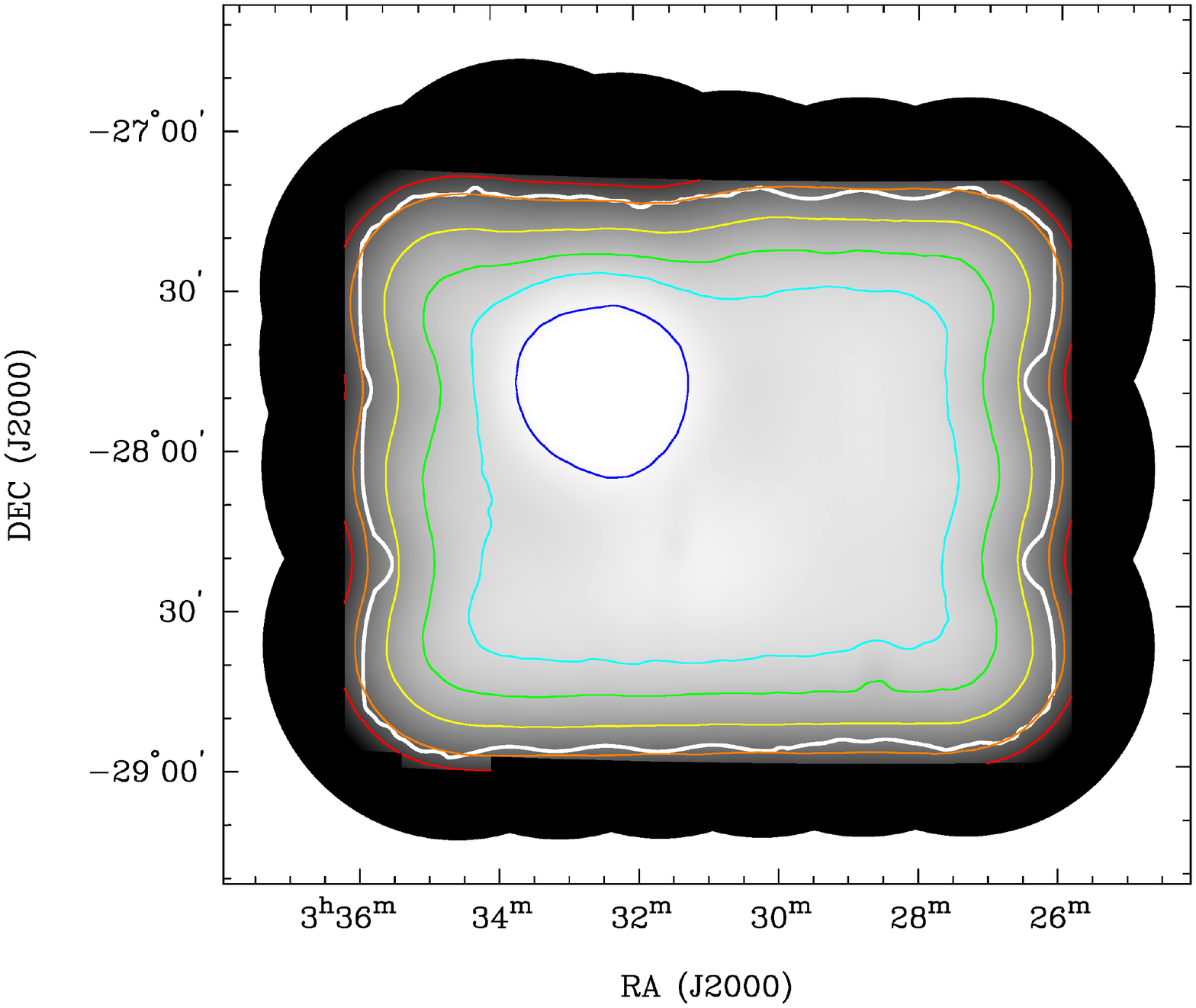}
 \hspace{5pc}
 \includegraphics[trim = 10mm 50mm 10mm 40mm, height=62mm, angle=-90]{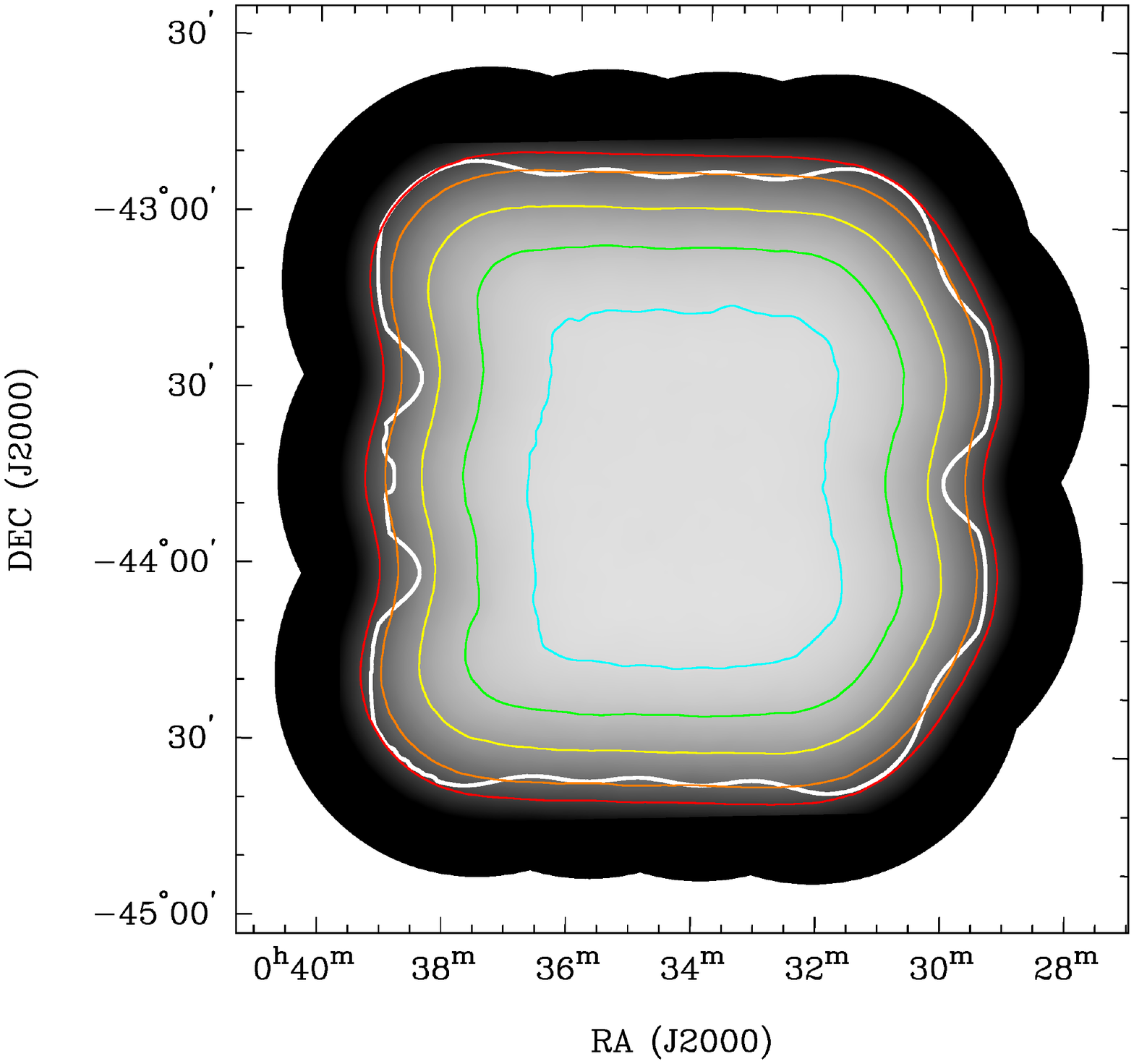}
 \begin{minipage}{140mm}
 \caption{Spatial rms noise maps of the CDF-S (left column) and ELAIS-S1 (right column)
 	  fields for total intensity ($I_{\trm{\tiny MFS}}$; top row), circular polarization
	  ($V_{\trm{\tiny MFS}}$; middle row) and linear polarization ($L_{\trm{\tiny RM}}$;
	  bottom row). Shading levels are scaled equally between panels in each row; only
	  the top two rows are shaded equally. Thin contours in the upper and middle panels
	  indicate rms levels of 30, 40,
	  50, 75, 100, and 150 $\mu$Jy~beam$^{\ms -1}$, as calculated by SExtractor. Thin
	  contours in the lower panels indicate rms levels of 17 (CDF-S panel only), 22, 27,
	  37, 57, and 77 (ELAIS-S1 panel only) $\mu$Jy~beam$^{\ms -1}$, as calculated using
	  a combination of SExtractor and Equations~(20)$-$(23) from \citet{2012MNRAS.424.2160H}.
	  The bold white contours, co-located within the panels in each column, indicate the
	  survey area boundaries (see \S~\ref{ch4:SecInstSubArea}).
}
 \label{ch4:fig:rmsmaps}
\end{minipage}
\end{figure*}

\begin{figure}
 \centering
 \includegraphics[trim=10mm 35mm 30mm 70mm, height=55mm]{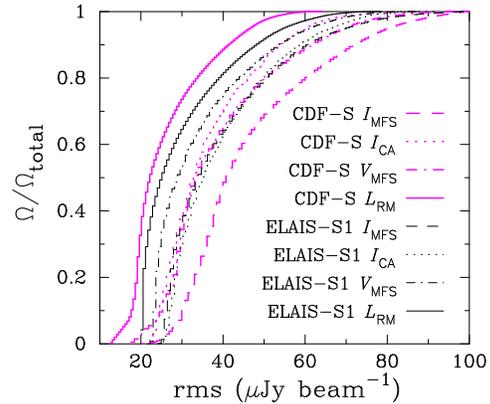}
 \caption{Fraction of sky area in ATLAS survey areas at or below a given rms noise level
	  (calculated from Fig.~\ref{ch4:fig:rmsmaps}).}
 \label{ch4:fig:rms}
\end{figure}

In Fig.~\ref{ch4:fig:rmsmaps} we display rms noise maps for each of the mosaics used
to detect and catalogue radio components, namely $I_{\trm{\tiny MFS}}$, $V_{\trm{\tiny MFS}}$,
and $L_{\trm{\tiny RM}}$; these noise maps are used in \S~\ref{ch4:SecExtSubFlood} to evaluate
local SNRs at any spatial location.
In Fig.~\ref{ch4:fig:rms} we present cumulative histograms of the rms noise distributions
exhibited by each of these mosaics; for completeness, we also include the noise
distributions from the $I_{\trm{\tiny CA}}$ mosaics.
The linear polarization mosaics are both more sensitive, and more uniform in their
sensitivity, than the total intensity images because there are fewer sources and
fewer imaging artefacts in the former.

\section{Instrumental Effects}\label{ch4:SecInst}

In this section we describe three systematic effects $-$ time-average smearing,
bandwidth smearing, and instrumental polarization $-$ and our methods to model
their wide-field behaviours across the ATLAS mosaics. Consideration of 3D-smearing
\citep{1999ASPC..180..383P} is not required because the ATCA is a coplanar array.
We conclude this section by defining the survey area boundary for each ATLAS field.

\subsection{Time-Average Smearing}\label{ch4:SecInstSubTAS}

Time-average smearing is the well-known effect whereby visibilities are smeared in the $uv$-plane
due to the rotation of the sky during a correlator cycle time. The result is a decrement in
the observed peak surface brightness of sources; integrated surface brightnesses are
conserved.

For our correlator cycle time of 10 seconds, we used the theoretical assumptions from
\citet{1999ASPC..180..371B} to estimate a loss in peak flux density of no worse than
1.5\% at the edges of individual pointing images. Consequently, we did not correct for the
marginal degree of time-average smearing in our data.

\subsection{Bandwidth Smearing}\label{ch4:SecInstSubBS}

Bandwidth smearing, or chromatic aberration, is the well-known effect whereby
visibilities are smeared in the $uv$-plane due to the finite bandwidth of receiver
channels. The result is a decrement in the observed peak surface brightness of
sources; this is accompanied by source broadening in a radial direction from the pointing
phase centre, such that integrated surface brightnesses remain conserved.
The bandwidth smearing effect is proportional to the radial offset from the
phase centre in units of projected synthesised beamwidths, and to the
fractional bandwidth $\Delta\nu_{\trm{\scriptsize eff}}/\nu$, where $\nu$ is the
reference frequency for setting delay terms when gridding in the $uv$-plane and
$\Delta\nu_{\trm{\scriptsize eff}}$ is the effective passband width. In the following
we present our prescription for handling bandwidth smearing from a non-circular
beam. While this prescription is trivial, we are unaware of any previous
studies that have accounted for non-circular beams.

For a source
at position angle $\zeta$ East of North with respect to the phase centre, the projected
beam FWHM for an elliptical beam with major axis FWHM $B_\trm{\scriptsize maj}$, minor
axis FWHM $B_\trm{\scriptsize min}$, and position angle $\psi$ East of North is given by
\begin{equation}\label{ch4:eqn:BWSbeam}
	B_{\trm{\scriptsize proj}}(\zeta) = \frac{B_\trm{\scriptsize maj} B_\trm{\scriptsize min}}
		    {\sqrt{\left[ B_\trm{\scriptsize maj}\sin\left(\zeta-\psi\right) \right]^2 + 
		           \left[ B_\trm{\scriptsize min}\cos\left(\zeta-\psi\right) \right]^2}} \;.
\end{equation}
Assuming a Gaussian beam and rectangular passband, the bandwidth smearing effect for an
individual pointing is then given by \citep{1998AJ....115.1693C}
\begin{equation}\label{ch4:eqn:BWS}
	\frac{S_{\trm{\scriptsize peak}}}{S_{\trm{\scriptsize peak}}^{\ms 0}} = 
	\Bigg\{
	1 + \frac{2\ln2}{3}
	\left[ \frac{\Delta\nu_{\trm{\scriptsize eff}}}{\nu}
	\frac{d}{B_{\trm{\scriptsize proj}}(\zeta)} \right]^2
	\Bigg\}^{\ms -\frac{1}{2}} \,,
\end{equation}
where the ratio $S_{\trm{\scriptsize peak}}/S_{\trm{\scriptsize peak}}^{\ms 0}$
represents the peak surface brightness attenuation (smearing) for a source at radial
distance $d$ from the phase centre with respect to an unsmeared source at $d=0$.

\begin{figure*}
\centering
 \includegraphics[trim = 0mm 0mm 0mm 0mm, clip, width=65mm, angle=-90]{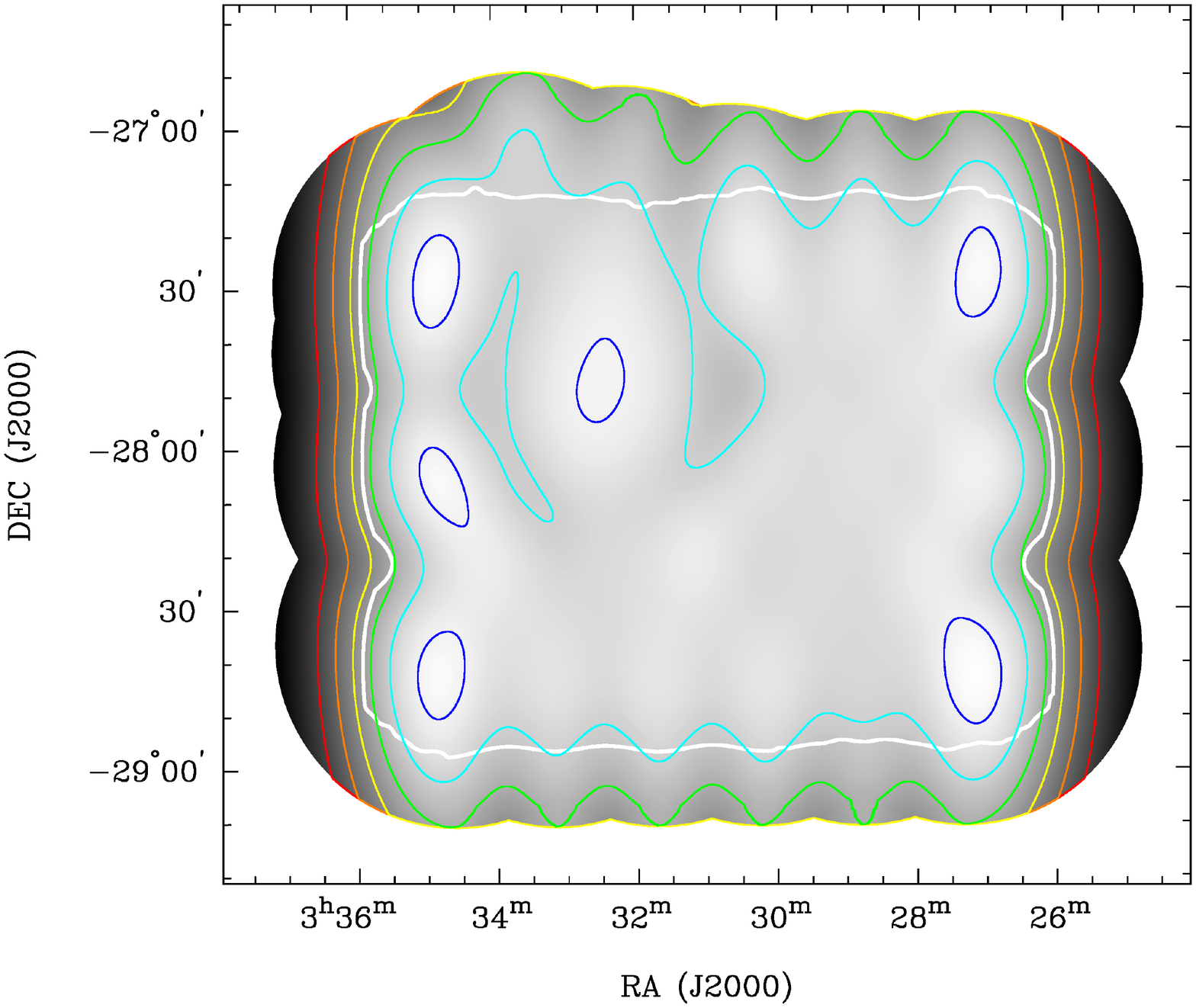}
 \hspace{1pc}
 \includegraphics[trim = 0mm 0mm 0mm 0mm, clip, width=65mm, angle=-90]{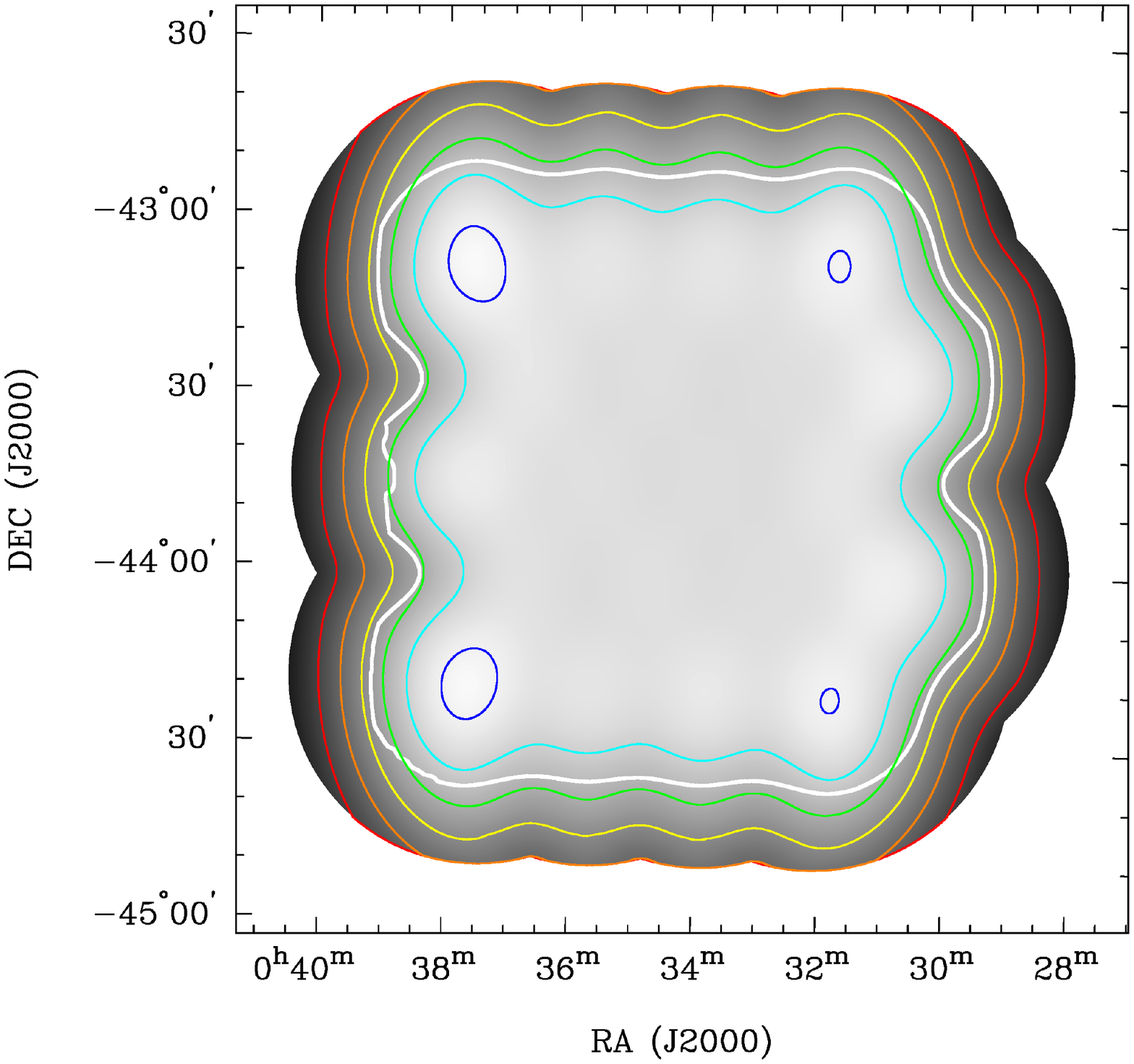}\\
 \includegraphics[trim = 0mm 0mm 0mm 0mm, clip, width=65mm, angle=-90]{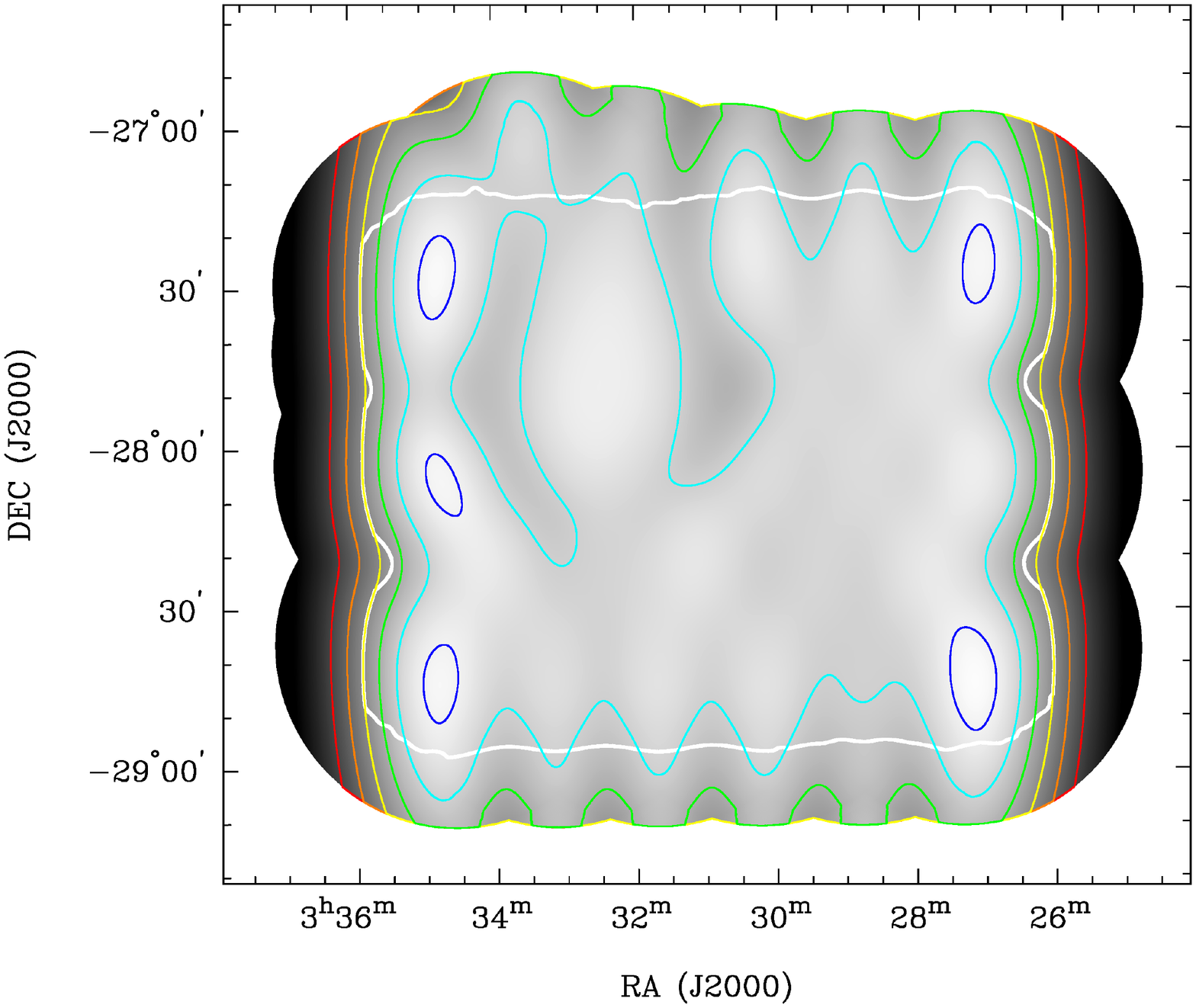}
 \hspace{1pc}
 \includegraphics[trim = 0mm 0mm 0mm 0mm, clip, width=65mm, angle=-90]{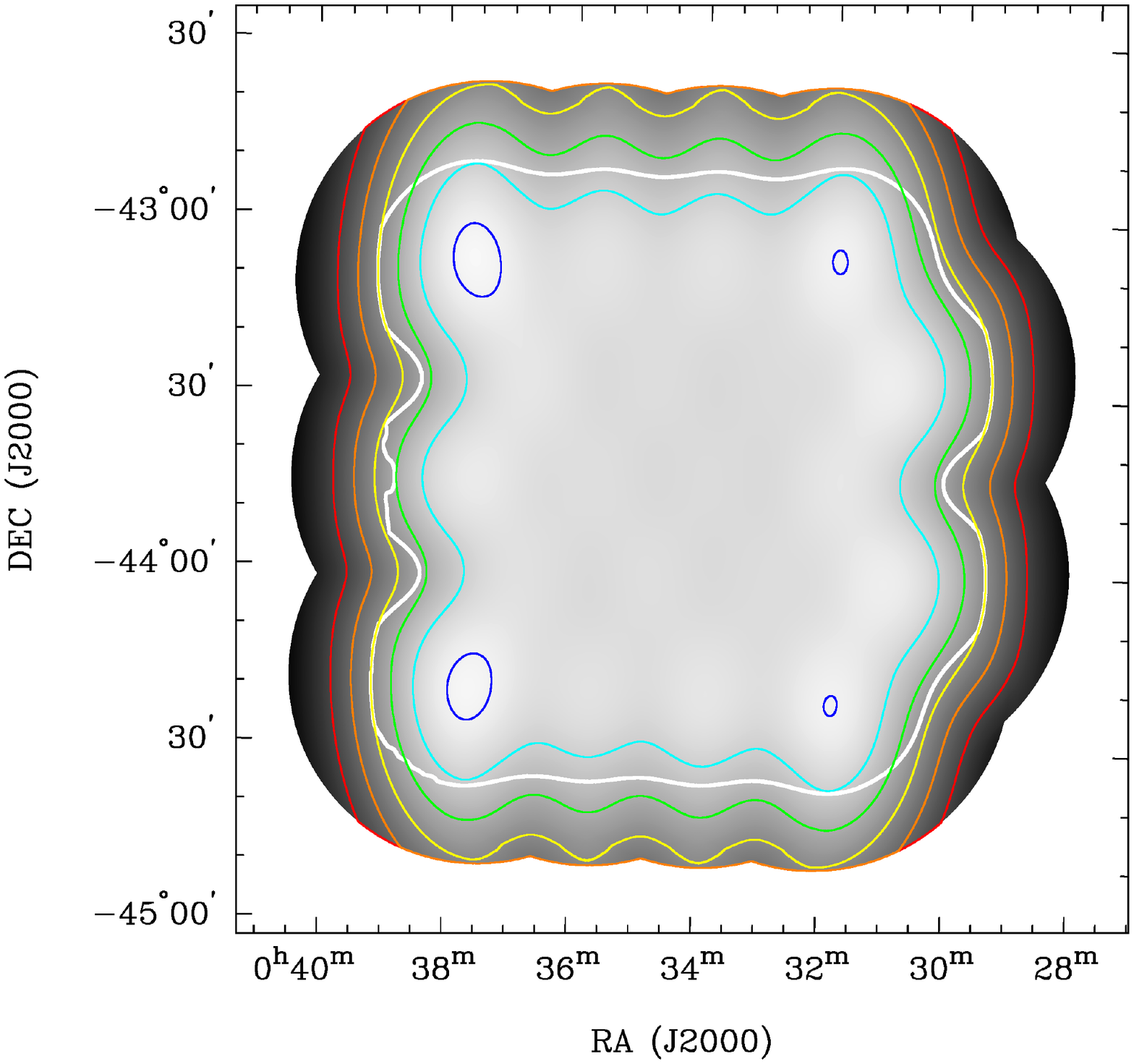}
\begin{minipage}{140mm}
 \caption{Top: Spatial bandwidth smearing maps of the CDF-S (left column) and ELAIS-S1 (right
	  column) fields for total intensity ($I_{\trm{\tiny MFS}}$; top row) and linear polarization ($L_{\trm{\tiny RM}}$;
	  bottom row). Shading levels are identical in each panel. The thin contours
	  indicate peak surface brightness attenuation levels ($\varpi$; see \S~\ref{ch4:SecInstSubBS}) of
	  70\% (outermost), 75\%, 80\%, 85\%, 90\%, and 95\% (innermost).
	  The bold white contours, co-located within the upper and lower panels in
	  each column, indicate the survey area boundaries (see \S~\ref{ch4:SecInstSubArea}).
}
 \label{ch4:fig:bsmaps}
\end{minipage}
\end{figure*}

\begin{figure}
 \centering
 \includegraphics[trim=10mm 35mm 30mm 70mm, height=55mm]{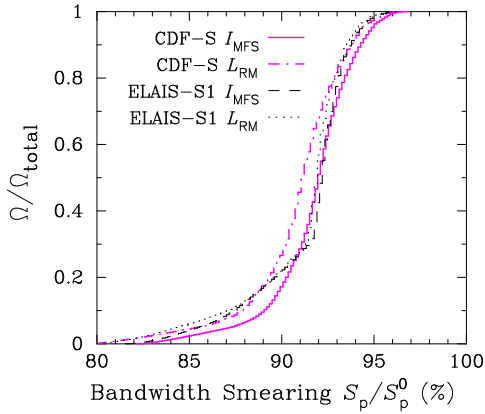}
 \caption{Fraction of sky area in ATLAS survey areas at or below a given bandwidth
	  smearing level (calculated from Fig.~\ref{ch4:fig:bsmaps}).}
 \label{ch4:fig:bs}
\end{figure}

To model the amount of bandwidth smearing at any spatial position within the
$I_{\trm{\tiny MFS}}$ mosaic for each ATLAS field, we first used Equation~(\ref{ch4:eqn:BWS})
with $\Delta\nu_{\trm{\scriptsize eff}}=8.44$~MHz (rather than the nominal 8~MHz; see
\S~\ref{ch4:SecDatSubFlag}) and $\nu=1.387$~GHz to produce simulated images quantifying the smearing
exhibited over individual pointings. We then mosaicked these simulated images together using the
same weighting factors that were used to construct the $I_{\trm{\tiny MFS}}$ mosaics.
We followed a similar procedure to model bandwidth smearing
within the $L_{\trm{\tiny RM}}$ mosaics. First, we modelled the smearing effect
within all Stokes $Q$ and $U$ images for each pointing and channel, using
$\Delta\nu_{\trm{\scriptsize eff}}=8.44$~MHz and setting $\nu$ to each channel's respective
frequency. We then mosaicked all simulated pointing images for each channel together
to produce simulated bandwidth smearing channel mosaics, using the same weighting factors
that were applied to construct the PC Stokes $Q_{\ms i}$ and $U_{\ms i}$ mosaics. Next, we combined
the Stokes $Q$ and $U$ mosaics of simulated bandwidth smearing together within each channel, weighting each
mosaic by the same factors applied to construct $\sigma_{\ms Q,U,i}$ (see \S~\ref{ch4:SecDatSubRM}).
Finally, we stacked these combined channel mosaics together, weighting each channel
by $\sigma_{\ms Q,U,i}$ in the same way that $L_{\trm{\tiny RM}}$ was constructed (see
\S~\ref{ch4:SecDatSubRM}). The resulting mosaics, which map the bandwidth smearing ratio
$\varpi(x,y) \equiv S_{\trm{\scriptsize peak}}(x,y)/S_{\trm{\scriptsize peak}}^{\ms 0}$
over all spatial positions within the CDF-S and ELAIS-S1 $I_{\trm{\tiny MFS}}$ and
$L_{\trm{\tiny RM}}$ mosaics, are presented in Fig.~\ref{ch4:fig:bsmaps}.
In Fig.~\ref{ch4:fig:bs} we present cumulative histograms of the bandwidth smearing
distributions from Fig.~\ref{ch4:fig:bsmaps}.
Bandwidth smearing maps were not required for $V_{\trm{\tiny MFS}}$ (see \S~\ref{ch4:SecExtSubFlood}). 

We checked the accuracy of our individual pointing and combined mosaic bandwidth
smearing solutions by following the procedure outlined by \citet{2000A&AS..146...41P}.
We measured peak and integrated surface brightnesses for a series of strong
point sources visible in multiple overlapping pointings, noting offsets from their respective
pointing centres. We found good agreement between observed and predicted decrements
in peak surface brightness, verifying our modelled solutions.

The bandwidth smearing ratio $\varpi$ is typically greater than 90\% over
the ATLAS mosaics, as indicated in Fig.~\ref{ch4:fig:bsmaps} and Fig.~\ref{ch4:fig:bs}. Unlike
for an individual pointing, bandwidth smearing in a mosaic is not negligible, even at locations
situated over pointing centres. This is because many adjacent pointings overlap and contribute
to the smearing at any position. Locations that experience the maximum ratio between
contributing numbers of on- and off-axis pointings will experience minimal bandwidth
smearing in a mosaic. For example, note the lessened smearing over the corner pointings
in Fig.~\ref{ch4:fig:bsmaps}. Note also the lessened smearing over the GOODS-South
region (refer to Fig.~\ref{ch4:fig:cdfspoint}) in the CDF-S panels of Fig.~\ref{ch4:fig:bsmaps},
where pointings are spaced more tightly than elsewhere, in turn reducing the relative
impact of adjacent off-axis pointings.

\subsection{Instrumental Polarization}\label{ch4:SecInstSubLeak}

To model spurious polarized emission over the $L_{\trm{\tiny RM}}$ mosaics,
caused by leakages of Stokes $I$ into Stokes $Q$ and $U$ \citep{saultMemo1,1999ASPC..180..111C}
within individual channels for each pointing, we needed to account for two forms
of instrumental polarization. The first was an `absolute' contribution that was
position-independent, applying uniformly over the full field of view for each pointing,
while the second was a `relative' contribution that was position-dependent. To estimate
the former, we considered gain errors resulting from our standard complex leakage calibrations
(see \S~\ref{ch4:SecDatSubFlag}), which nominally corrected the raw ATLAS data for couplings
between linear-feed outputs for each antenna. By assuming that the $10^{-3}$ variability
exhibited by these calibration solutions represented an absolute level of instrumental
polarization across each pointing, we estimated that the position-independent
leakages from Stokes $I$ to Stokes $Q$ or $U$ in each channel were
$\sim10^{-3}/\sqrt{2}=0.07\%$.

Position-dependent leakages are caused by a number of telescope design properties,
the most dominant of which are reflector geometry and aperture
blockage by feed support struts. Because ATCA antennas are alt-az mounted,
the instrumental polarization response rotates with parallactic angle against
the field of view throughout an observation. A model of the ATCA's off-axis
polarization response (neglecting complicated antenna deformations, for example
due to pointing elevation or wind-speed) has been included in the {\tt MIRIAD}
package \citep[see also][]{saultMemo2}. In principle, the ATCA's primary beam
polarization response may be corrected by using the {\tt MIRIAD} task {\tt OFFAXIS},
which removes rotated instrumental leakages from visibility data as a function of
time. However, we were unable to verify the performance of this task, which
predicted unrealistic leakage corrections for strong Stokes $I$ sources within
the ATLAS fields.

\begin{figure*}
\centering
 \includegraphics[bb = 41 70 583 713, trim = 0mm 0mm 0mm 0mm, clip, width=65mm, angle=-90]{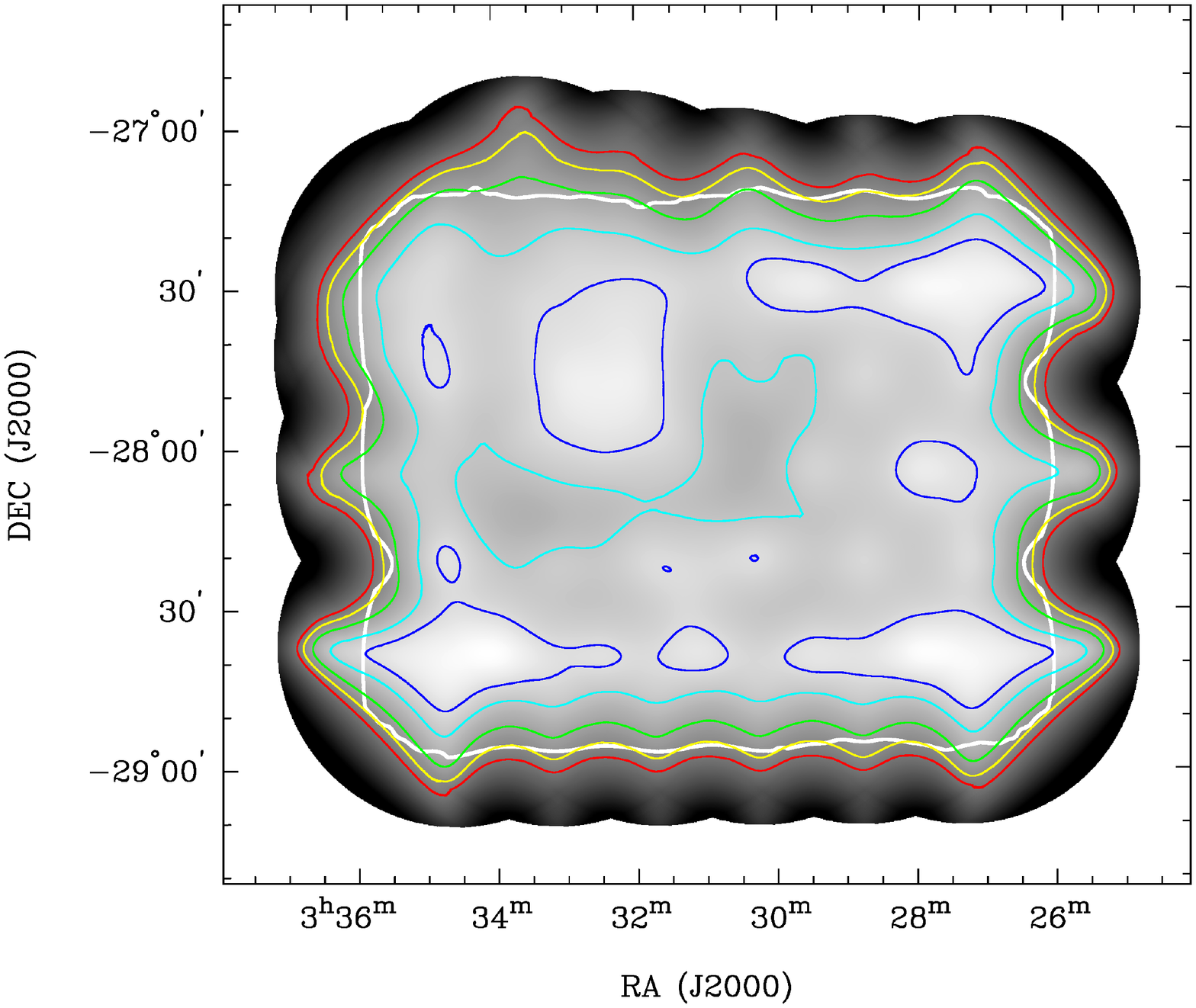}
 \hspace{1pc}
 \includegraphics[bb = 41 102 583 680, trim = 0mm 0mm 0mm 0mm, clip, width=65mm, angle=-90]{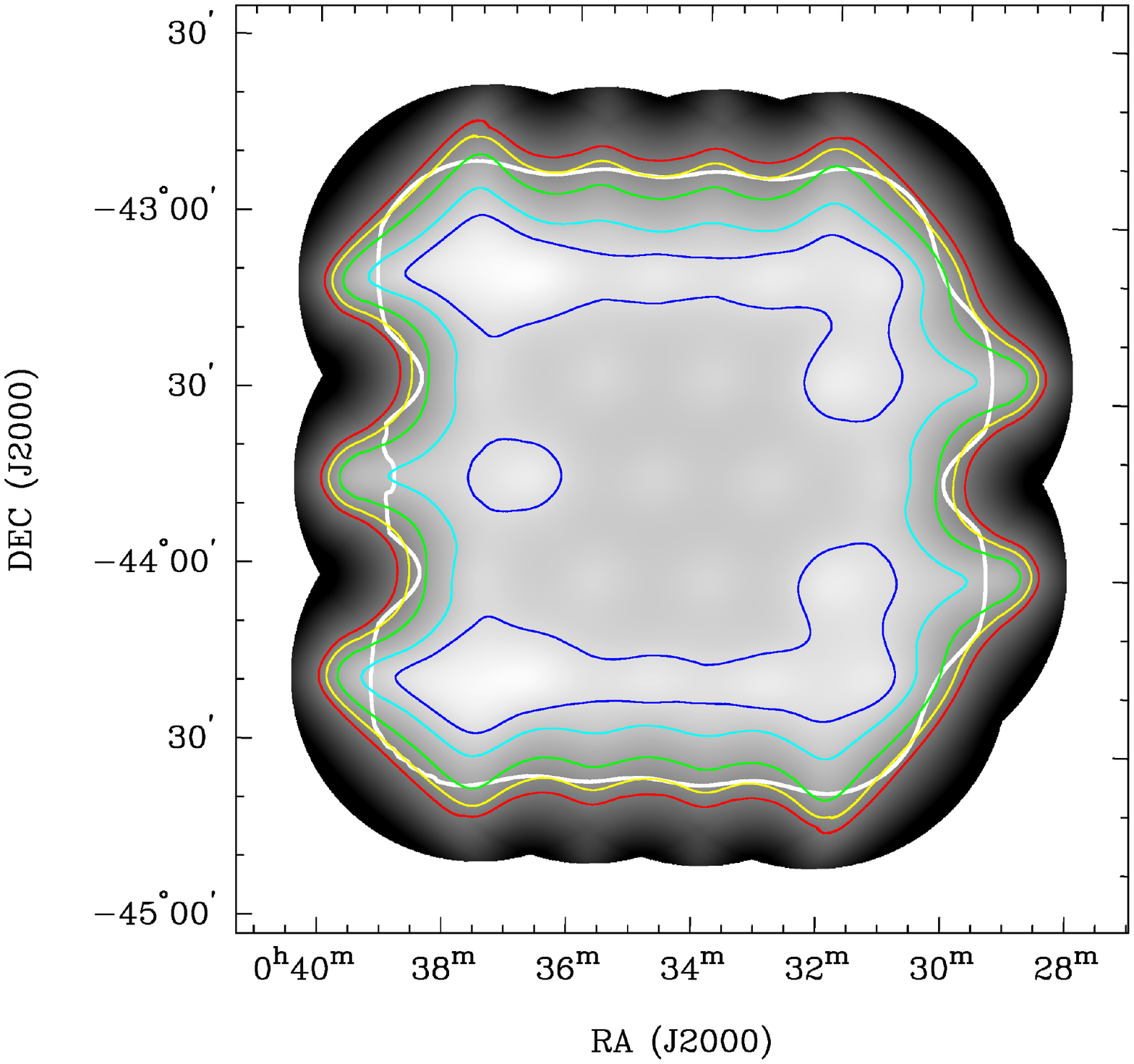}
\begin{minipage}{140mm}
 \caption{Spatial maps of $K_{\trm{\tiny LEAK}}$, indicating the fraction of Stokes $I$ surface brightness that
 	  may leak and appear as spurious polarized emission within the CDF-S (left) and ELAIS-S1 (right)
	  linear polarization fields, due to instrumental artefacts.
 	  Contours indicate leakage levels of 0.3\% (innermost), 0.4\%, 0.6\%, 0.8\%, and 1.0\% (outermost).
	  The bold white contours indicate the survey area boundaries (see \S~\ref{ch4:SecInstSubArea}).}
 \label{ch4:fig:leakmaps}
\end{minipage}
\end{figure*}

\begin{figure}
 \centering
 \includegraphics[trim=10mm 35mm 30mm 70mm, height=55mm]{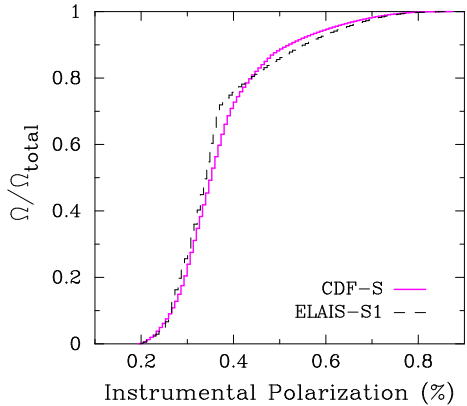}
 \caption{Fraction of sky area in ATLAS linear polarization survey areas
 	  ($L_{\trm{\tiny RM}}$) at or below a given instrumental polarization
	  leakage level (calculated from Fig.~\ref{ch4:fig:leakmaps}).}
 \label{ch4:fig:leak}
\end{figure}

Instead, we used the (less complicated) task {\tt OFFPOL} to simulate images of the
instrumental response exhibited by Stokes $Q$ and $U$ for each pointing in each channel
of each ATLAS field. These images quantified the position-dependent
fractions of Stokes $I$ surface brightness leaked into Stokes $Q$ and $U$
over the course of full synthesis observations, relative to the absolute leakage
level described above. The leakages exhibited at the phase centres for each these
simulated images were zero. To account for the missing absolute levels
of instrumental polarization, we added $0.07\%$ in quadrature to
each Stokes $Q$ and $U$ leakage image. We then mosaicked all pointing images
for each channel together to produce channel mosaics, using the same weighting
factors that were applied to construct the PC Stokes $Q_{\ms i}$ and $U_{\ms i}$ mosaics.
Next, for each channel we combined the Stokes $Q$ and $U$ leakage mosaics together in
quadrature, weighting each mosaic by the same factors applied to construct $\sigma_{\ms Q,U,i}$
(see \S~\ref{ch4:SecDatSubRM}).  Finally, we stacked these combined leakage mosaics together,
using the same weighting scheme as applied to $L_{\trm{\tiny RM}}$ and $I_{\trm{\tiny CA}}$
(see \S~\ref{ch4:SecDatSubRM}), resulting in what we term the $K_{\trm{\tiny LEAK}}$ mosaics.
These mosaics, which map the fraction of Stokes $I$ surface brightness that may appear
as spurious linearly polarized emission at any spatial position within the CDF-S and
ELAIS-S1 $L_{\trm{\tiny RM}}$ mosaics, are presented in Fig.~\ref{ch4:fig:leakmaps}.
In Fig.~\ref{ch4:fig:leak} we present cumulative histograms of the instrumental leakage
distributions from Fig.~\ref{ch4:fig:leakmaps}.
Both figures indicate that instrumental polarization is almost always less than 0.8\%
over the ATLAS mosaics, though never smaller than 0.2\%. As with bandwidth
smearing (\S~\ref{ch4:SecInstSubBS}), polarization leakage levels are found to be
diminished within the edge pointings of Fig.~\ref{ch4:fig:leakmaps}.

We checked the accuracy of our individual pointing and combined mosaic instrumental
polarization solutions by attempting to detect spurious polarized emission from
strong unpolarized sources that were visible in multiple overlapping pointings.
We found good agreement between our predictions and the observed detections or
upper bounds, verifying our modelled solutions.

The leakage correction scheme described above is rudimentary and only formally valid
for polarized sources with rotation measures near 0~rad~m$^{\ms -2}$. This is because
polarization leakage is not expected to vary with $\lambda^{\ms 2}$. We justify our use
of the scheme above to model polarization leakage in ATLAS sources at any RM
by noting that typical ATLAS RMs are {\footnotesize $\lesssim$}~$70$~rad~m$^{\ms -2}$
in magnitude, while the FWHM of the RMSF is 293~rad~m$^{\ms -2}$ (see \S~\ref{ch4:SecDatSubRM}).
Thus leakage near 0~rad~m$^{\ms -2}$ will contaminate all polarized ATLAS sources.
Furthermore, as will be described below, the magnitude of the polarization leakage
corrections for ATLAS sources are small, particularly for faint sources which are of
principal interest in this study. Any systematic overcorrection for leakage in ATLAS
sources with $|\tnm{RM}|>0$~rad~m$^{\ms -2}$ is likely to be negligible.

In preparation for image analysis, we produced corrected maps of linearly polarized
intensity for each ATLAS field, which we denote by $L_{\trm{\tiny RM}}^{\trm{\tiny CORR}}$,
by performing a scalar correction at each spatial pixel,
\begin{equation}\label{ch4:eqn:leakcorr}
	L_{\trm{\tiny RM}}^{\trm{\tiny CORR}}(x,y) =
	L_{\trm{\tiny RM}}(x,y) - I_{\trm{\tiny CA}}(x,y) \;
	K_{\trm{\tiny LEAK}}(x,y) \,,
\end{equation}
with $I_{\trm{\tiny CA}}$ from \S~\ref{ch4:SecDatSubRM}. We used $I_{\trm{\tiny CA}}$
rather than $I_{\trm{\tiny MFS}}$ in Equation~(\ref{ch4:eqn:leakcorr}) because
the former was produced in an equivalent manner to $L_{\trm{\tiny RM}}$ and
$K_{\trm{\tiny LEAK}}$, thus suitably reflecting the effective Stokes $I$
surface brightness that may have leaked into $L_{\trm{\tiny RM}}$.
We note that the noise properties of $L_{\trm{\tiny RM}}$
\citep[as described by][]{2012MNRAS.424.2160H} render Equation~(\ref{ch4:eqn:leakcorr})
an approximation for removing underlying levels of spurious emission. However,
we estimate that any systematic errors resulting from the use of Equation~(\ref{ch4:eqn:leakcorr})
are small, and that they are accounted for by the conservative absolute calibration error set
for our analysis in \S~\ref{ch4:SecExtSubFlood}.

\begin{figure}
 \centering
 \includegraphics[trim=10mm 35mm 40mm 80mm, height=57mm]{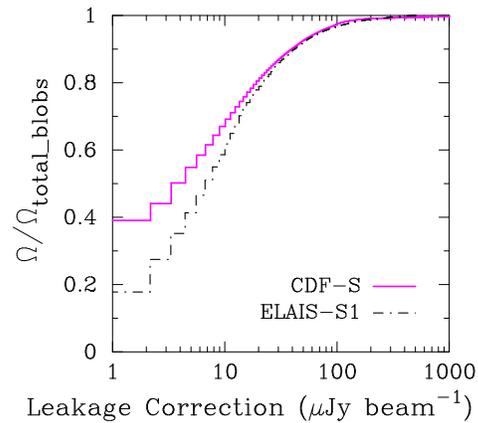}
 \caption{Fraction of polarized blob pixels within each ATLAS field at or below a given
 	  surface brightness correction for spurious instrumental polarized emission;
	  namely $L_{\trm{\tiny RM}}-L_{\trm{\tiny RM}}^{\trm{\tiny CORR}}$
	  for pixels agglomerated within linearly polarized blobs, not including
	  field pixels.
}
 \label{ch4:fig:leakcorr}
\end{figure}

To evaluate the impact of using Equation~(\ref{ch4:eqn:leakcorr}) on the data, we extracted
all pixels exhibiting significant polarized emission from $L_{\trm{\tiny RM}}^{\trm{\tiny CORR}}$
(blob extraction is described in \S~\ref{ch4:SecExtSubFlood}), and compared their brightnesses
with their uncorrected values from $L_{\trm{\tiny RM}}$. In Fig.~\ref{ch4:fig:leakcorr}
we plot the difference between uncorrected and corrected pixel brightness values.
Fig.~\ref{ch4:fig:leakcorr} indicates that the surface brightness corrections for
$>80$\% of blob pixels were smaller than the typical $\sim25~\mu$Jy~beam$^{-1}$
rms noise levels in the polarization mosaics. Less that 5\% of corrections
were greater than $100~\mu$Jy~beam$^{-1}$; these were associated
with the small number of strong $\sim$Jy total intensity components in the
ATLAS fields.

\subsection{Survey Area Boundaries}\label{ch4:SecInstSubArea}

We defined survey area boundaries for the CDF-S and ELAIS-S1 fields by enforcing
that the following conditions were met within both the $I_{\trm{\tiny MFS}}$ and
$L_{\trm{\tiny RM}}$ mosaics: rms noise $\le100$~$\mu$Jy~beam$^{-1}$, bandwidth
smearing $\varpi \ge 80\%$, instrumental polarization leakage $\le 1\%$, and
mosaicked primary beam response $\ge 40\%$. The resultant survey areas for the CDF-S
and ELAIS-S1 fields, which we denote $F^{\trm{\tiny AREA}}$, were 3.626~deg$^2$ and
2.766~deg$^2$, respectively. These areas were largely constrained by the bandwidth
smearing condition within the $L_{\trm{\tiny RM}}$ mosaics; see the lower panels
in Fig.~\ref{ch4:fig:bsmaps}. We note that less than 0.3 spurious $5\sigma$ detections
are expected by chance over the survey area for the CDF-S $I_{\trm{\tiny MFS}}$
mosaic [using Equation~(3) from \citealt{2012MNRAS.425..979H}]; even fewer are
expected over the other mosaics.

\section{Radio Component Extraction}\label{ch4:SecExt}

In the following sections we describe how radio components were detected and
extracted from the ATLAS total intensity and polarization mosaics, taking into
account the instrumental systematics described in \S~\ref{ch4:SecInst}, how
unresolved and resolved components were identified and assigned flux densities
given by their peak or integrated surface brightness measurements, respectively,
and how flux densities in both total intensity and linear polarization were
debiased to account for noise-induced systematics. We use
the term {\it component} to refer to an isolated region
of emission that is best described by a single 2D elliptical Gaussian. Blended
regions of contiguous emission may consist of multiple individual components.
Following the terminology from \citet{2012MNRAS.425..979H}, a {\it blob} is an agglomerated
island of pixels above a SNR cutoff, which may encapsulate a single component
or a blended region of emission. In \S~\ref{ch4:SecClass} we use the term {\it source}
to refer to single or multiple components belonging to the same astronomical object.

\subsection{{\tt BLOBCAT} and Follow-Up Fitting}\label{ch4:SecExtSubFlood}

Radio component detection and extraction were performed independently in total
intensity and linear polarization using a combination of two packages. First, the
{\tt BLOBCAT} package \citep{2012MNRAS.425..979H} was used to detect and catalogue blobs
in these images, flagging all blobs likely to consist of multiple blended
components for follow-up and assuming that the remainder represented
individual components. For total intensity images, the {\tt MIRIAD} task
{\tt IMFIT} was then used to decompose the flagged blobs into
individual components. For linear polarization images, a combination of
{\tt BLOBCAT} and {\tt IMFIT} was used to decompose blobs into individual
components. As ATLAS DR2 is the first survey to make use of {\tt BLOBCAT},
we now describe these procedures in some detail.

{\tt BLOBCAT} exhibits accurate measurement performance in both total intensity
and linear polarization \citep[see][]{2012MNRAS.425..979H}. The software enables rms
noise maps and bandwidth smearing maps to be included within the blob detection and
cataloguing procedure. However, no capabilities are provided within {\tt BLOBCAT} to
handle instrumental polarization maps. We therefore removed spurious instrumental
polarized emission from our $L_{\trm{\tiny RM}}$ mosaics prior to analysis
with {\tt BLOBCAT} using Equation~(\ref{ch4:eqn:leakcorr}).

We ran {\tt BLOBCAT} over the defined survey areas within the $I_{\trm{\tiny MFS}}$ and
$L_{\trm{\tiny RM}}^{\trm{\tiny CORR}}$ mosaics and their respective rms noise and
bandwidth smearing maps for each ATLAS field.
We set the SNR detection thresholds to $5\sigma$ in total intensity
and $6.25\sigma_{\trm{\tiny RM}}$ in linear polarization. The latter is
equivalent to a statistical significance (Type-I error) of $\alpha=10^{-7}$, or a standard
Gaussian detection threshold of $\pm5.33\sigma$; see Equation~(30) from
\citet{2012MNRAS.424.2160H} with $M=28$. We do not consider the effects of uncertainties associated
with these detection thresholds, which are $\sim0.4\sigma$ due to uncertainties in our
estimates of rms noise (see \S~\ref{ch4:SecDatSubNoise}). To ensure realistic errors were
calculated for the catalogue entries, we specified a number of input arguments to
{\tt BLOBCAT}; see \citet{2012MNRAS.425..979H} for full error propagation details.
We specified absolute positional uncertainties of 0\farcsecd01 in both R.A. and Decl.
for the phase calibrators\footnote{http://www.vla.nrao.edu/astro/calib/manual/index.shtml}
PKS~B0237$-$233 and PKS~B0022$-$423 for the CDF-S and ELAIS-S1 fields, respectively. We
characterised the relative positional uncertainties between the ATLAS mosaics and the assumed
positions of the phase calibrators by specifying typical values for the standard error of
the mean (SEM) of the phase variations resulting from the self-calibration step described
in \S~\ref{ch4:SecDatSubMFS}. In \S~\ref{ch4:SecExtSubPosErr} we describe how the SEM values were
calculated from the observed self-calibration rms phase variations given in \S~\ref{ch4:SecDatSubMFS}.
We set the absolute flux density error conservatively to 5\%, taking into account
2\% error in the ATCA's absolute flux density scale \citep{reynolds} and other sources of
uncertainty such as time-average smearing, uncertainties in modelled instrumental
systematics, and unflagged RFI. We set the input argument for pixellation error, which
encapsulates uncertainties in peak surface brightness measurements due to image
pixellation, to 1\% for each mosaic. We set the clean bias correction
parameter to zero for each mosaic.

For each blob we retained a subset of entries from {\tt BLOBCAT}'s full output catalogue
\citep[see \S~2.6 from][]{2012MNRAS.425..979H}. We have used these entries to construct the
ATLAS DR2 component catalogue presented in Appendix~A. The retained items for each blob were
their identification number, number of agglomerated pixels $n_{\trm{\tiny pix}}$, weighted
centroid position and associated errors, detection SNR $A_{\trm{\tiny S}}$, local rms noise value
$\sigma_{\trm{\tiny S}}$, local bandwidth smearing value $\varpi$, peak surface brightness
corrected for bandwidth smearing $S_{\trm{\tiny peak}}$ and associated error
$\sigma_{S_\trm{\tiny peak}}$, integrated surface brightness $S_{\trm{\tiny int}}$
and associated error $\sigma_{S_\trm{\tiny int}}$, estimated size in units of sky
area covered by an unresolved component with the same peak surface brightness
$R^{\trm{\tiny EST}}$, and fraction of survey area (or visibility area) over which
the blob could have been detected due to rms noise and bandwidth smearing
fluctuations $V^{\trm{\tiny AREA}}$.
{\tt BLOBCAT} does not account for polarization bias in its measurements of peak polarized surface
brightness. {\tt BLOBCAT}'s integrated polarized surface brightnesses are unaffected
by polarization bias \citep[see][]{2012MNRAS.425..979H}. In \S~\ref{ch4:SecExtSubDB} we account
for biases in measurements of $S_{\trm{\tiny peak}}$ due to noise boosting in total
intensity, and a combination of boosting and polarization bias in linear polarization.

We manually inspected all blobs identified near regions of strong total intensity
emission\footnote{
The noise estimation algorithm described in \S~\ref{ch4:SecDatSubNoise} performs
sub-optimally in regions where the rms noise changes rapidly over spatial scales
much smaller than the mesh size. In the few such regions of the ATLAS images, we
carefully inspected the data to account for potentially underestimated
rms noise values and in turn overestimated detection significances.}
within both the $I_{\trm{\tiny MFS}}$ and $L_{\trm{\tiny RM}}^{\trm{\tiny CORR}}$
mosaics. We identified $\sim30$ total intensity blobs in each ATLAS field that were clearly
associated with image artefacts, and removed these from the catalogue. We also removed
$\sim10$ linearly polarized blobs in each ATLAS field that did not exhibit total
intensity counterparts; these were unlikely to be signs of Galactic foreground
emission as the ATLAS fields are located more than $50\degree$ below the Galactic plane.
To identify blobs likely to consist of multiple components, we flagged all
catalogue entries with $R^{\trm{\tiny EST}}\ge1.4$ and $n_{\trm{\tiny pix}}\ge500$.
We selected these values following manual testing to identify the most suitable
criteria for conservative automatic identification of blended-component blobs.
We note that the $R^{\trm{\tiny EST}}\ge1.4$ criterion is likely to be suitable for
{\tt BLOBCAT} analyses in general, whereas the $n_{\trm{\tiny pix}}\ge500$ criteria
depends on the relationship between image resolution and pixel size (this ratio
is $\sim10$ for ATLAS DR2 images).
We attempted to fit multiple Gaussian components to each of the total intensity
flagged blobs using {\tt IMFIT}. For each blob, we first identified
positions at which up to 6 individual Gaussian components could be situated.
We then ran {\tt IMFIT} with these initial conditions and inspected the output
fits and fitting residuals. Catalogue entries for each flagged blob were replaced
by entries for each {\tt IMFIT} component identified; component identification
numbers were assigned by suffixing {\tt C}$j$ to the original blob number for
each $j$'th component extracted. No more than 6 components
were required for any individual blob; often, only 2 components were required.
Components with $\textrm{SNR}<5\sigma$ were excluded from the catalogue. For
blobs best fit by a single Gaussian component, their original {\tt BLOBCAT}
catalogue entries were retained, unless image artefacts such as sidelobe
ridge-lines were seen to be affecting them. We followed the same general
procedure to decompose flagged blobs in linear polarization, with some
minor differences. We applied a $4\sigma_{\trm{\tiny RM}}$ cutoff threshold for
fitting linearly polarized components with {\tt IMFIT}, to prevent polarization
non-Gaussianities from interfering with {\tt IMFIT}'s least-squares fitting
algorithm; all fits and fitting residuals were carefully inspected for biases.
Some of the flagged polarized blobs were found to consist of isolated components
that were joined by a small bridge of low-SNR emission. For these blobs, we used
{\tt BLOBCAT}, rather than {\tt IMFIT}, to fit each clearly-separated component;
image masking was applied to isolate the emission from each individual component
prior to refitting. Identification numbers were assigned to each extracted component
by suffixing {\tt C}$j$ or {\tt F}$j$ to the original blob number for fits obtained
with {\tt IMFIT} or {\tt BLOBCAT}, respectively. All refit polarized components with
$\textrm{SNR}<6.25\sigma{\trm{\tiny RM}}$ were excluded from the catalogue.

All {\tt IMFIT} measurements of total intensity and linear polarization were
carefully compared with their original {\tt BLOBCAT} measurements for consistency.
We are confident that no systematic differences are present between the two samples,
taking into account the different regimes where the two extraction methods are
known to become inaccurate; see \citet{2012MNRAS.425..979H} for a formal comparison 
between {\tt BLOBCAT} and {\tt IMFIT}. A total of 1268 (113) and 1148 (59) components
were extracted in total intensity and linear polarization, the latter in parentheses,
within the CDF-S and ELAIS-S1 survey areas, respectively. Of these, 244 (6) and
373 (5) were extracted using {\tt IMFIT}, while (18) and (7) were extracted
following image masking using {\tt BLOBCAT}, respectively.

Finally, we ran {\tt BLOBCAT} over the survey areas within the CDF-S
and ELAIS-S1 $V_{\trm{\tiny MFS}}$ mosaics and their associated rms noise maps,
searching for blobs with positive or negative surface brightness.
All $\sim20$ circularly polarized blobs identified in each field were consistent
with likely instrumental leakage from $I_{\trm{\tiny CA}}$ to $V_{\trm{\tiny MFS}}$
at or below the 0.5\% level.

\subsection{Image Frame Position Errors}\label{ch4:SecExtSubPosErr}

Formal position errors for each blob were calculated by combining three errors
as described in \citet{2012MNRAS.425..979H}: the absolute uncertainties defined in
\S~\ref{ch4:SecExtSubFlood}, the positional uncertainties of the image frames about the
assumed locations of the secondary calibrators, and the measurement errors from {\tt BLOBCAT}
or {\tt IMFIT}. In this section we describe the calculation of SEM values for each ATLAS
field, which are needed to calculate the image frame errors.

\begin{figure*}
\begin{minipage}{140mm}
 \centering
 \vspace{7pc}
 \includegraphics[trim = 110mm 25mm 0mm 80mm, height=41mm]{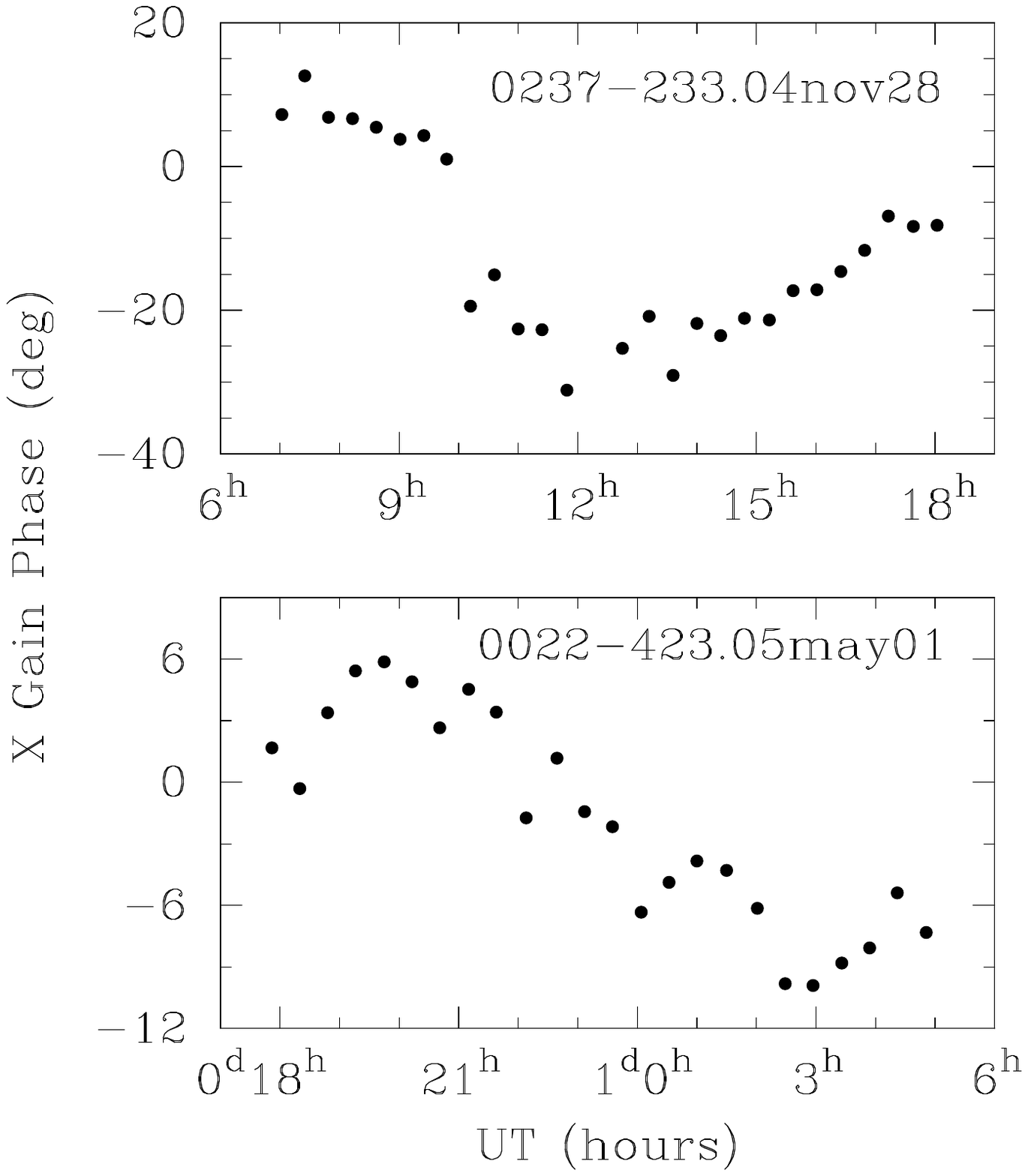}
 \hspace{1pc}
 \includegraphics[trim = 110mm 25mm 0mm 80mm, height=41mm]{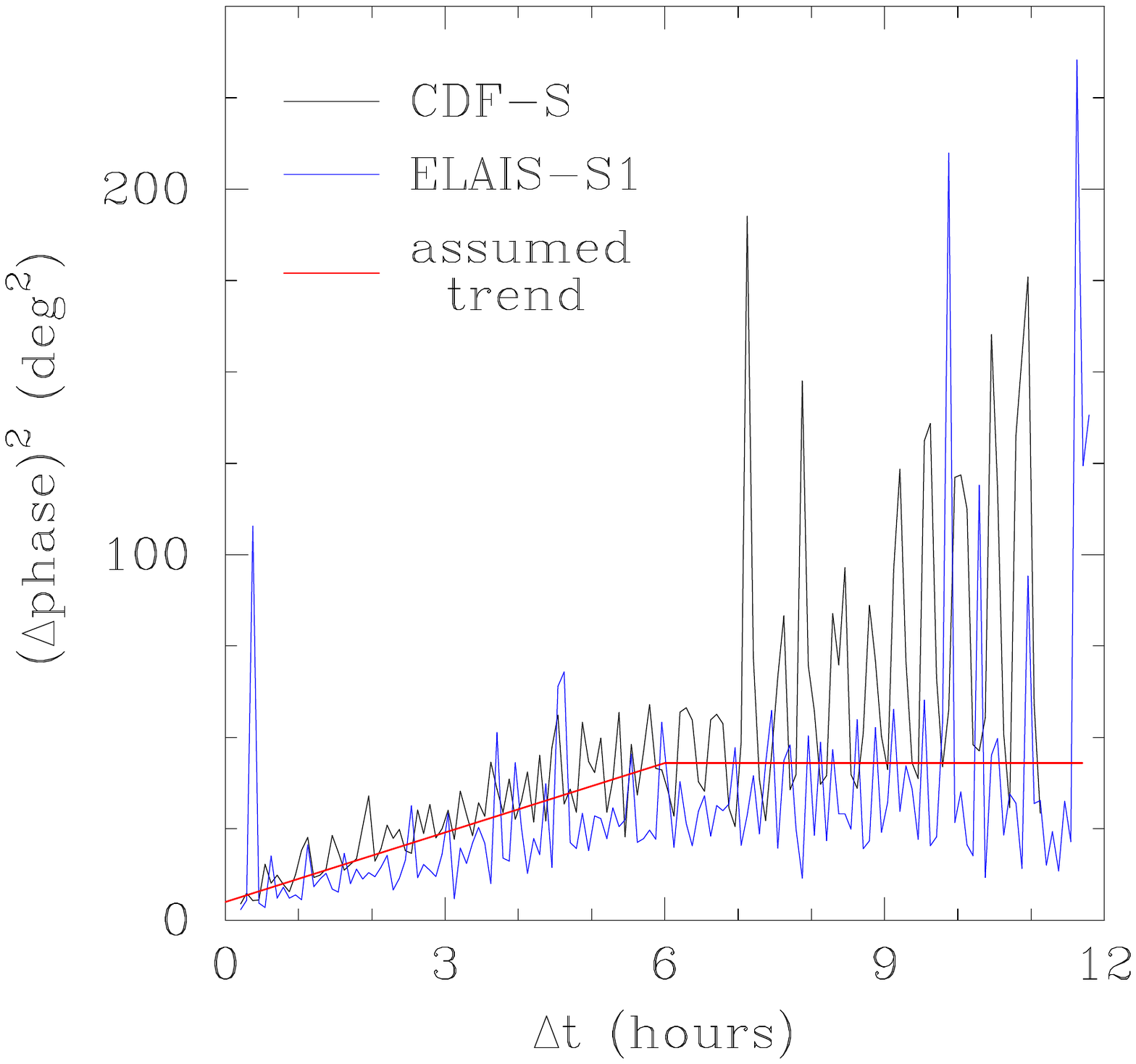}
 \caption{Upper left panel: Phase variations observed towards the CDF-S gain calibrator PKS~0237$-$233 using the
 	  X feed on antenna 5 throughout an observation on 2004 Nov 25. Lower left panel: Phase variations
	  observed towards the ELAIS-S1 gain calibrator PKS~0022$-$423 using the X feed on antenna 4 throughout
	  an observation on 2005 May 1. Right panel: Phase variation structure function (Equation~\ref{ch4:eqn:sf}),
	  constructed by combining all ATLAS DR2 observations from all antennas for PKS~0237$-$233 or PKS~0022$-$423
	  in 5~min bins.
	  The red lines represent our assumed trend (not a fit) for phase variations in the CDF-S and ELAIS-S1 data,
	  which flatten at $\Delta t \approx 6$~hours. The oscillations with period $\Delta t \approx 25$~mins are
	  due to aliasing; gain calibrators were typically observed every 25~mins.}
 \label{ch4:fig:sem}
\end{minipage}
\end{figure*}

In \S~\ref{ch4:SecDatSubMFS}, typical rms values for the variation in the phase corrections
resulting from self-calibration were found to be $9.5^\circ$ and $7.0^\circ$ for pointings in
the CDF-S and ELAIS-S1 fields, respectively. If the samples used to calculate these rms values were
uncorrelated, then the SEM for each field could be calculated by dividing the rms values by the
square root of the number of self-calibration intervals. However, we found that phase variations
throughout our 1.4~GHz ATCA observations were correlated. To characterise this correlation and
subsequently calculate a more appropriate SEM for each ATLAS field, we utilised the phase variation
structure function which we defined as
\begin{equation}\label{ch4:eqn:sf}
	\tnm{SF}_{\!\tnm{phase}}\left( \Delta t \right) = \tnm{median} \big\{ \left[ \tnm{phase}\left(t\right) -
	\tnm{phase}\left(t+\Delta t\right) \right]^2 \big\} \,.
\end{equation}
We calculated this structure function for phase variations seen towards the CDF-S and ELAIS-S1 gain calibrators,
as displayed in the right panel of Fig.~\ref{ch4:fig:sem} and explained in the caption. For reference, phase
variations for each of the gain calibrators are displayed for a single antenna and single observation in the
left panels of Fig.~\ref{ch4:fig:sem}.
The phase variation structure function is observed to flatten at $\Delta t \approx 6$~hours.
The large-amplitude oscillations at large $\Delta t$, which are most significant for the CDF-S data for
$\Delta t>7$~hours about an approximately flat mean of $\sim90$~deg$^2$, likely demonstrate that
coherent structures of scale length $\Delta t \approx 6$~hours are sequentially encountered by the array,
each with slightly different mean phase.

For an outer scale of fluctuations at $\Delta t \approx 6$~hours,
the effective timescale to observe statistically independent phases (i.e. uncorrelated samples at the
Nyquist rate) is $\Delta t \approx 3$~hours. For quasi-sinusoidal fluctuations with period 12~hours, the
structure function should rise to an outer scale at $\sim6$~hours, consistent with Fig.~\ref{ch4:fig:sem}.
Such fluctuations are consistent with semi-diurnal oscillations in the ionosphere due to atmospheric
tides \citep{1970attg.book.....C,AG3885}.

We corrected the self-calibration rms phase values for correlation and calculated suitable SEM values as
follows. We modelled the ATLAS observations using a synthesis timescale of 10~hours, characterised by a
Gaussian autocorrelation function denoted by $\rho_h$ with $\tnm{FWHM}=3$~hours sampled at the 3~min
self-calibration timescale. The autocorrelation function was therefore discretised into $\kappa=201$
samples, with 60 samples per FWHM. We corrected the observed rms values following \citeauthor{anderson}
[\citeyear{anderson}; see Equation~(51) in Chapter~8 of their work, adjusted to represent sample variance
following their Equation~(48)] using
\begin{equation}\label{ch4:eqn:rms}
	\tnm{rms}_{\tnm{\tiny true}} = \tnm{rms}_{\tnm{\tiny obs}} \left[
	1 - \frac{2}{\kappa-1}\sum_{h=1}^{\kappa-1}
	\left(1-\frac{h}{\kappa}\right)|\rho_h| \right]^{-\frac{1}{2}}\,.
\end{equation}
The value inside the square brackets was found to be 0.71. We then calculated the SEM following
\citeauthor{anderson} [\citeyear{anderson}; see Equation~(32) in Chapter~8 of their work] using
\begin{equation}\label{ch4:eqn:sem}
	\tnm{SEM} = \frac{\tnm{rms}_{\tnm{\tiny true}}}{\sqrt{\kappa}} \left[
	1 + 2\sum_{h=1}^{\kappa-1}
	\left(1-\frac{h}{\kappa}\right)|\rho_h| \right]^{\frac{1}{2}}\,.
\end{equation}
The value inside the square brackets was found to be 59. The SEM values for the phase variations
resulting from self-calibration were thus calculated as $6.1^\circ$ and $4.5^\circ$ for the
CDF-S and ELAIS-S1 fields, respectively. For reference, we find that by computing Equation~(19)
from \citet{2012MNRAS.425..979H} with a $\sim10\arcsec$ beam and $\tnm{SEM}\approx5^\circ$, we estimate that
the positional uncertainty of an image frame about an (assumed) position of a phase calibrator will be
$\sim0\farcsecd2$. We note that in the formalism above, we assume that any phase differences between the
target field and phase calibrator (for example due to elevation differences) are accounted for because
the ATLAS synthesis observations are long enough to sample many different elevations.

For completeness, we note that correlation timescales of $\sim3$~hours are unlikely to be caused by
tropospheric delay fluctuations due to water vapour \citep{1997A&AS..122..535L,1999RaSc...34..817C},
though DC offsets due to clouds with scale sizes up to $\sim100$~km \citep{clouds} and $\sim3$~hour
timescales to advect over a point may be relevant to some of the ATLAS observations.

\subsection{Deconvolution}\label{ch4:SecExtSubDeconv}

In the absence of noise, the peak surface brightness of an unresolved radio component
(assumed to be of 2D elliptical Gaussian morphology), measured in Jy~beam$^{-1}$,
is equal in magnitude to its integrated surface brightness,
measured in Jy. The observed spatial extent of a Gaussian radio
component, relative to the synthesised beam, may therefore be deduced from its
ratio of integrated to peak surface brightness \citep[e.g.][]{2000A&AS..146...41P,
2003A&A...403..857B,2005AJ....130.1373H,2010ApJS..188..384S}, namely
\begin{equation}\label{ch4:eqn:resunres}
	\frac{S_{\trm{\tiny int}}}{S_{\trm{\tiny peak}}} =
	\frac{\theta_\trm{\scriptsize maj} \, \theta_\trm{\scriptsize min}}
	{B_\trm{\scriptsize maj} \, B_\trm{\scriptsize min}}
\end{equation}
where $\theta_\trm{\scriptsize maj}$ and $\theta_\trm{\scriptsize min}$
are the component's observed (not deconvolved) major and minor axis FWHMs, respectively.
If images were noise-free, then the ratio $S_{\trm{\tiny int}}/S_{\trm{\tiny peak}}$
would be unity for unresolved components and $>1$ for resolved components, following
from perfect component extraction and measurement. However, noise in real images
causes some unresolved components to exhibit $S_{\trm{\tiny int}}<S_{\trm{\tiny peak}}$
and others $S_{\trm{\tiny int}}>S_{\trm{\tiny peak}}$, such that not
all components with $S_{\trm{\tiny int}}/S_{\trm{\tiny peak}}>1$ may be
unambiguously classified as being resolved.

\begin{figure*}
 \centering
 \includegraphics[trim=10mm 30mm 10mm 30mm, width=130mm]{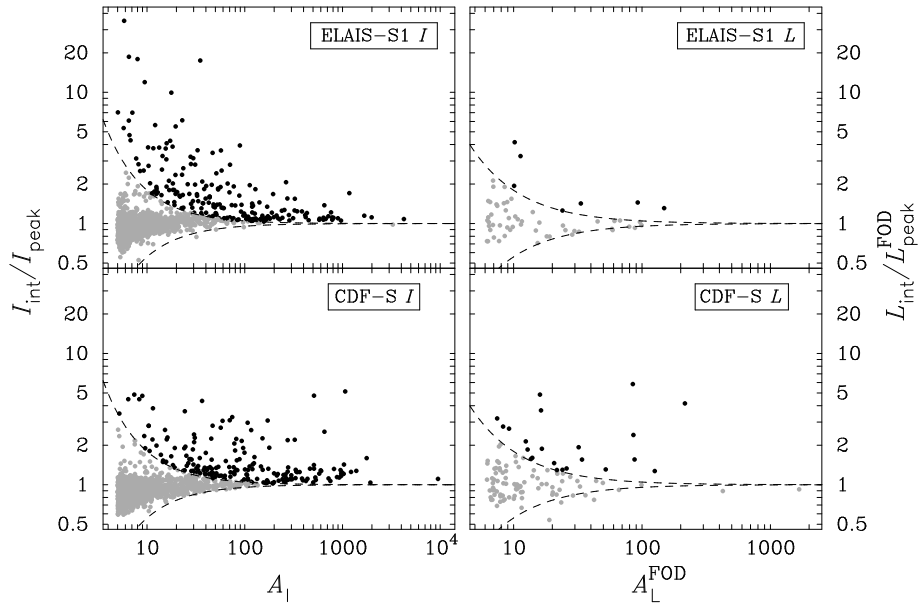}
 \begin{minipage}{140mm}
 \caption{	Ratio of integrated to peak surface brightness as a function
 		of detection SNR for total intensity (left column) and linear
		polarization (right column) components in the ELAIS-S1 (top row)
		and CDF-S (bottom row) fields. In linear polarization, detection
		SNRs and peak surface brightness measurements have been corrected
		for Ricean bias using a first-order debiasing scheme. The loci
		in each panel, given by Equation~(\ref{ch4:eqn:RURlocus}) and mirrored
		below the $S_{\trm{\tiny int}}=S_{\trm{\tiny peak}}$ line, enable
		classification of components as unresolved (grey points) or resolved
		(black points).}
 \label{ch4:fig:IPR}
 \end{minipage}
\end{figure*}

To classify each ATLAS component as unresolved or resolved, we first examined
the distribution of $S_{\trm{\tiny int}}/S_{\trm{\tiny peak}}$ for
all components in each survey field as a function of their detection SNR,
$A_{\trm{\tiny S}}$, as shown in Fig.~\ref{ch4:fig:IPR}.
To prevent polarization bias from shifting the positions of linearly polarized
components in Fig.~\ref{ch4:fig:IPR}, we performed two first-order debiasing (FOD)
corrections \citep{vla161}. We corrected the detection SNRs using
$A_{\trm{\tiny L}}^{\trm{\tiny FOD}}\approx(A_{\trm{\tiny L}}^{\ms 2}-1^{\ms 2})^{\ms 1/2}$,
and the peak polarized surface brightness measurements using
$L_{\trm{\tiny peak}}^{\trm{\tiny FOD}}\approx
[L_{\trm{\tiny peak}}^{\ms 2} -
\sigma_{\trm{\tiny RM}}^{\ms 2}/\varpi^{\ms 2}]^{\ms 1/2}$.
No bias corrections were required for $L_{\trm{\tiny int}}$ \citep[see][]{2012MNRAS.425..979H}.
We note that, as discussed by \citet{2012MNRAS.424.2160H}, the FOD scheme is
designed for application to data exhibiting Ricean statistics and not
$L_{\trm{\tiny RM}}$ as relevant here. However,
at the SNRs relevant to our data, the probability density functions
characterising the \citet{rice} distribution and the $M=28$ distribution for
$L_{\trm{\tiny RM}}$ are very similar \citep[see][]{2012MNRAS.424.2160H}. We therefore
assume approximate validity of the FOD scheme in application to the analysis
described in this section. We found no significant shifting
of points in Fig.~\ref{ch4:fig:IPR} when debiasing corrections were neglected entirely,
demonstrating that the distribution of points is not highly sensitive to
polarization bias.

To identify unresolved components within the total intensity panels of
Fig.~\ref{ch4:fig:IPR}, we defined a locus enveloping $\sim$99\% of
components with $I_{\trm{\tiny int}}/I_{\trm{\tiny peak}}<1$.
We then mirrored this locus above the $I_{\trm{\tiny int}}=I_{\trm{\tiny peak}}$ line,
assuming the presence of a similar distribution of unresolved components with
$I_{\trm{\tiny int}}/I_{\trm{\tiny peak}}>1$. We assumed that these
loci also characterised the linear polarization data; separate polarization
loci were not constructed. We defined the upper locus using the function
\citep{2010ApJS..188..384S}
\begin{equation}\label{ch4:eqn:RURlocus}
	\frac{S_{\trm{\tiny int}}}{S_{\trm{\tiny peak}}} =
	a^{-b/\left(A_{\trm{\tiny S}}\right)^{\ms c}}\,,
\end{equation}
with $a=0.35$, $b=7.0$, $c=1.1$, and where in linear polarization we replaced
$S_{\trm{\tiny peak}}$ by $L_{\trm{\tiny peak}}^{\trm{\tiny FOD}}$ and
$A_{\trm{\tiny S}}$ by $A_{\trm{\tiny L}}^{\trm{\tiny FOD}}$. All components
above the upper locus were classified as resolved and assigned flux densities
given by $I_{\trm{\tiny int}}$ or $L_{\trm{\tiny int}}$, while all components
below it were classified as unresolved and assigned flux densities given by
the magnitudes of $I_{\trm{\tiny peak}}$ or $L_{\trm{\tiny peak}}$.
A total of 189 (22) and 204 (7) components were classified as resolved
in total intensity and linear polarization, the latter in parentheses, within
the CDF-S and ELAIS-S1 survey areas, respectively. For each resolved component,
we used Equation~(\ref{ch4:eqn:resunres}), with $S_{\trm{\tiny peak}}$ replaced by
$L_{\trm{\tiny peak}}^{\trm{\tiny FOD}}$ in linear polarization, to estimate
a deconvolved angular size as
\begin{equation}\label{ch4:eqn:deconvest}
	\Theta \approx
	\sqrt{\theta_\trm{\scriptsize maj} \, \theta_\trm{\scriptsize min}
	- B_\trm{\scriptsize maj} \, B_\trm{\scriptsize min}} \,.
\end{equation}
We calculated upper bounds to the deconvolved angular sizes of unresolved
components by equating Equation~(\ref{ch4:eqn:resunres}) with
Equation~(\ref{ch4:eqn:RURlocus}) and then evaluating Equation~(\ref{ch4:eqn:deconvest}).
We note that direct measurements of $\theta_\trm{\scriptsize maj}$ and
$\theta_\trm{\scriptsize min}$ were not used, nor required, for the analysis presented
above; a characteristic angular size
$\theta\approx\sqrt{\theta_\trm{\scriptsize maj}\,\theta_\trm{\scriptsize min}}$
was evaluated for each component using Equation~(\ref{ch4:eqn:resunres}), which was
then deconvolved using Equation~(\ref{ch4:eqn:deconvest}). We estimated the
uncertainty in measurements of $\Theta$ for resolved components by following
standard error propagation, resulting in
\begin{equation}\label{ch4:eqn:deconvestERR}
	\sigma_{\ms \Theta} \approx \sqrt{
	\frac{B_\trm{\scriptsize maj} \, B_\trm{\scriptsize min}}
	{4\left(S_{\trm{\tiny int}}/S_{\trm{\tiny peak}}-1\right)}
	\left[\left(\frac{\sigma_{S_\trm{\tiny peak}}}{S_{\trm{\tiny peak}}}\right)^{\!2} +
	      \left(\frac{\sigma_{S_\trm{\tiny int}}}{S_{\trm{\tiny int}}}\right)^{\!2}
	      \right]} \,.
\end{equation}
We set angular size uncertainties for unresolved components to zero.

In reality, it is not possible for a component to be truly unresolved
(components have real physical dimensions). The flux densities of components
classified as unresolved by the scheme above
will therefore be systematically underestimated, due to their assignment using
$S_{\trm{\tiny peak}}$. We do not correct for this flux density bias on an
individual component basis, which depends on the SNR and flux density of each
component as well as the distribution of intrinsic angular sizes, and which
will become increasingly significant at faint flux densities where majority
of components are classified as unresolved using Equation~(\ref{ch4:eqn:RURlocus}).
However, we do account for this bias in a collective sense in \S~\ref{ch4:SecCNCSubRB}
when considering component number-counts in flux density bins.

\subsection{Total Intensity and Linear Polarization Deboosting}\label{ch4:SecExtSubDB}

For a given observed flux density, the probability of detecting a faint unresolved
component located on a noise peak is greater than the probability of detecting a
strong unresolved component located in a noise trough, because faint radio components
are more numerous. This results in a bias between true and observed flux densities
known as flux density boosting \citep[following the terminology of][]{2010ApJ...719..763V},
which depends on the SNR of the detection, the noise distribution in which the detection
was made, and the slope of the radio component differential number-counts, $\gamma$,
where $dN/dS \propto S^{-\gamma}$. Flux density boosting of individual components
leads to \citet{1913MNRAS..73..359E} bias in their observed differential number-counts;
we discuss Eddington bias later in \S~\ref{ch4:SecCNCSubEB}.

To account for flux density boosting we used Bayes' theorem to quantify the bias
\citep{1938MNRAS..98..190J,1940MNRAS.100..354E}, obtaining the posterior distribution
\begin{equation}\label{ch4:eqn:deboostBayes}
	f\!\left(S_{\trm{\tiny TRUE}}|S_{\trm{\tiny OBS}}\right)
	\propto f\!\left(S_{\trm{\tiny OBS}}|S_{\trm{\tiny TRUE}}\right)
	f\!\left(S_{\trm{\tiny TRUE}}\right)\,,
\end{equation}
where $S_{\trm{\tiny OBS}}$ is the observed flux density, $S_{\trm{\tiny TRUE}}$
is the true flux density, $f\!\left(S_{\trm{\tiny OBS}}|S_{\trm{\tiny TRUE}}\right)$ is
the likelihood of measuring $S_{\trm{\tiny OBS}}$ given $S_{\trm{\tiny TRUE}}$,
and $f\!\left(S_{\trm{\tiny TRUE}}\right)$ is a prior which is proportional to
the differential number counts $dN/dS$. We obtained maximum-likelihood (ML) solutions
to Equation~(\ref{ch4:eqn:deboostBayes}), described as follows, to correct component
flux densities for boosting in total intensity and linear polarization;
we use the term deboosting to describe these corrections. We note that deboosting
is not required for resolved components because noise fluctuations about their true
peak surface brightnesses are largely accounted for by extraction algorithms
such as those used in {\tt BLOBCAT} and {\tt IMFIT}; we have not applied any of
the deboosting corrections described in this section to resolved components.

\begin{figure*}
 \centering
 \includegraphics[trim = 10mm 30mm 40mm 30mm, height=90mm]{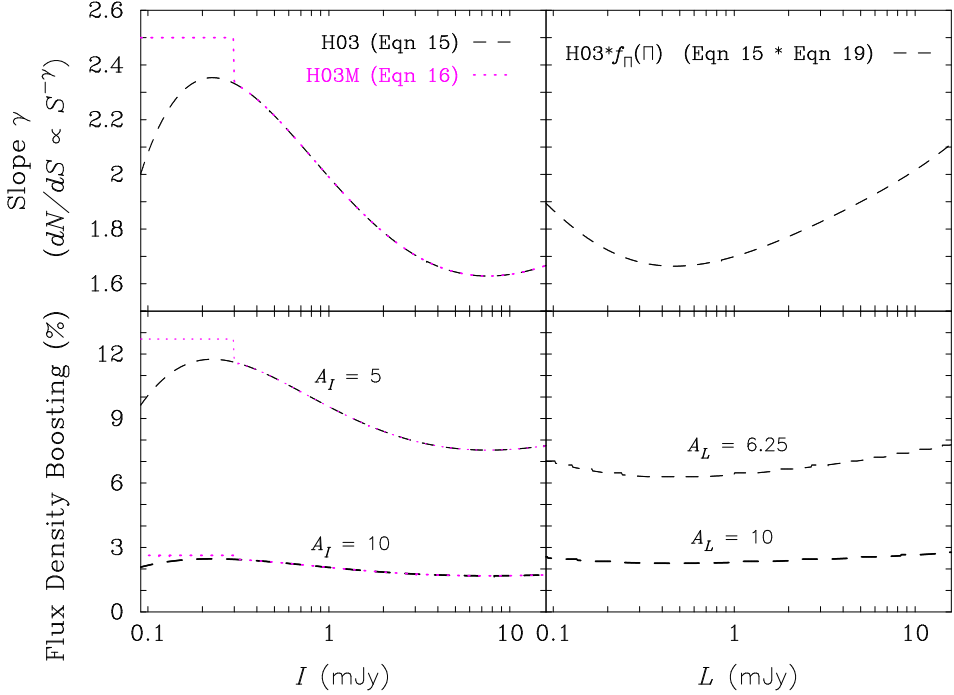}
 \begin{minipage}{140mm}
 \caption{	Flux density boosting as a function of observed flux density in
 		total intensity (left column) and linear polarization (right column)
		for unresolved components with observed SNRs as indicated. The dashed and
		dotted curves represent different underlying number-count distributions.
		Polarization results utilising the H03M model are not shown, as they
		are identical to the H03 results over the flux density range shown.
	 }
 \label{ch4:fig:IEB}
 \end{minipage}
\end{figure*}

In total intensity, we deboosted observed flux densities using the ML solution
\citep{1968ApJ...152..647J,1998PASP..110..727H}
\begin{equation}\label{ch4:eqn:EBcorrCmpI}
	I_{\trm{\tiny ML}} =
	\frac{I}{2}\left(1 + \sqrt{ 1 -
	\frac{4\gamma_{\trm{\tiny I}}}{
	A_{\trm{\tiny I}}^{\ms 2}}}\;\right) \;,
\end{equation}
which implicitly takes into account the presence of bandwidth smearing (provided that
the assumptions described at the end of this section are met). To model
the differential component number-counts curve and in turn obtain its slope
$\gamma_{\trm{\tiny I}}$, we used the sixth-order empirical fit to the Phoenix and
FIRST surveys presented by \citet{2003AJ....125..465H}. This curve, which we
denote by H03, is given by
\begin{equation}\label{ch4:eqn:H03}
	\log\left[\frac{dN_{\trm{\tiny H03}}/dI}{I^{\ms -2.5}}
	\right] = \sum_{j=0}^{6} a_{j} \left[ \log\left(
	\frac{I}{\trm{mJy}} \right)\right]^{j}\,,
\end{equation}
with $a_{\ms 0}=0.859$, $a_{\ms 1}=0.508$, $a_{\ms 2}=0.376$, $a_{\ms 3}=-0.049$,
$a_{\ms 4}=-0.121$, $a_{\ms 5}=0.057$, and $a_{\ms 6}=-0.008$. In Fig.~\ref{ch4:fig:IEB}
we plot the boosting ratio $I/I_{\trm{\tiny ML}}$ resulting from
Equation~(\ref{ch4:eqn:EBcorrCmpI}) with $\gamma_{\trm{\tiny I}}$ obtained from H03.
There are suggestions that the H03 number-counts fall off too quickly at faint
flux densities \citep[e.g.][]{2010MNRAS.409.1172S,2010ApJS..188..178M,2011MNRAS.415.3641V}.
To illustrate the potential boosting effects of an exaggerated population of
faint components, we have constructed a modified H03 distribution, which we
denote by H03M, by inserting a Euclidean
slope between 30$-$300$\mu$Jy, namely
\begin{equation}\label{ch4:eqn:H03M}
	\frac{dN_{\trm{\tiny H03M}}}{dI}\left(I\right)
	= \left\{
	\begin{array}{l l}
		dN_{\trm{\tiny H03}}/dI\left(I\right)
		& \quad \textrm{if $I\ge300~\mu$Jy}\\
		dN_{\trm{\tiny H03}}/dI\left(\textrm{$300~\mu$Jy}\right)
		& \quad \textrm{if $30\le I < 300~\mu$Jy}\\
		dN_{\trm{\tiny H03}}/dI\left(10\times I\right)
		& \quad \textrm{if $I<30~\mu$Jy}\,.\\
	\end{array} \right.
\end{equation}
In Fig.~\ref{ch4:fig:IEB} we plot the boosting ratio $I/I_{\trm{\tiny ML}}$
using $\gamma_{\trm{\tiny I}}$ obtained from H03M; differences
between the H03 and H03M solutions are minimal, being limited to faint
components with low-SNR.

In linear polarization, we obtained the posterior distribution from
Equation~(\ref{ch4:eqn:deboostBayes}) by assuming that observational errors
were described by the \citet{rice} distribution (note comments in
\S~\ref{ch4:SecExtSubDeconv}), giving
\begin{eqnarray}
	f\!\left(L_{\trm{\tiny TRUE}}|L,\sigma_{\trm{\tiny RM}},\varpi,
	\gamma_{\trm{\tiny L}}\right)
	&\propto& \left(\frac{L_{\trm{\tiny TRUE}}}{L}
	\right)^{-\gamma_{\trm{\tiny L}}} \frac{L}{\widetilde{\sigma}_{\trm{\tiny RM}}^{\ms 2}} \, \times \nonumber\\
	&& \exp\left(-\frac{L_{\trm{\tiny TRUE}}^{\ms 2} + L^{\ms 2}}
	{2\widetilde{\sigma}_{\trm{\tiny RM}}^{\ms 2}} \right)\;, \label{ch4:eqn:EBcorrCmpL1}
\end{eqnarray}
where we define $\widetilde{\sigma}_{\trm{\tiny RM}}=\sigma_{\trm{\tiny RM}}/\varpi$.
The ML solution for each linearly polarized component with observed flux density
$L$ (corrected for bandwidth smearing) was then found by solving for
$L_{\trm{\tiny ML}}$ in
\begin{equation}\label{ch4:eqn:EBcorrCmpL2}
	\left(\frac{L_{\trm{\tiny ML}}}{\widetilde{\sigma}_{\trm{\tiny RM}}}\right)^{\ms 2} - 
	\frac{L_{\trm{\tiny ML}} L}{\widetilde{\sigma}_{\trm{\tiny RM}}^{\ms 2}} \;
	\frac{\mathcal{I}_{\ms 1} \left(L_{\trm{\tiny ML}} L\right /
	\widetilde{\sigma}_{\trm{\tiny RM}}^{\ms 2})}
	{\mathcal{I}_{\ms 0} \left(L_{\trm{\tiny ML}} L /
	\widetilde{\sigma}_{\trm{\tiny RM}}^{\ms 2}\right)} +
	\gamma_{\trm{\tiny L}} = 0\;,
\end{equation}
where $\mathcal{I}_{\ms k}$ are modified Bessel functions of the first kind of order $k$.
To enable evaluation of $\gamma_{\trm{\tiny L}}$, the slope of the linearly polarized
differential component number-counts $dN/dL \propto L^{-\gamma_{\trm{\tiny L}}}$, we
constructed a model for $dN/dL$ by convolving the total intensity H03 distribution from
Equation~(\ref{ch4:eqn:H03}) with a probability distribution for fractional linear
polarization $f_{\ms \Pi}(\Pi) \equiv f_{\ms \Pi}(L/I)$ given by
\begin{equation}\label{ch5:eqn:fracpol}
	f_{\ms \Pi}\left(\Pi\right) =
	\frac{1}{\Pi \sigma_{\ms 10}\ln\!\left(10\right)\sqrt{2\pi}}
	\exp\Bigg\{\frac{-\left[\log_{\ms 10}\!\left(\Pi/\Pi_{\ms 0}\right)\right]^{\ms 2}}
	{2 \sigma_{\ms 10}^{\ms 2}}\Bigg\} \,,
\end{equation}
where $\Pi_{\ms 0}=4.0\%$ and $\sigma_{\ms 10}=0.3$. The motivation for using
Equation~(\ref{ch5:eqn:fracpol}) is described in detail in Paper II. We denote the
resulting $dN/dL$ model by $\tnm{H03}*f_{\ms \Pi}(\Pi)$. The calculated $L/L_{\trm{\tiny ML}}$
boosting ratios for components with low SNR detections over a range of flux densities are
displayed in Fig.~\ref{ch4:fig:IEB}.

Formally, Equations~(\ref{ch4:eqn:EBcorrCmpI}) and (\ref{ch4:eqn:EBcorrCmpL2}) are only valid
for constant slopes (i.e. for $\gamma_{\trm{\tiny S}}$ independent of flux density);
however, in practice, their solutions are valid provided that their input slopes do not
exhibit large changes as functions of flux density. Separately, we note that the
solutions above assume that all components are observed as truly unresolved; while this
assumption is not met by our data (real sources have non-zero extents), we estimate that
any resulting systematic errors due to slope miscalculation are much smaller than the
flux density uncertainties for each component.

\section{Cross-Identification and Classification}\label{ch4:SecClass}

To construct a catalogue of radio sources we implemented two cross-identification
and classification schemes. The first scheme was used to cross-match radio sources
comprising individual or multiple total intensity components with infrared and optical
counterparts, and to classify each of these sources according to their multiwavelength
properties. The second scheme was used to cross-match linearly polarized radio
components with their total intensity counterparts, to obtain polarization upper limits
for total intensity components and sources lacking catalogued polarized counterparts,
and where possible to classify these associations based on their polarized morphologies.
The two schemes are described as follows.

\subsection{Total Intensity Radio$-$Infrared$-$Optical Associations}\label{ch4:SecClassSubRIOCI}

\subsubsection{Cross-Identification}\label{ch4:SecClassSubRIOCI1}

We followed a similar procedure to that described by \citet{2006AJ....132.2409N}
and \citet{2008AJ....135.1276M} for ATLAS DR1 to cross-identify radio components
with SWIRE infrared sources and to identify radio sources comprising multiple
components. We began by utilising the DR1 radio component catalogues, searching them
for matches at the position of each DR2 total intensity component. If a DR1 component
was found within $5^{\prime\prime}$ of a DR2 position, then the SWIRE identification
found in DR1 was applied to the DR2 component. In the case of identifications made
by \citet{2008AJ....135.1276M} with the SWIRE Data Release 4 (SDR4) catalogue, which was
never fully published, a re-identification with the SDR3 catalogue was made by
searching for a SDR3 source within $1^{\prime\prime}$ of the SDR4 position. A small
number of SDR4 sources were not found in SDR3, in which case we assumed no
DR1-assisted cross-match was available for the DR2 radio component. For each remaining
DR2 component without an infrared cross-identification, we searched the SDR3
catalogue for an associated source within $7^{\prime\prime}$ using a nearest-neighbour
match; we did not take into account the infrared colours of potential matches. We then
examined a number of components by eye using radio contours superimposed
on the SWIRE 3.6~$\mu$m image, assessing the suitability of each cross-identification,
or lack thereof. We used this radio-contoured infrared image to identify nearby components
that were clearly physically associated with each other. In 36 cases, nearby components
were found to form a triple radio source in which a core of radio emission from a host galaxy
lay roughly midway between two radio lobes; we comment on our use of the terms {\it core}
and {\it lobe} below. Of these 36 cases, we found that 34
comprised 3 components, 1 comprised 4 components, and 1 comprised
6 components. In 78 cases, pairs of components were found to form a double radio source
comprising twin radio lobes with no detected emission from a core. In 41 cases, pairs of
components were found to exhibit core-lobe morphology. For each of these 155 multi-component
cases, we grouped the components together and assigned them to a common radio source;
each of the remaining 2066 radio components were assigned to a single-component radio
source. The ATLAS DR2 source catalogue thus comprised a total of 2221 sources. We found
that 149 of the multi-component sources and 1774 of the single-component sources were
matched with SWIRE sources, leaving a total of 298 radio sources without identifiable
infrared counterparts. Of these un-matched sources, only 2 of them (sources C5 and C318)
were the result of incomplete SWIRE coverage (97\%) over the ATLAS DR2 CDF-S survey area.
Finally, using the pre-matched infrared-optical data (\S~\ref{ch4:SecObsSubSubOpt}),
we associated 409 optical sources with DR2 radio sources.

We note that not all radio components were examined by eye as part of our cross-identification
procedure, and that a large number of components (perhaps $\sim200$ or more) are likely to
remain unassociated with true multi-component sources. Statistics regarding associations
between radio components and infrared sources are thus incomplete.
Furthermore, because the sky density of SWIRE sources ($\sim$60,000~deg$^{-2}$) is much
higher than that of ATLAS DR2 radio sources ($\sim$350~deg$^{-2}$), there is a chance
that some of our radio-infrared cross-identifications are incorrect. We have
not carried out an error analysis to estimate an upper limit to the false-positive
cross-identification rate for our data. However, we note that this upper limit was estimated
in DR1 as being $\sim$5\%; see \citet{2006AJ....132.2409N} and \citet{2008AJ....135.1276M} for
details. This rate is likely to be representative of our cross-matched DR2 data.
The issues above do not impact upon the key ATLAS results presented in Paper II.

\subsubsection{Classification}\label{ch4:SecClassSubRIOCIclass}

We classified each source according to whether their energetics were likely
to be driven by an AGN, star formation (SF) within a star-forming galaxy (SFG),
or emission associated with an individual star. Similar to \citet{2011ApJ...740...20P},
we define AGN sources as those with energetics dominated
in at least one wavelength band by a supermassive black hole. We have not split
sources containing an AGN into subclasses such as \citet{1974MNRAS.167P..31F}
type I (FRI; limb-darkened) and type II (FRII; also known as classical double or
triple radio sources due to their limb-brightened morphologies). Given that the
resolutions of our ATLAS data
($\sim10^{\prime\prime}$) often limited our ability to identify regions of emission
associated with AGN jets (FRI sources) compared with lobes formed about jet-termination
hotspots (FRII sources), we have systematically used the term {\it lobe} to describe
both jets and lobes in sources with radio double or triple morphologies.
For completeness, we note that our use of the term {\it core} in radio triple
sources is generic in that it does not indicate physical association with a compact,
flat-spectrum region of emission. Because spectral indices are not considered in
this work, restarted AGN jets or lobes may contribute or even dominate the emission
observed in the regions we have designated as cores. We provide our working definition
of SFGs further below.

We used four selection criteria to identify AGNs $-$ radio morphologies, 24~$\mu$m to
1.4~GHz flux density ratios, mid-infrared colours, and optical spectral characteristics
$-$ with the latter also used to identify SFGs and stars. We describe each of these
criteria below.

{\it Radio morphology.}$-$We classified each source exhibiting a lobe-core-lobe,
lobe-lobe, or core-lobe radio morphology as an AGN; 150 sources were
identified as AGNs by this criterion.

{\it Infrared-radio ratio.}$-$The linear and tight correlation between global
far-infrared (FIR) and radio emission from star-forming systems
\citep[e.g.][and references therein]{2010ApJ...717....1L,2010ApJS..186..341S},
known as the FIR-radio correlation (FRC), may be used to identify radio-loud AGN
due to their departure from this relationship \citep[e.g.][]{1991MNRAS.251P..14S}.
Following \citet{1985ApJ...298L...7H}, the FRC is commonly referred to by the
parameter $q$, which is the logarithm of the ratio between FIR to radio flux density.
\citet{2004ApJS..154..147A} found that
$q_{\ms 24} = \log_{\ms 10}[S_{24.0\,\mu\trm{\scriptsize m}}/S_{1.4\,\trm{\scriptsize GHz}}] = 0.8$
for flux density measurements at 24~$\mu$m and 1.4~GHz; we use this relationship as a
surrogate for the FRC. We classified each source
with a radio flux density more than ten times that expected from the FRC as an AGN,
namely for sources with $q_{\ms 24}\le-0.2$, including those sources with SWIRE
non-detections (limits are given in \S~\ref{ch4:SecObsSubSubIR}) meeting this criterion;
878 sources were classified as AGNs by this approach. Given the relative lack of multiwavelength
data included in this work, no corrections were made to convert observed 24~$\mu$m and
1.4~GHz flux densities to rest-frame values \citep[e.g.][]{2011ApJ...740...20P}, nor
were full K-corrections performed \citep[][and references therein]{1964ApJ...140..969K,
2010ApJS..186..341S}. However, we note that our $q_{\ms 24}\le-0.2$ scheme ensures that only
sources departing strongly from the FRC are classified as AGNs. It is therefore unlikely
that the corrections above would significantly alter our AGN classifications.
We note that \citet{2004ApJS..154..147A} and others
\citep[e.g.][]{2010ApJ...714L.190S,2011ApJ...731...79M} found no significant
evolution of the FRC with redshift \cite[though see][]{2010A&A...518L..31I}.

{\it Mid-infrared colours.}$-$Following the observation of a large sample of extragalactic
sources with the {\it Spitzer Space Telescope}, \citet{2004ApJS..154..166L} recognised
that the distribution of IRAC colours exhibited by AGNs extended into a region of
parameter space largely devoid of other source classes. \citet{2005ApJ...621..256S}
extended this work, investigating the parameter space occupied by continuum-dominated
sources for redshifts ranging between $z\sim0-2$ and investigating the colours
of SFG candidates dominated by polycyclic aromatic hydrocarbon (PAH) and sources
dominated by old-population (10~Gyr) starlight emission. Here we focus on the
continuum-dominated sources, displayed as blue points in the top two panels of
Fig.~10 from \citet{2005ApJ...621..256S}. We followed \citet{2011ApJ...740...20P}
to construct a locus for identifying AGNs, which we defined as the union of
$\log_{\ms 10}[S_{8.0\,\mu\trm{\scriptsize m}}/S_{4.5\,\mu\trm{\scriptsize m}}]>0$,
$\log_{\ms 10}[S_{5.8\,\mu\trm{\scriptsize m}}/S_{3.6\,\mu\trm{\scriptsize m}}]>0$,
and $\log_{\ms 10}[S_{8.0\,\mu\trm{\scriptsize m}}/S_{4.5\,\mu\trm{\scriptsize m}}]<
11\log_{\ms 10}[S_{5.8\,\mu\trm{\scriptsize m}}/S_{3.6\,\mu\trm{\scriptsize m}}]/9+0.3$,
and classified each source falling within its boundaries as an AGN; 238 sources
were classified as AGNs by this approach.

{\it Optical spectrum.}$-$Each optical spectrum was
classified visually by \citet{2012MNRAS.426.3334M} as an AGN, SFG, or star.
\citet{1999PASA...16..247S} reported that a similar `eyeball' classification
scheme for spectra obtained with the 2dF (Two-degree Field) spectrograph
\citep{2002MNRAS.333..279L} was robust and could
be used with confidence. Given this visual classification system, in this work
we define SFGs (somewhat loosely) as galaxies with SF rates sufficient to
produce an optical spectrum exhibiting (1) emission lines and line ratios
characteristic of SF, such as a strong and narrow H$\alpha$ line, and (2) a
distinct lack of features typically associated with AGN activity
\citep[see AGN/SF classification details in][]{1999PASA...16..247S}.
The latter criterion maintains consistency with our definition of AGN sources
above. Using the optical data, we classified 279 sources as AGNs, 126 as SFGs,
and 4 as stars. Of these, we found that 12 SFGs and 2 stars had been classified
as an AGN by one of the previously described AGN selection criteria, with an
additional 2 SFGs classified as an AGN by two of the previous criteria.
Given the high quality of the spectral classifications and the statistical
nature of the previous AGN diagnostics, we reclassified each of these
sources according to their optical classifications.

In summary, of the 2221 catalogued ATLAS DR2 sources, 1169 were classified as AGNs,
126 as SFGs, and 4 as radio stars. Of the AGN sources, 858 were recognised as such
by only one of the four diagnostics above, 255 were recognised by two, 47 by three,
and only 9 sources were recognised as an AGN by all four diagnostic criteria.
We note that our classifications are biased in favour of AGNs, due to the
overheads required to classify stars and SFGs using optical spectroscopy.
Therefore, in general our data are unsuited to the investigation of relationships
between star formation and AGN activity.

\subsection{Linear Polarization$-$Total Intensity Associations}\label{ch4:SecClassSubPA}

\subsubsection{Cross-Identification}\label{ch4:SecClassSubPACI}

To enable the investigation of fractional polarization trends, we visually
cross-matched each linearly polarized component with a total intensity counterpart.
In most cases it was possible to match an individual linearly polarized
component with an individual total intensity component and, in turn, their
associated multiwavelength counterpart from \S~\ref{ch4:SecClassSubRIOCI}. However, in some cases,
one-to-one matches were prevented due to ambiguities posed by the blending of
adjacent components in total intensity or linear polarization. For example,
we encountered situations in which a linearly polarized component was
positioned mid-way between two blended total intensity components, such that
it was unclear whether the polarized emission was caused by one of the total
intensity components, or both. Given such complexities in our data, we avoided
the use of a simple nearest-neighbour scheme for cross-matching, as this would
have led to overestimates of fractional polarization for any mis-matched components.
Instead, we grouped together all linearly polarized and total intensity components
contributing to an ambiguous cross-match, so that the fractional polarization could
be obtained for the group rather than for any potentially incorrect subset of the group.
In total, we found that 130 of the 2221 catalogued ATLAS DR2 sources exhibited linearly
polarized emission, 118 of which had available infrared cross-identifications.
Statistics of one-to-one and group associations are presented below.

In Fig.~\ref{ch4:fig:ILassoc} we display the four types of cross-matches encountered
in our data.
\begin{figure*}
\begin{minipage}{140mm}
\centering
 \includegraphics[trim = 0mm 50mm 0mm 51mm, clip, width=65mm]{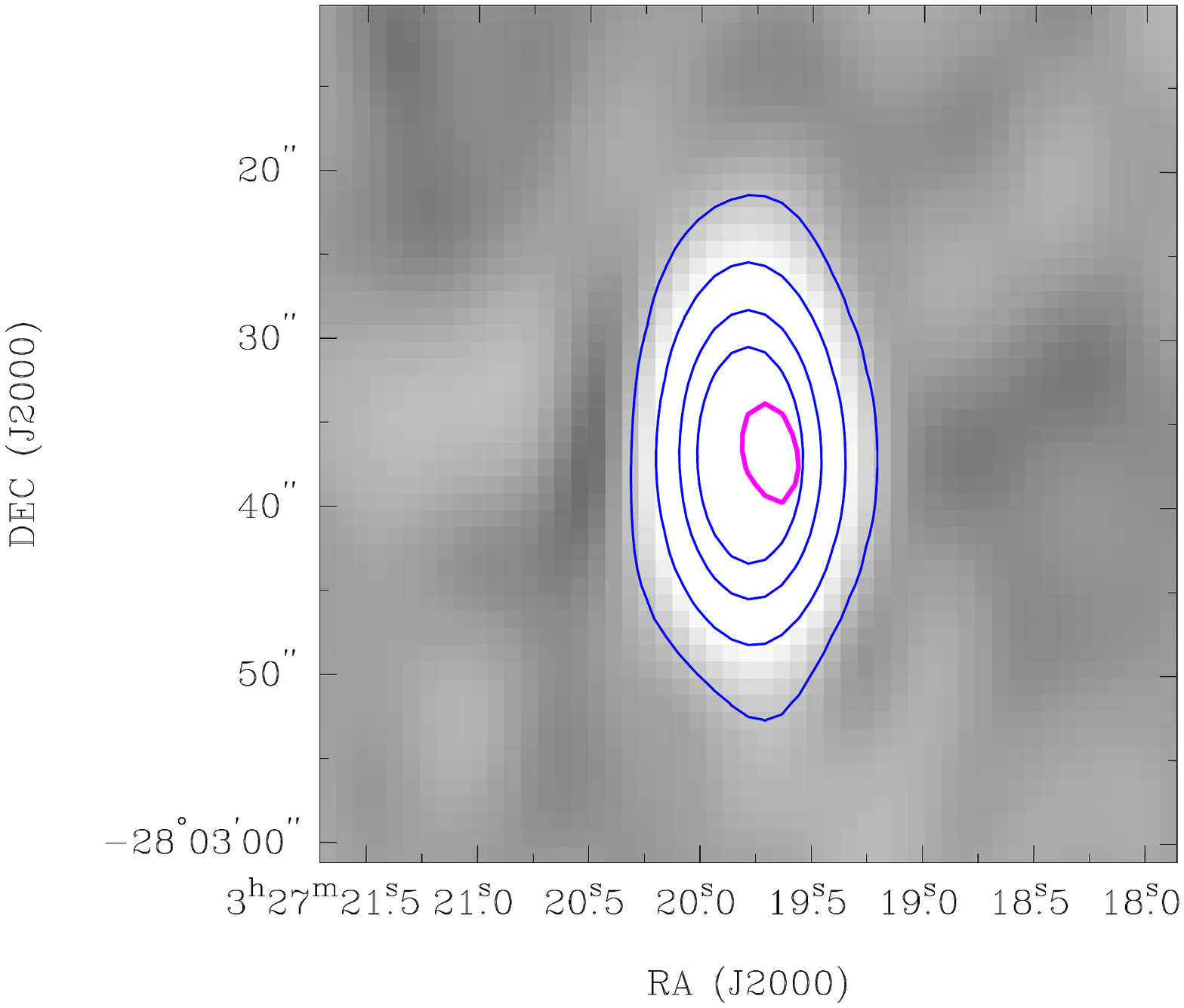}
 \hspace{0.5pc}
 \includegraphics[trim = 0mm 50mm 0mm 51mm, clip, width=65mm]{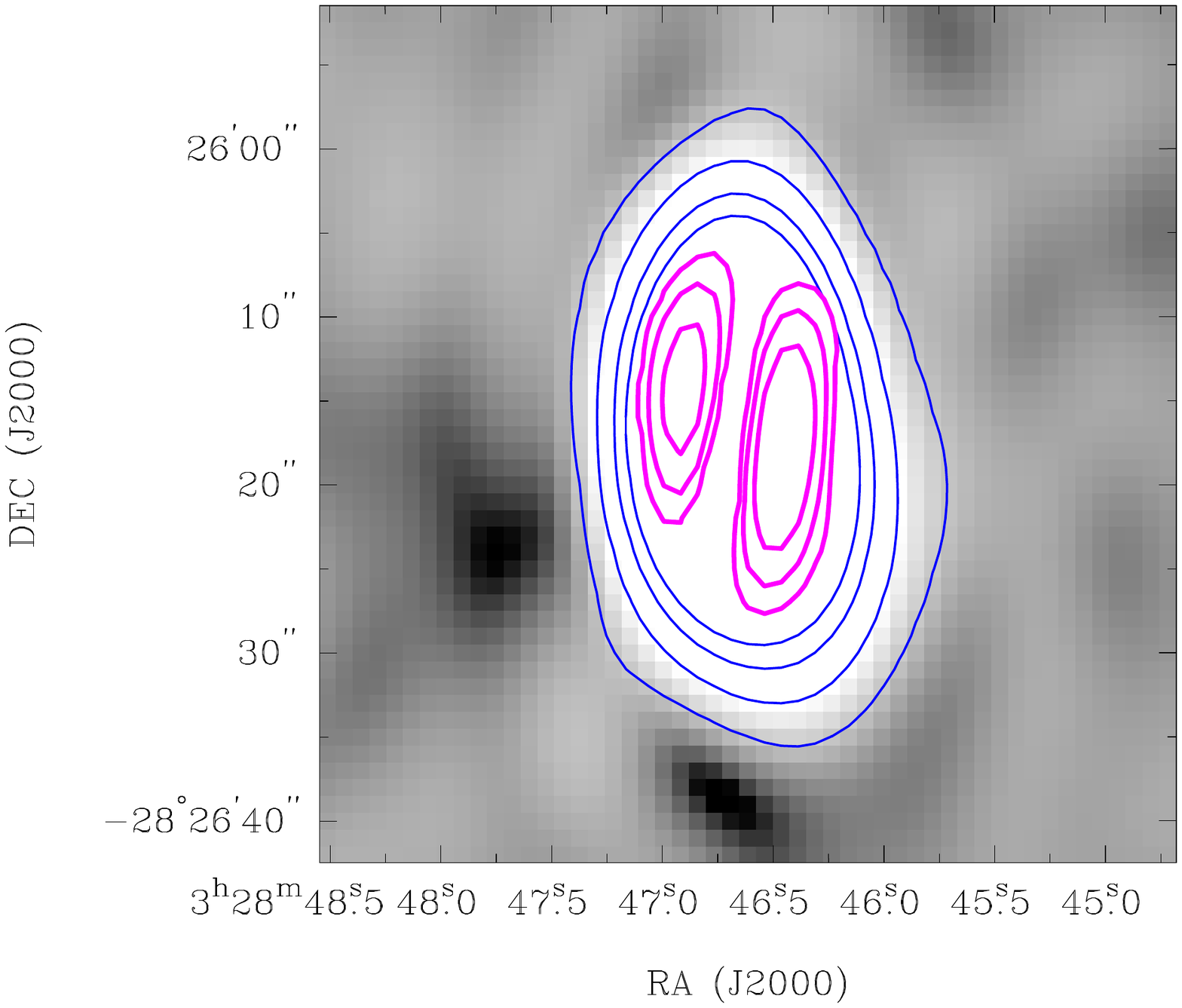}\\
 \includegraphics[trim = 0mm 50mm 0mm 51mm, clip, width=65mm]{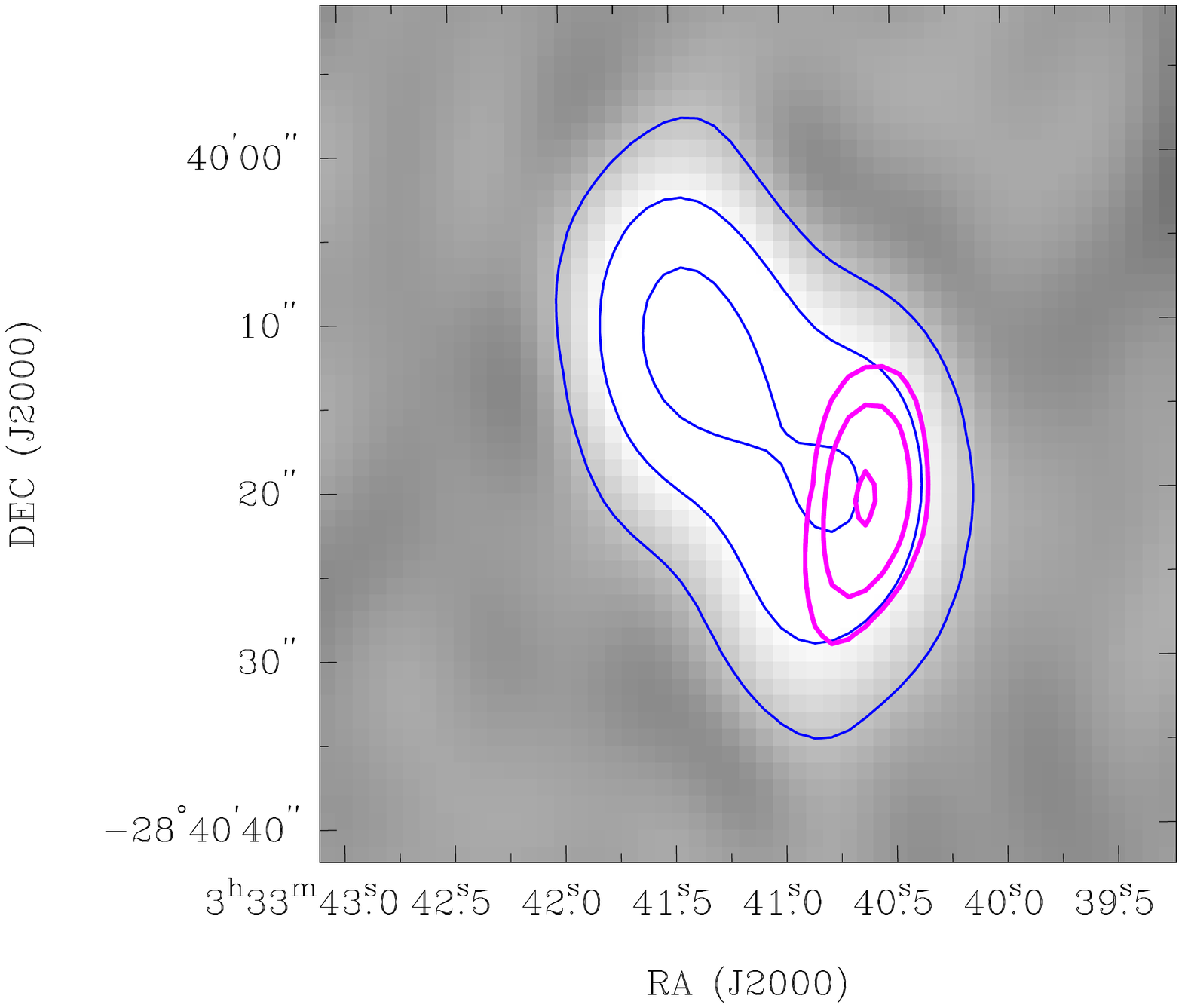}
 \hspace{0.5pc}
 \includegraphics[trim = 0mm 50mm 0mm 51mm, clip, width=65mm]{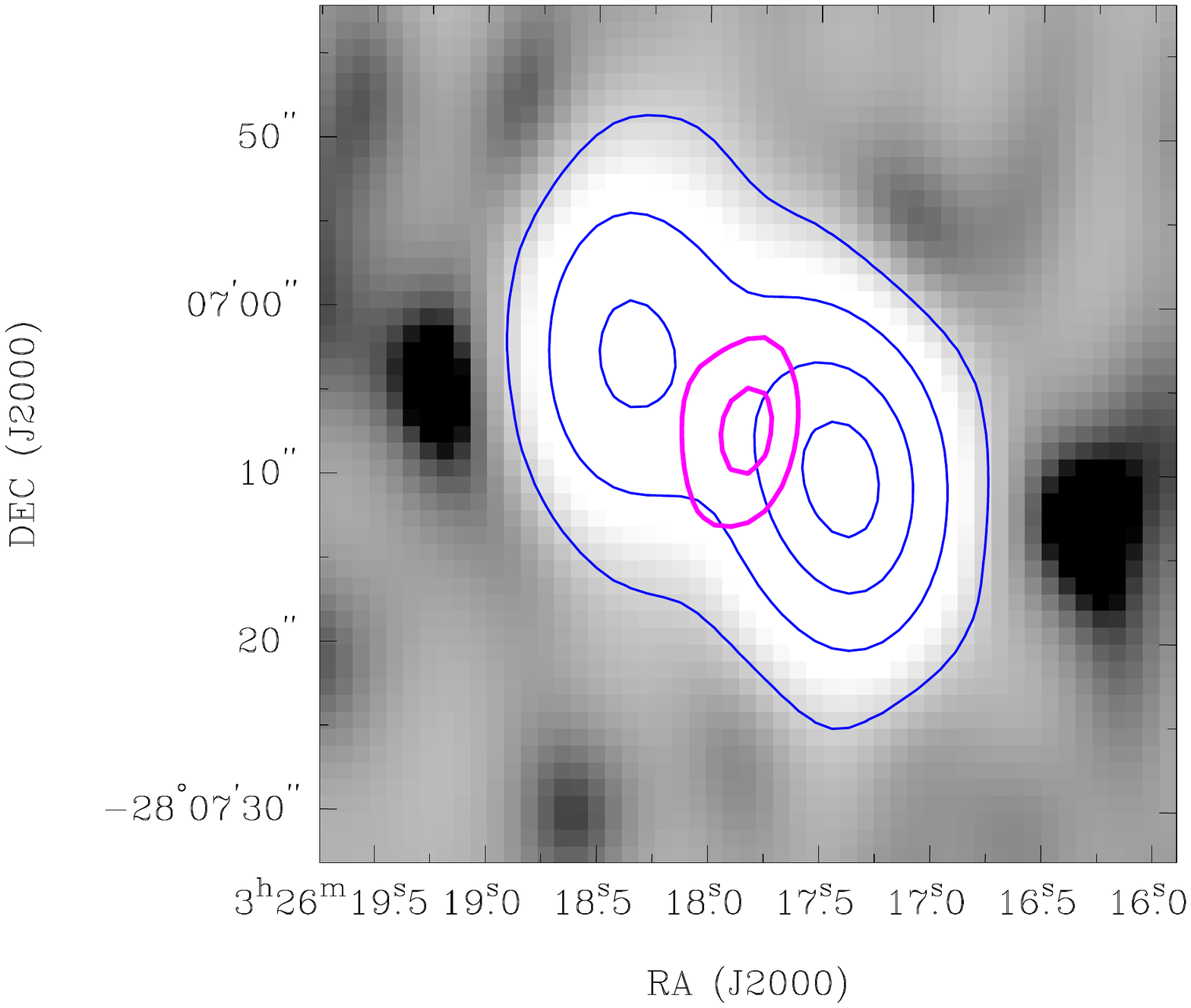}
 \caption{	Examples of one-to-one (left column panels) and complex
 		(right column panels) cross-identifications encountered
		between linearly polarized components and total intensity
		counterparts.
		Background total intensity images in each panel are shaded
		logarithmically, saturating black below -0.2~mJy~beam$^{-1}$
		and white above 1~mJy~beam$^{-1}$.
		Total intensity contours (blue) represent 5, 25, 60, and
		$100\sigma$. Linear polarization contours (magenta) represent
		6.25, 10, and $15\sigma_{\trm{\tiny RM}}$.
		Respective beam sizes are given in \S~\ref{ch4:SecDatSubMFS}
		and \S~\ref{ch4:SecDatSubPCI}. See \S~\ref{ch4:SecClassSubPACI}
		for panel details.
 }
 \label{ch4:fig:ILassoc}
\end{minipage}
\end{figure*}
The top-left panel shows a one-to-one match between a linearly polarized
component and a total intensity component. The bottom-left panel shows a
two-component total intensity source exhibiting limb-brightened
linearly polarized emission, which we interpret as an unambiguous
one-to-one match between the polarized component and the western total intensity
component; the eastern component is undetected in polarization. The top-right
panel shows the only example in our data where a single total intensity component
was found to be enveloping two separate linearly polarized components;
all three components were assigned to a group. We note that this example
is likely to be highlighting a total intensity component that should
have been decomposed into two separate components during the
{\tt BLOBCAT}/{\tt IMFIT} extraction phase, rather than a perfectly Gaussian
total intensity component with unusual polarization substructure. The
bottom-right panel shows an example of an ambiguous match between a
linearly polarized component and two total intensity components; all
three components were grouped together to prevent potential overestimation
of the fractional polarization for either total intensity component.

To enable the investigation of fractional polarization trends using all
available radio data, not just using the one-to-one and group associations
identified above, we calculated upper limits to the linearly polarized
flux densities of all total intensity components lacking a polarization
counterpart. As discussed extensively by \citet{2010ApJ...719..900K},
we note the distinction between an upper limit and an upper bound\footnote{
An upper bound describes an inference range for a flux density measurement;
an upper limit, on the other hand, describes the minimum flux density required
to ensure detection at a specified false-positive (Type I) error rate (i.e. a
SNR cutoff) and false-negative (Type II) error rate for a given noise distribution,
and thus calibrates the detection process irrespective of the observed
flux density.}. We followed the procedure outlined by \citet{2010ApJ...719..900K}
to evaluate polarization upper limits at the positions of unpolarized
total intensity components. By combining the Type I error rate of $\alpha=10^{-7}$
from \S~\ref{ch4:SecExtSubFlood}, a Type II error rate conservatively defined as
$\beta=0.9$, and the probability density function (PDF) for $L_{\trm{\tiny RM}}$
with $M=28$ given by Equation~(28) from \citet{2012MNRAS.424.2160H}, we evaluated
that the upper limit definition required to meet these statistical criteria was
\begin{equation}\label{ch4:eqn:Lupper}
	L_{\trm{\tiny UL}}=7.46\,\sigma_{\trm{\tiny RM}}\,.
\end{equation}
For unpolarized sources, we assigned polarization upper limits by selecting
the weakest limit (i.e. largest in magnitude) associated with any
constituent total intensity component.

\subsubsection{Classification}\label{ch4:SecClassSubPAClass}

\begin{figure*}
\vspace{2pc}
\begin{minipage}{140mm}
\centering
 \includegraphics[trim = 0mm 50mm 0mm 50mm, clip, width=66mm]{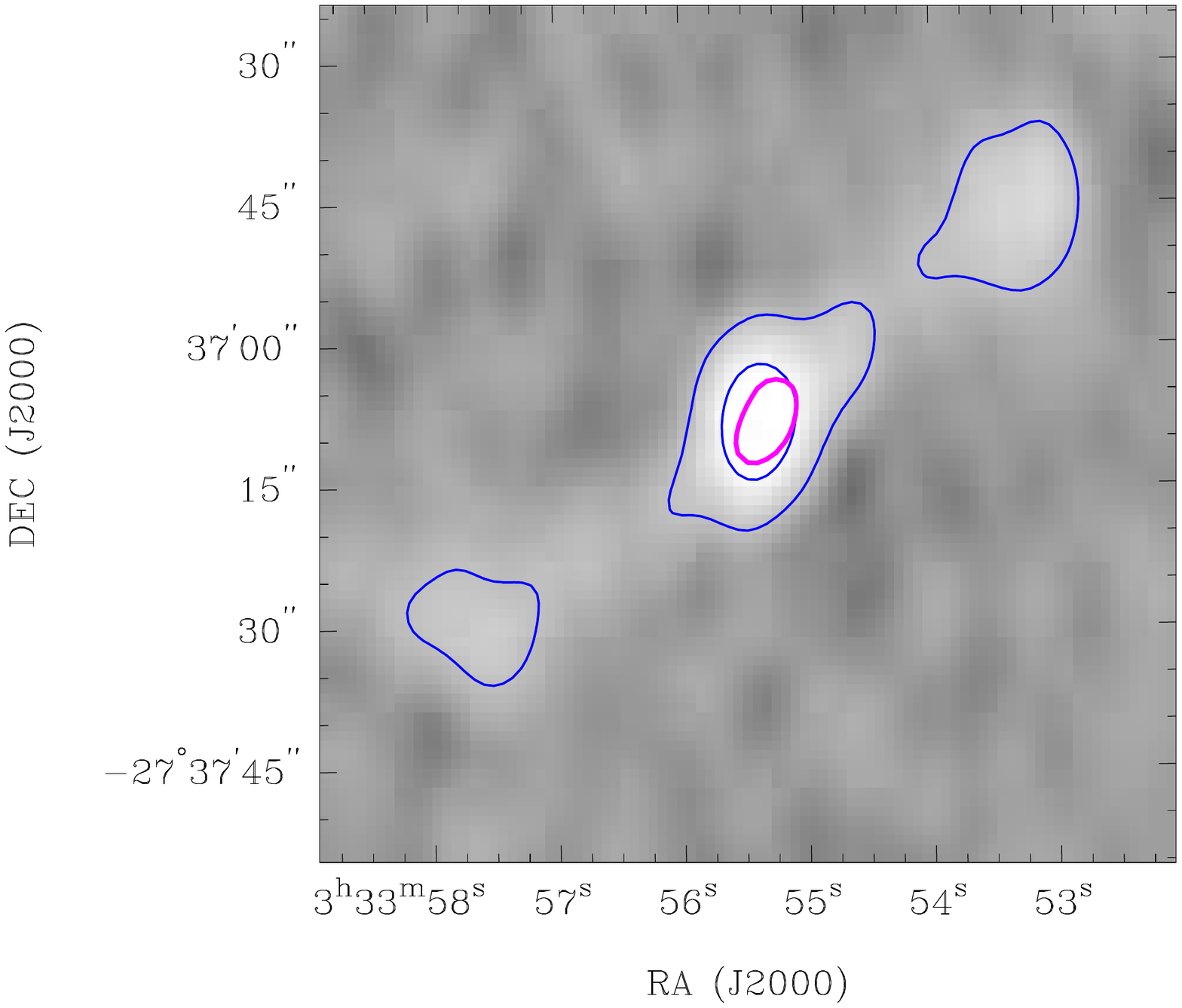}
 \hspace{0.5pc}
 \includegraphics[trim = 0mm 50mm 0mm 50mm, clip, width=66mm]{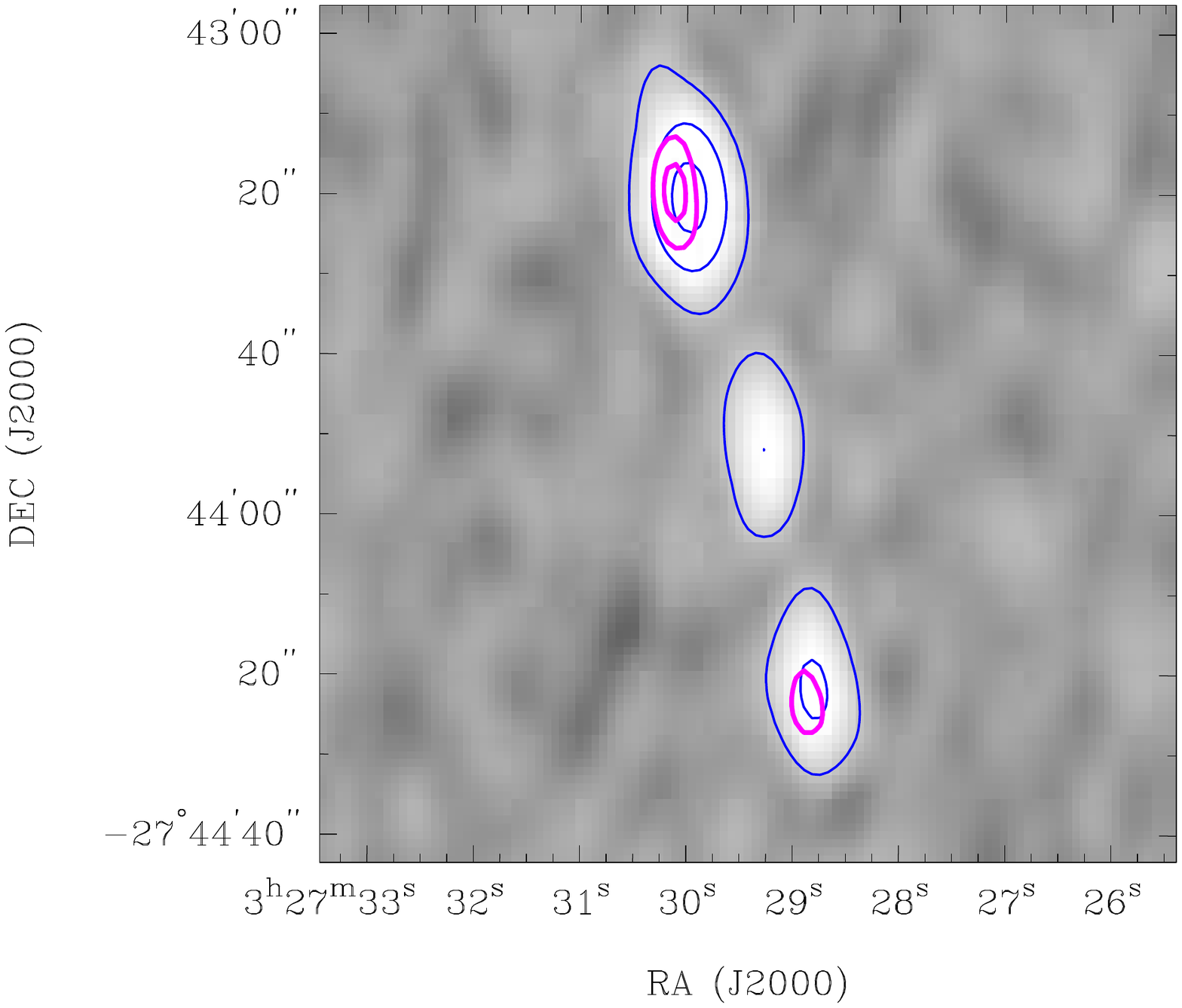}
 \caption{	Examples of linear polarization$-$total intensity classifications
 		(see also Fig.~\ref{ch4:fig:ILassoc} and Fig.~\ref{ch4:fig:ILdepol}).
 		Panel shading and contours are identical to Fig.~\ref{ch4:fig:ILassoc}.
 		The left panel shows a Type~5 association. The right panel shows a Type~6
		source with two Type~0 lobes.
 }
 \label{ch4:fig:ILassocTypes}
\end{minipage}
\end{figure*}

\begin{figure*}
\begin{minipage}{140mm}
\centering
 \includegraphics[trim = 0mm 50mm 0mm 40mm, clip, width=65mm]{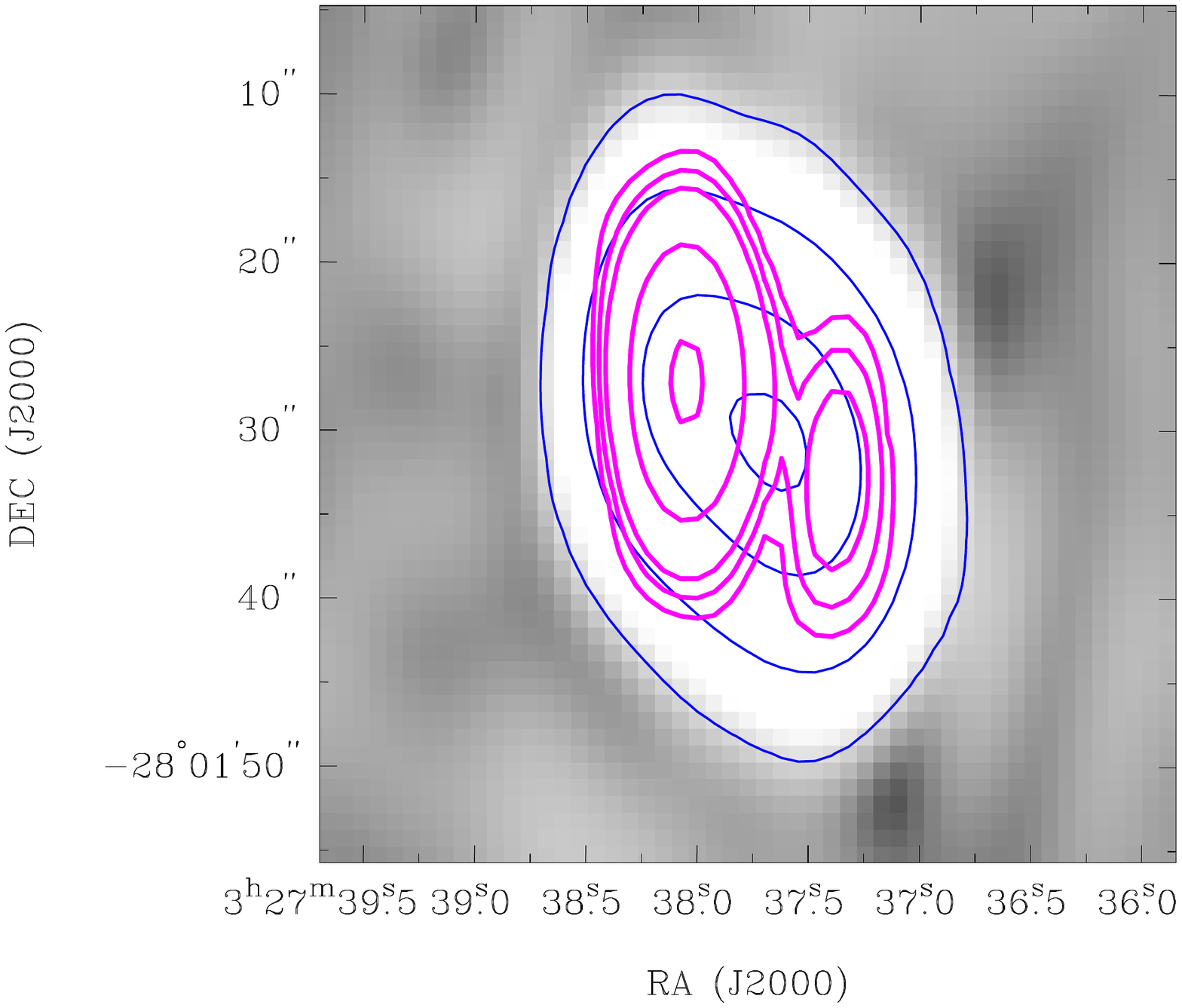}
 \hspace{0.5pc}
 \includegraphics[trim = 0mm 50mm 0mm 40mm, clip, width=65mm]{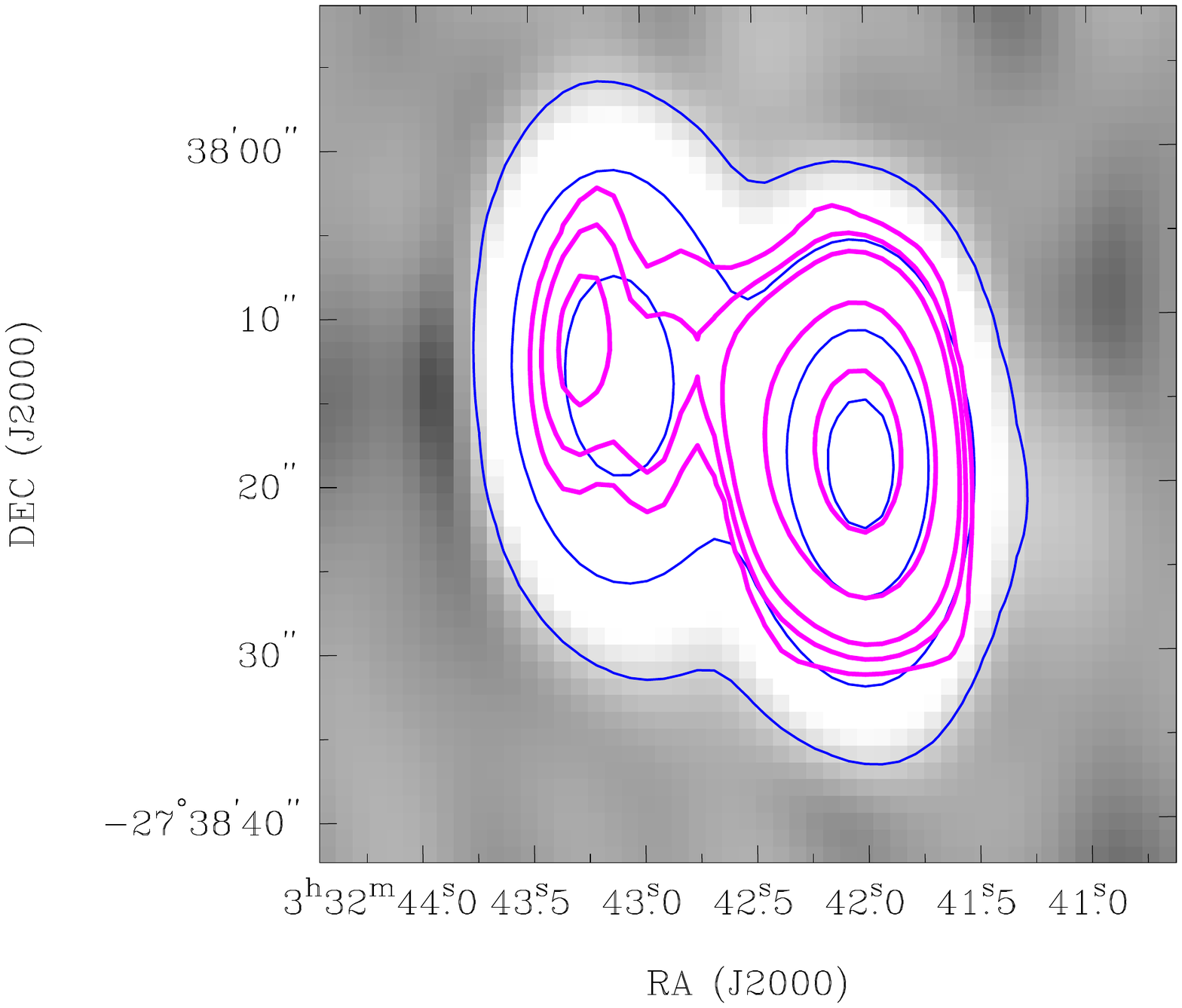}
 \caption{	Classical double sources in ATLAS that appear to exhibit asymmetric
 		depolarization. Each lobe was classified as Type 1, and thus each source
		as Type 7. The left panel displays source C7. This source is
		best fit by two Gaussian components in total intensity, the centroids
		of which correspond to the components observed in linear polarization.
		Flux densities for the Eastern and Western lobes are $I=30.6$~mJy
		and $L=2.3$~mJy, and $I=64.5$~mJy and
		$L=0.5$~mJy, respectively. The right panel displays
		source C8. Flux densities for the Eastern and Western lobes
		are $I=38.8$~mJy and $L=0.3$~mJy, and
		$I=55.4$~mJy and $L=2.6$~mJy, respectively.
		Background shading levels in each panel are identical to
		Fig.~\ref{ch4:fig:ILassoc}. Total intensity contours
		(blue) represent 10, 100, 500, and $1000\sigma$. Linear
		polarization contours (magenta) represent 6.25, 10, 15, 40,
		and $90\sigma_{\trm{\tiny RM}}$.
 }
 \label{ch4:fig:ILdepol}
\end{minipage}
\end{figure*}

We visually classified each one-to-one and group association from above,
each source comprising two such associations (no source had more than two), and each
unpolarized component according to the following scheme, which we designed to
account for differing (de-)polarized morphologies. Examples of each classification
type are displayed in Fig.~\ref{ch4:fig:ILassoc}, Fig.~\ref{ch4:fig:ILassocTypes},
and Fig.~\ref{ch4:fig:ILdepol}, as described below.
\begin{itemize}

\setlength{\itemindent}{2.6em}
\item[{\it Type 0}] $-$ A one-to-one or group association identified as a lobe
of a double or triple radio source. Both lobes of the source are clearly
polarized, having linearly polarized flux densities within a factor of 3.
(These criteria do not formally reference the ratio between lobe total intensity
flux densities, which we note here are within a factor of 3 for all
double or triple ATLAS DR2 sources; cf. \citealt{1998MNRAS.300..257M}.) To illustrate,
two Type~0 associations are displayed in the right panel of Fig.~\ref{ch4:fig:ILassocTypes},
one for each lobe.

\setlength{\itemindent}{4.1em}
\item[{\it Types 1/2}] $-$ A one-to-one or group association identified as a lobe
of a double or triple radio source that does not meet the criteria for Type 0.
A lobe classified as Type 1 indicates that the ratio of polarized flux densities
between lobes is greater than 3. A lobe
classified as Type 2 indicates that the opposing lobe is undetected in polarization
and that the polarization ratio may be less than 3, in which case it is possible
that more sensitive observations may lead to re-classification as Type 0. Sources
with lobes classified as Type 1 exhibit asymmetric depolarization in a manner
qualitatively consistent with the Laing-Garrington effect \citep{1988Natur.331..149L,
1988Natur.331..147G}, where one lobe appears more fractionally polarized than the opposite
lobe. To illustrate, Type~1 associations are suitable for the pair of lobes displayed in each panel
of Fig.~\ref{ch4:fig:ILdepol}. A Type~2 classification is appropriate for the detected lobe
shown in the bottom-left panel of Fig.~\ref{ch4:fig:ILassoc}.

\setlength{\itemindent}{2.6em}
\item[{\it Type 3}] $-$ A group association representing a source, involving a
linearly polarized component situated midway between two total intensity components.
It is not clear whether such associations represent two polarized lobes, a polarized
lobe adjacent to a depolarized lobe, or a polarized core. An example is displayed in
the bottom-right panel of Fig.~\ref{ch4:fig:ILassoc}.

\item[{\it Type 4}] $-$ An unclassified one-to-one or group association representing
a source. Examples of the former and latter are displayed in the top-left and top-right
panels of Fig.~\ref{ch4:fig:ILassoc}, respectively.

\item[{\it Type 5}] $-$ A one-to-one association clearly identified as the
core of a triple radio source (where outer lobes are clearly distinct from
the core). An example is displayed in the left panel of Fig.~\ref{ch4:fig:ILassocTypes}.

\item[{\it Type 6}] $-$ A source comprising two Type 0 associations, or
a group association representing a non-depolarized double or triple radio source
where blended total intensity and linear polarization components have prevented
clear subdivision into two Type 0 associations. For example, a Type~6 source
is displayed in the right panel of Fig.~\ref{ch4:fig:ILassocTypes}.

\item[{\it Type 7}] $-$ A source comprising one or two Type 1 associations. For example,
each panel of Fig.~\ref{ch4:fig:ILdepol} displays a Type~7 source.

\item[{\it Type 8}] $-$ A source comprising one Type 2 association. For example,
a Type~8 source is displayed in the bottom-left panel of Fig.~\ref{ch4:fig:ILassoc}.

\item[{\it Type 9}] $-$ An unpolarized component or source.

\end{itemize}

From a total of 172 catalogued linearly polarized components, 138 were found
to exhibit clear one-to-one associations with individual total intensity
components. The remaining 34 polarized components required grouping in order to
be associated with total intensity counterparts. We classified 58 one-to-one
associations as Type 0, 4 as Type 1, 25 as Type 2, 48 as Type 4, and 3 as Type 5.
We note that all 3 sources containing Type 5 core associations exhibited unpolarized
lobes. Of the group associations comprising a total of 34 polarized components,
2 groups were classified as Type 0, 14 as Type 3, 1 as Type 4, and 8 as Type 6.
We classified 29 sources comprising two Type 0 associations as Type 6.
We classified 2 sources as Type 7, each of which exhibited linearly polarized
emission from both the polarized and depolarized lobe, and 25 sources as Type 8.

As described above, only 1 group association was classified as Type 4 (see top-right
panel of Fig.~\ref{ch4:fig:ILassoc}). While it is possible
that the two polarized components within this group are in reality a single
extended region of polarized emission, broken in two by a depolarization canal
\citep[e.g.][]{2006MNRAS.371L..21F}, a more likely explanation as commented on above
is that rather than there being a single total intensity
component, two adjacent polarized lobes are in fact present.

\section{Component Number-Count Biases}\label{ch4:SecCNC}

We used the DR2 component catalogue to construct 1.4~GHz differential component
counts in total intensity and linear polarization for each ATLAS field. We did
not use the DR2 source catalogue to construct differential source counts because
of concern regarding the multi-component association process (recall \S~\ref{ch4:SecClassSubRIOCI1}).
The resulting component counts will be presented in Paper II; here we
describe our method.

The differential component counts were calculated by dividing the
number of sky density normalised components (i.e. units of sr$^{-1}$)
observed within each flux density bin by the bin width, then multiplying
each bin value by two bias correction factors. The effective number of
components in each $i$'th flux density bin was thus calculated as
\begin{equation}\label{ch4:eqn:Neff}
	N_{\ms \tnm{\tiny eff},i} =
	r_{\ms i}\,e_{\ms i}
	\sum_{\ms j=1}^{\ms J_i}
	\left(V^{\trm{\tiny AREA}}_{\ms j} \, F^{\trm{\tiny AREA}}
	\right)^{\ms -1}\;,
\end{equation}
where $F^{\trm{\tiny AREA}}$ denotes the relevant field area from
\S~\ref{ch4:SecInstSubArea}, and the visibility area term
$V^{\trm{\tiny AREA}}_{\ms j}$ accounts for the potentially limited survey area
over which each $j$'th of $J_i$ components in each bin could have been detected due to
spatial variations in image sensitivity and bandwidth smearing (see \S~\ref{ch4:SecExtSubFlood}).
Only bins with visibility area factors greater than 0.1 were
accepted for the number count results presented in Paper~II.
The correction factors $r_{\ms i}$ and $e_{\ms i}$ were used to account for
resolution bias and Eddington bias, respectively, as described in
\S~\ref{ch4:SecCNCSubRB} and \S~\ref{ch4:SecCNCSubEB} below.
The differential counts, representing the number of components
per unit sky area per unit flux density, were then normalised by the standard
Euclidean slope of $S^{-2.5}$ \citep{1966MNRAS.133..421L,1968ARA&A...6..249R}.

\subsection{Resolution Bias}\label{ch4:SecCNCSubRB}

We use the term resolution bias to collectively describe two effects: (1)
incompleteness in number-count bins resulting from a lack of sensitivity to
resolved components with low surface brightness, and (2) the redistribution of
counts between bins resulting from systematic undervaluation of flux densities
for components classified as unresolved. An analytic scheme to account for the first
effect has been attempted by \citet{2001A&A...365..392P} and \citet{2005AJ....130.1373H}.
The second effect was identified in an empirical investigation by \citet{2008ApJ...681.1129B};
an analytic formalism to describe this effect is not presently available. In this section we
present a new analytic method that both improves upon the scheme described by
\citet{2001A&A...365..392P} and \citet{2005AJ....130.1373H} and accounts for
the bias described by \citet{2008ApJ...681.1129B}.

\subsubsection{Effect 1: Sensitivity to Resolved Components}\label{ch4:SecCNCSubRB1}

We begin by discussing incompleteness to resolved components, which may be
manifested in two ways.

First, a lack of short baselines can limit the maximum observable angular size of
components. For a minimum projected baseline of 30~m, at 1.4~GHz the ATCA becomes
progressively insensitive\footnote{Joint deconvolution schemes can recover larger
scales than those from single pointing schemes \citep{1996A&AS..120..375S}; computational
limitations prevented joint deconvolution of the ATLAS data.} to components larger
than $5^\prime$, at which point only 50\% of a component's true flux density can be
detected \citep[e.g.][]{forster}. Given that no millijansky sources are expected to
exhibit such large angular sizes (according to any of the distributions described below)
and that ATLAS observations include projected baselines down to 30~m (see
\S~\ref{ch4:SecObsSubRadio}), we assume that no limitations have been imposed on
observable component angular sizes by ATLAS $uv$-plane coverage.

Second, components with flux densities sufficient to be included in a number-count bin
may be resolved to the extent that their peak surface brightnesses may fall below the
SNR detection threshold, preventing them from being catalogued and counted and thus
resulting in bin incompleteness.
To correct for this second type of incompleteness to resolved components in the
total intensity and linear polarization number-counts for each ATLAS field, we
estimated the fraction of missing components at any given flux density by
comparing the maximum detectable angular size with an underlying true size
distribution.

We estimated the maximum intrinsic (i.e. deconvolved) angular size,
$\Theta_\tnm{\scriptsize max}(S)$, that a component with flux density $S$ could attain
while still meeting the detection threshold by modifying Equation~(\ref{ch4:eqn:resunres})
and deconvolving using Equation~(\ref{ch4:eqn:deconvest}), deriving
\begin{eqnarray}
	\left[\Theta_\trm{\scriptsize max}(S)\right]^{2} &=&
	\Bigg\{	\int_{0}^{S/A_\tnm{\scriptsize S}}
	\sqrt{\frac{S \, B_\trm{\scriptsize maj} \, B_\trm{\scriptsize min}}
	{A_\tnm{\scriptsize S} \, z}}\,
	f_{\widetilde{\sigma}}\!\left(z\right) dz \, \times \nonumber \\
	& & \left[\int_{0}^{S/A_\tnm{\scriptsize S}}
	f_{\widetilde{\sigma}}\!\left(z^{\ms\prime}\right)
	dz^{\ms\prime}\right]^{-1} \Bigg\}^2 -
	B_\trm{\scriptsize maj} \, B_\trm{\scriptsize min} \, , \label{ch4:eqn:maxang}
\end{eqnarray}
where $A_\tnm{\scriptsize S}$ is the SNR threshold given by $5.0$ in total
intensity or $6.25$ in linear polarization, we have defined
$\widetilde{\sigma}(x,y)=\sigma(x,y)/\varpi(x,y)$ (or using
$\sigma_{\trm{\tiny RM}}$ in polarization),
and where $f_{\widetilde{\sigma}}$ is a probability distribution
for $\widetilde{\sigma}$ [in practice this is a normalised histogram of
$\widetilde{\sigma}(x,y)$ values]. The integrals in Equation~(\ref{ch4:eqn:maxang})
enable $\Theta_\tnm{\scriptsize max}(S)$ to be calculated as a weighted average,
taking into account spatial variations in both sensitivity and bandwidth
smearing (i.e. variations in $\widetilde{\sigma}$) over each survey area.
The upper limit to each integral gives the maximum value of $\widetilde{\sigma}(x,y)$
for any given flux density $S$, above which not even an ideally unresolved
component could be observed above the detection threshold. Therefore, at faint
flux densities, the weighted average of observed angular sizes (square root
term) is not computed using the full distribution of $\widetilde{\sigma}$, but
rather a renormalised distribution in which the term in square brackets has value
less than unity.

\begin{figure*}
\centering
 \includegraphics[trim=20mm 30mm 10mm 30mm, height=85mm]{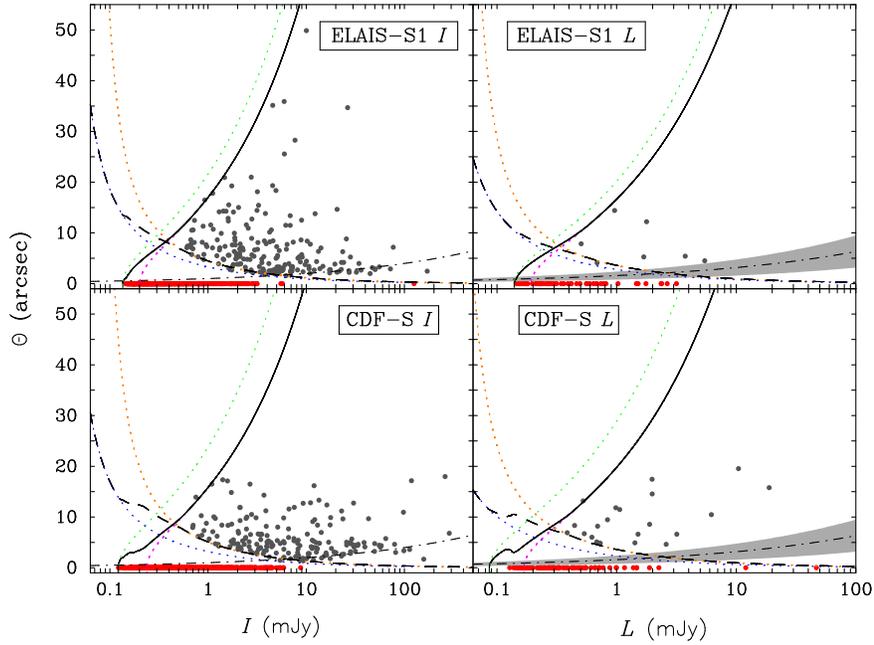}
 \begin{minipage}{140mm}
 \caption{	Deconvolved angular size as a function of flux density for ATLAS
 		total intensity (left column) and linearly polarized (right column)
		components in the ELAIS-S1 (top row) and CDF-S (bottom row) fields.
 		For visual clarity, angular sizes of unresolved components (red points)
		are displayed at zero, rather than at their upper bounds. The solid
		curve in each panel indicates the maximum angular size above which
		a component's peak surface brightness will drop below the survey
		detection threshold, as defined by Equation~(\ref{ch4:eqn:maxang}).
		The dashed curve in each panel indicates the minimum angular size
		required of a component to be classified as resolved and
		deconvolved, as defined by Equation~(\ref{ch4:eqn:minang}). The
		dot-dashed curve in each panel indicates the median of the assumed
		underlying true size distribution, with associated shaded regions in
		the right panels indicating filling-factor uncertainties; see text in
		\S~\ref{ch4:SecCNCSubRB1} for details. The dotted curves indicate
		limiting behaviours of the solid and dashed curves,
		given by Equations~(\ref{ch4:eqn:maxangLow})
		(green), (\ref{ch4:eqn:maxangHigh}) (magenta), (\ref{ch4:eqn:minangLow})
		(blue), and (\ref{ch4:eqn:minangHigh}) (orange).
	}
 \label{ch4:fig:AngSize}
 \end{minipage}
\end{figure*}

In Fig.~\ref{ch4:fig:AngSize} we plot the deconvolved angular sizes
of ATLAS DR2 components and indicate the locus defined by Equation~(\ref{ch4:eqn:maxang})
for each survey area (solid curves).
For clarity, we characterise the limiting behaviour of Equation~(\ref{ch4:eqn:maxang}) at
low and high flux densities by defining two simplified versions of this Equation,
which we plot as dotted curves about the solid curve in each panel
of Fig.~\ref{ch4:fig:AngSize}. The first of these uses the minimum effective noise
$\widetilde{\sigma}_\tnm{\scriptsize min}=\min\left[\widetilde{\sigma}(x,y)\right]$
to characterise the maximum angular size $\Theta_{\tnm{\scriptsize max}^\prime}(S)$
at all flux densities, given by
\begin{equation}\label{ch4:eqn:maxangLow}
	\left[\Theta_{\trm{\scriptsize max}^\prime}(S)\right]^{2} =
	\frac{S \, B_\trm{\scriptsize maj} \, B_\trm{\scriptsize min}}
	{A_\tnm{\scriptsize S} \, \widetilde{\sigma}_\tnm{\scriptsize min}} -
	B_\trm{\scriptsize maj} \, B_\trm{\scriptsize min} \, ,
\end{equation}
where $\widetilde{\sigma}_\tnm{\scriptsize min}$ is 24 (14) and 27
(22)~$\mu$Jy~beam$^{-1}$ in the CDF-S and ELAIS-S1 total intensity (polarization)
fields, respectively. The second definition uses the full effective noise
distribution at all flux densities to evaluate a weighted maximum angular size
$\Theta_{\tnm{\scriptsize max}^{\prime\prime}}(S)$, given by
\begin{equation}\label{ch4:eqn:maxangHigh}
	\left[ \Theta_{\trm{\scriptsize max}^{\prime\prime}}(S)\right]^{2} =
	\left[ \int_{0}^{\infty}
	\sqrt{\frac{S \, B_\trm{\scriptsize maj} \, B_\trm{\scriptsize min}}
	{A_\tnm{\scriptsize S} \, z}}\,
	f_{\widetilde{\sigma}}\!\left(z\right) dz
	\right]^2 -
	B_\trm{\scriptsize maj} \, B_\trm{\scriptsize min} \; .
\end{equation}
Equation~(\ref{ch4:eqn:maxang}) limits to Equation~(\ref{ch4:eqn:maxangLow}) at faint flux
densities and to Equation~(\ref{ch4:eqn:maxangHigh}) at higher flux densities. We note that
if $\Theta_\trm{\scriptsize max}(S)$ were defined using a fixed minimum noise
value, as in Equation~(\ref{ch4:eqn:maxangLow}), then maximum angular sizes would in general
be overestimated at all flux densities (or underestimated if the maximum noise value was
selected). Similarly, if $\Theta_\trm{\scriptsize max}(S)$ were defined without taking
into account the visibility area associated with component detection at faint
flux densities, as in Equation~(\ref{ch4:eqn:maxangHigh}), then the maximum angular
sizes estimated at faint flux densities would be significantly underestimated; the
relevant dotted curves in Fig.~\ref{ch4:fig:AngSize} fall to zero angular size at flux
densities higher than the faintest observed ATLAS components, indicating that
Equation~(\ref{ch4:eqn:maxangHigh}) may not be used to estimate
$\Theta_\trm{\scriptsize max}(S)$ at faint $S$.

To model the underlying true size distribution for {\it components} in total intensity,
we modified the integral angular size distribution presented by \citet{1990ASPC...10..389W}
for 1.4~GHz {\it sources}. The \citet{1990ASPC...10..389W} distribution gives the
fraction of sources with largest angular size (LAS) greater than $\Theta$, and is
parameterised as
\begin{equation}\label{ch4:eqn:windT}
	h\!\left(>\Theta,S\right) = 2^{-\left(\Theta/\Theta_\tnm{\tiny med,I}\right)^{0.62}}\;,
\end{equation}
where $\Theta_\tnm{\tiny med,I}$ is the median LAS as a function of flux density given by
$2\farcsecd0\left(S_{\ms 1.4 \tnm{\tiny GHz}}/1\;\tnm{\small mJy}\right)^{0.3}$.
The density function corresponding to Equation~(\ref{ch4:eqn:windT}) is
\begin{equation}\label{ch4:eqn:windTpdf}
	f_{\ms \Theta}\!\left(\Theta,S\right) =
	\frac{0.62 \, \ln2}{\Theta_\tnm{\scriptsize med,I}}
	\left(\frac{\Theta}{\Theta_\tnm{\scriptsize med,I}}\right)^{\!-0.38}
	h\!\left(>\Theta,S\right) \, .
\end{equation}
The LAS of a source characterises its largest angular extent. The LAS for a
single-component source is given by its deconvolved angular size. The LAS for
a multi-component source is given by the maximum angular separation between
its components, or if greater, the largest deconvolved angular size of any of
its components. We note that, in principle, there are no resolution bias
constraints preventing the detection of multi-component sources with arbitrarily
large LASs, provided that their individual components are each smaller than
$\Theta_{\trm{\scriptsize max}}$ and thus individually detectable. We modelled
the size distribution for total intensity components by retaining the
parameterisation presented in Equation~(\ref{ch4:eqn:windT}), but with a
modified relationship for $\Theta_\tnm{\tiny med,I}$ given by
\begin{equation}\label{ch4:eqn:windTMedMod}
	\Theta_\tnm{\tiny med,I} = 1.0^{\prime\prime}\left(
	\frac{S}{1\;\tnm{\small mJy}}
	\right)^{\!0.3}\;,
\end{equation}
where $S$ here denotes component flux density. For a single-component source,
Equation~(\ref{ch4:eqn:windTMedMod}) predicts a median LAS that is half that
predicted by \citet{1990ASPC...10..389W}. Equation~(\ref{ch4:eqn:windTMedMod})
is plotted in the left-column panels of Fig.~\ref{ch4:fig:AngSize}; this model
appears to be consistent with the observed ATLAS components.

\begin{figure*}
 \vspace{2pc}
 \centering
 \includegraphics[trim=10mm 30mm 10mm 60mm, height=60mm]{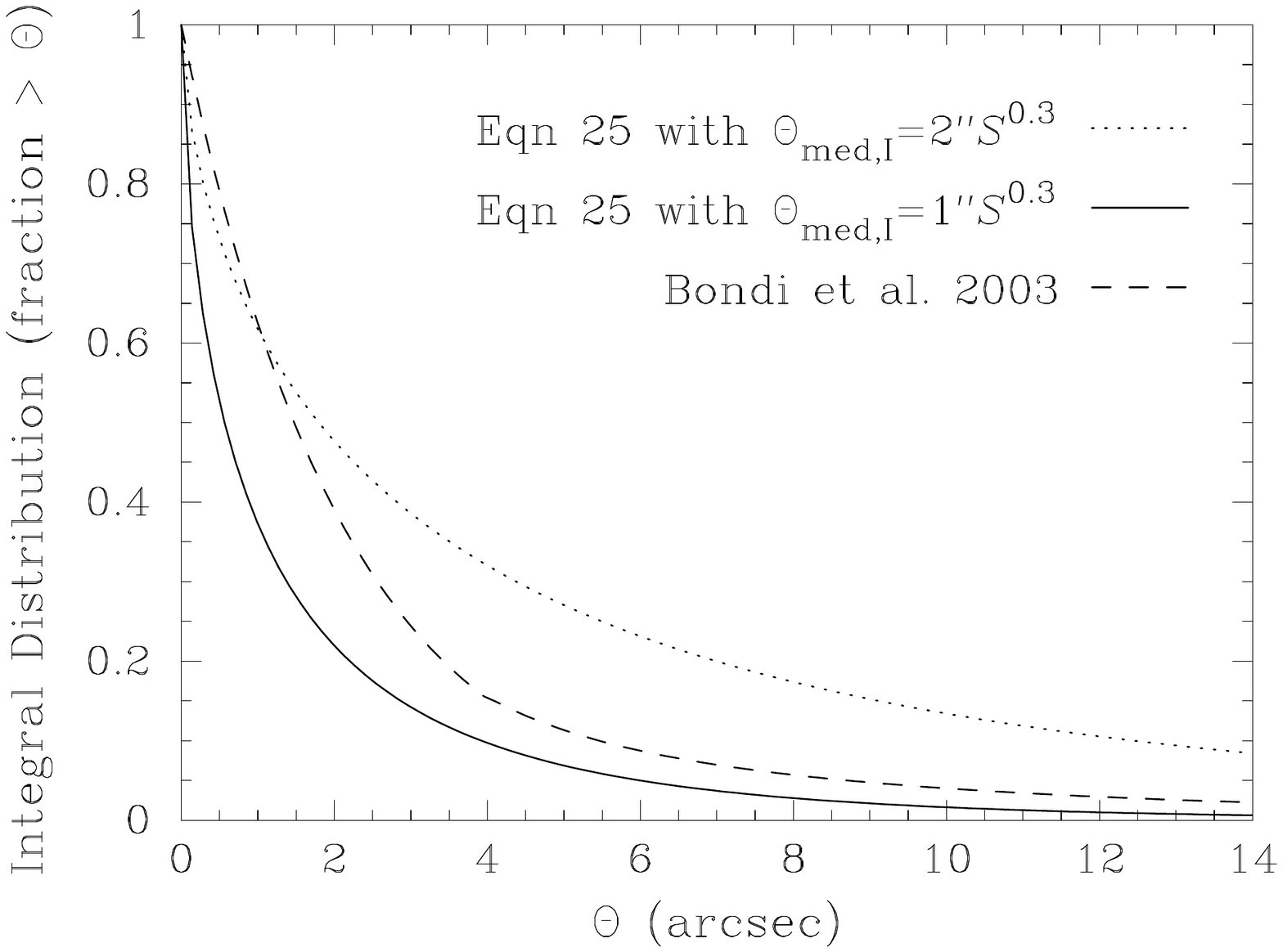}
 \begin{minipage}{140mm}
 \caption{	Integral angular size distributions in total intensity. The dashed
 		curve gives the distribution presented by \citet{2003A&A...403..857B}
		for sources with $0.4 \le S < 1.0$~mJy. The dotted curve gives
		the distribution for sources presented by \citet{1990ASPC...10..389W}.
		The solid curve gives our assumed distribution for ATLAS
		components, obtained following the \citet{1990ASPC...10..389W} source
		parameterisation but with assumed median angular sizes reduced by
		a factor of 2. The solid curve assumes components with $S=0.7$~mJy,
		while the dotted curve assumes sources with $S=0.7$~mJy.
}
 \label{ch4:fig:thetaDist}
\end{minipage}
\end{figure*}

We were motivated to develop Equation~(\ref{ch4:eqn:windTMedMod}) by considering
the angular size distribution presented by \citet{2003A&A...403..857B} for
sources in the VLA-VDF survey with flux density $0.4 \le S < 1.0$~mJy. The
VLA-VDF survey is similar to ATLAS with an observing frequency of 1.4~GHz,
1~deg$^2$ field of view, $6^{\prime\prime}$ synthesised beam FWHM, and
$\sim\!17$~$\mu$Jy~beam$^{-1}$ rms noise. Only 1 VLA-VDF source was found
to comprise multiple components in the flux density range above, with
all others forming single-component sources. We therefore assumed that the
\citet{2003A&A...403..857B} size distribution could be used to characterise
the true size distribution expected for ATLAS components. To demonstrate why
we modified Equation~(\ref{ch4:eqn:windTMedMod}) as such and why we didn't choose
to simply implement the original angular size distributions presented by
\citet{2003A&A...403..857B} or \citet{1990ASPC...10..389W}, we have
plotted each of the distributions in Fig.~\ref{ch4:fig:thetaDist}.
As shown, the original \citet{1990ASPC...10..389W} distribution over-predicts
a substantial tail of sources with angular sizes greater than those observed
and in turn modelled by \citet{2003A&A...403..857B}. Our modified distribution,
using Equation~(\ref{ch4:eqn:windT}) with $\Theta_\tnm{\scriptsize med,I}$ given by
Equation~(\ref{ch4:eqn:windTMedMod}), successfully reproduces the
\citet{2003A&A...403..857B} model for component sizes {\footnotesize
$\gtrsim$}$3^{\prime\prime}$. However, our modified distribution predicts a
greater proportion of components with sizes $<3^{\prime\prime}$ than the
\citet{2003A&A...403..857B} model. To suggest a possible explanation for this
discrepancy and provide a rudimentary justification for our assumed size
distribution, we note that \citet{2003A&A...403..857B} did not account for
bandwidth smearing across their mosaicked data. As a result, it is likely
that their results were biased against the detection of sources with small
angular sizes. For example, assuming a beam FWHM of 6\arcsec, a source with
0\farcsecd5 true angular size would be observed in the absence of bandwidth
smearing to have a size of 6\farcsecd02. But if smearing was present at the level
of 4\%, as may be representative of the VLA-VDF data (recall discussion of
bandwidth smearing in a mosaic from \S~\ref{ch4:SecInstSubBS}), then the observed
and deconvolved angular sizes would be $6\farcsecd02/\sqrt{0.96}$ and 1\farcsecd3,
respectively. Regardless of true size, no source with deconvolved angular size
$<1\farcsecd22$ could be observed in this scenario. This artificial size
inflation would diminish for sources with true angular sizes approaching the
beam FWHM. For example, a source with 3\arcsec\ true angular size would be
observed to have a deconvolved angular size of 3\farcsecd3 if uncorrected for
bandwidth smearing. While it is likely that the true underlying angular size
distribution for components (or single-component sources) lies somewhere
between the solid and dashed curves in Fig.~\ref{ch4:fig:thetaDist}, the key
requirement of our assumed distribution in this work is that it can
characterise populations of components with sizes $>3^{\prime\prime}$. From
the left column panels of Fig.~\ref{ch4:fig:AngSize} we find that
$\Theta_\trm{\scriptsize max}(S)$ does not fall below 3\arcsec\ until
$S<0.2$~mJy. Therefore, very few flux density bins are likely to be
significantly affected if our assumed true size distribution for
$\Theta<3^{\prime\prime}$ is in error.

\begin{figure}
 \centering
 \vspace{1pc}
 \includegraphics[trim = 20mm 20mm 10mm 30mm, height=60mm]{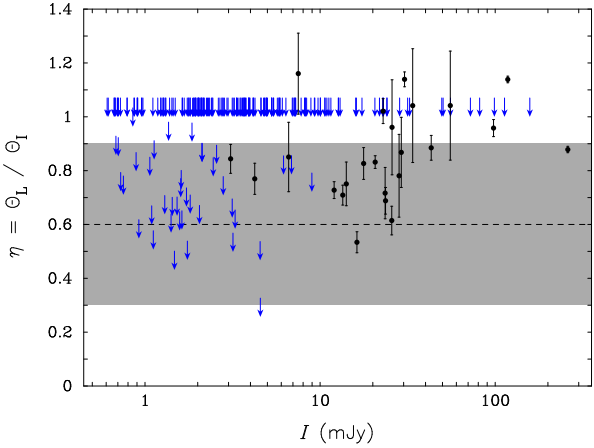}
 \caption{	Angular filling factor, $\eta$, for polarized emission within resolved
 		total intensity components in ATLAS. Upper bounds have been calculated
		using Equation~(\ref{ch4:eqn:minangSpot}). The dashed line and shaded region
		represent the fixed value of $\eta_\tnm{\tiny model}$ and its
		associated uncertainty range as given by Equation~(\ref{ch4:eqn:fillfrac}).
}
 \label{ch4:fig:fillfrac}
\end{figure}

To our knowledge, the true underlying size distribution for 1.4~GHz components
in linear polarization surveys such as ATLAS with $\sim10\arcsec$ resolution
has not yet been explored. Given the small fraction of polarized
components observed as resolved in our data (see Fig.~\ref{ch4:fig:AngSize}),
we were unable to directly investigate this distribution in a robust empirical manner.
Instead, to obtain the polarized size distribution, we first assumed that
angular sizes of polarized components could be related to their total intensity
angular sizes using a filling factor $\eta$, independent of flux density;
i.e. $\Theta_\tnm{\scriptsize L}=\eta\Theta_\tnm{\scriptsize I}$
where $\Theta_\tnm{\scriptsize L}$ and $\Theta_\tnm{\scriptsize I}$ are
a component's deconvolved linear polarization and total intensity angular sizes,
respectively. To estimate an appropriate model value for the angular filling factor,
we evaluated the ratio $\Theta_\tnm{\scriptsize L}/\Theta_\tnm{\scriptsize I}$
for all resolved total intensity components in ATLAS, as shown in Fig.~\ref{ch4:fig:fillfrac}.
Approximate errors for this ratio were calculated following standard error propagation as
\begin{equation}
	\sigma_{\ms \eta} \approx
	\frac{\Theta_\tnm{\scriptsize L}}{\Theta_\tnm{\scriptsize I}}
	\sqrt{
	\left(\frac{\sigma_{\ms \Theta_\tnm{\tiny L}}}{\Theta_\tnm{\scriptsize L}}\right)^{\!2} +
	\left(\frac{\sigma_{\ms \Theta_\tnm{\tiny I}}}{\Theta_\tnm{\scriptsize I}}\right)^{\!2}
	} \,,
\end{equation}
with angular size uncertainties from Equation~(\ref{ch4:eqn:deconvestERR}).
Angular filling factor upper bounds for the total intensity components with an unresolved
polarization counterpart, $\eta_{\trm{\scriptsize u}}$, were obtained by
combining Equations~(\ref{ch4:eqn:resunres}) and (\ref{ch4:eqn:RURlocus}) to estimate the
maximum value of $\Theta_{\trm{\scriptsize L}}$ that a polarized component could attain
before being classified as resolved, namely
\begin{equation}\label{ch4:eqn:minangSpot}
	\eta_{\trm{\scriptsize u}} = \min\left( 1  \, , \,
	\frac{1}{\Theta_{\trm{\scriptsize I}}} \sqrt{
	B_\trm{\scriptsize maj} \, B_\trm{\scriptsize min} \left\{
	a^{-b/\left[ L_{\trm{\tiny peak}}^{\trm{\tiny FOD}}
	/\widetilde{\sigma}(x,y) \right]^{\ms c}} -
	1 \right\}} \right)\;,
\end{equation}
where factors greater than 1 were not allowed. We note that the ATLAS components with
$\eta>1$ in Fig.~\ref{ch4:fig:fillfrac} are diffuse in total intensity and thus poorly
characterised by a 2D elliptical Gaussian, resulting in underestimated values of
$\Theta_{\trm{\scriptsize I}}$ and thus overestimated values of $\eta$.

To model the distribution of components and upper bounds shown in Fig.~\ref{ch4:fig:fillfrac},
we assumed that $\eta$ may be characterised by a constant value with an uncertainty
range, rather than a distribution of values as likely to be more appropriate in reality.
We conservatively modelled the angular filling factor as
\begin{equation}\label{ch4:eqn:fillfrac}
	\eta_\tnm{\tiny model} = 0.6 \pm 0.3 \,,
\end{equation}
as indicated by the dashed line and shading in Fig.~\ref{ch4:fig:fillfrac}.
To model the distribution of median angular sizes for linearly polarized components,
which we denote by $\Theta_\tnm{\tiny med,L}$, we multiplied the distribution of total
intensity median angular sizes from Equation~(\ref{ch4:eqn:windTMedMod}) by 0.3,
0.6, or 0.9 following Equation~(\ref{ch4:eqn:fillfrac}), and convolved each of
the three resulting curves by the fractional polarization distribution from
Equation~(\ref{ch5:eqn:fracpol}). The resulting predicted $\Theta_\tnm{\tiny med,L}$
and its uncertainty range are displayed in the right-column panels
of Fig.~\ref{ch4:fig:AngSize}; our polarization model appears to be
consistent with the observed size distribution of ATLAS components, taking
into account the increased presence of unresolved components towards faint
flux densities. Finally, we made the largely unjustified assumption that
the angular size distribution for polarized components could be modelled
using the same parameterisation presented for total intensity in
Equation~(\ref{ch4:eqn:windT}), with $\Theta_\tnm{\tiny med,I}$ replaced by
$\Theta_\tnm{\tiny med,L}$. The distribution of polarized components
exhibited in Fig.~\ref{ch4:fig:AngSize} does not refute this assumption,
though future high resolution studies are clearly required to support it.
We note that for simplicity, and to avoid placing too much emphasis on
the exact form of the total intensity angular size density function from
Equation~(\ref{ch4:eqn:windTpdf}), we did not estimate $\Theta_\tnm{\tiny med,L}$
above by first convolving Equation~(\ref{ch4:eqn:windTpdf}) with the
fractional polarization distribution from Equation~(\ref{ch5:eqn:fracpol}).
This more standard computational path should be utilised once the total intensity
angular size distribution for components is known with greater confidence.

\begin{figure*}
 \centering
 \includegraphics[trim=10mm 30mm 10mm 30mm, height=90mm]{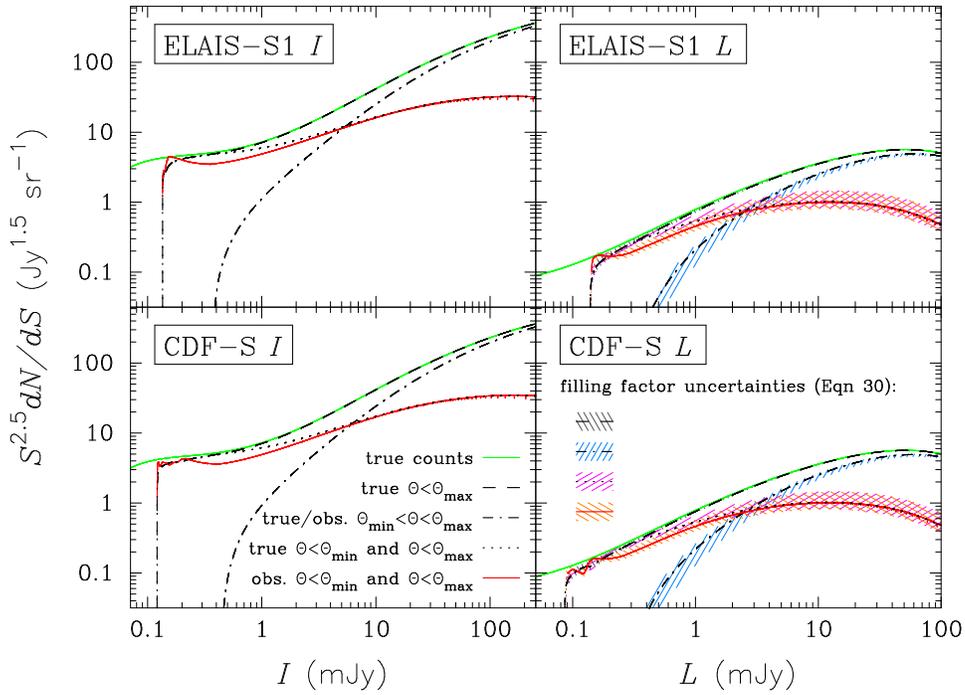}
 \begin{minipage}{140mm}
 \caption{	Modelled effects of resolution bias on differential component
 		counts in total intensity (left column) and linear polarization (right
		column). The green curves show the assumed true underlying counts, given
		by the H03 distribution in total intensity [Equation~(\ref{ch4:eqn:H03})]
		and the $\tnm{H03}*f_{\ms \Pi}(\Pi)$ distribution in linear polarization
		(see \S~\ref{ch4:SecExtSubDB}). The dashed curves show the counts for detectable
		components [Equation~(\ref{ch4:eqn:dashed})]. The dot-dashed curves show the
		counts for detectable components classified as resolved [Equation~(\ref{ch4:eqn:dotdash})].
		The dotted curves show the counts for detectable components classified as unresolved,
		for a scenario where measurement systematics are zero [Equation~(\ref{ch4:eqn:dotted})].
		The red curves show the counts for detectable components classified as unresolved, for
		a realistic scenario where measurement systematics are taken into account
		[Equation~(\ref{ch4:eqn:resconv})]. The shaded regions in the right panels represent
		the propagation of angular filling factor uncertainties [Equation~(\ref{ch4:eqn:fillfrac})].
}
 \label{ch4:fig:resbiasCnts}
 \end{minipage}
\end{figure*}

\begin{figure*}
 \centering
 \includegraphics[trim=10mm 30mm 10mm 30mm, height=90mm]{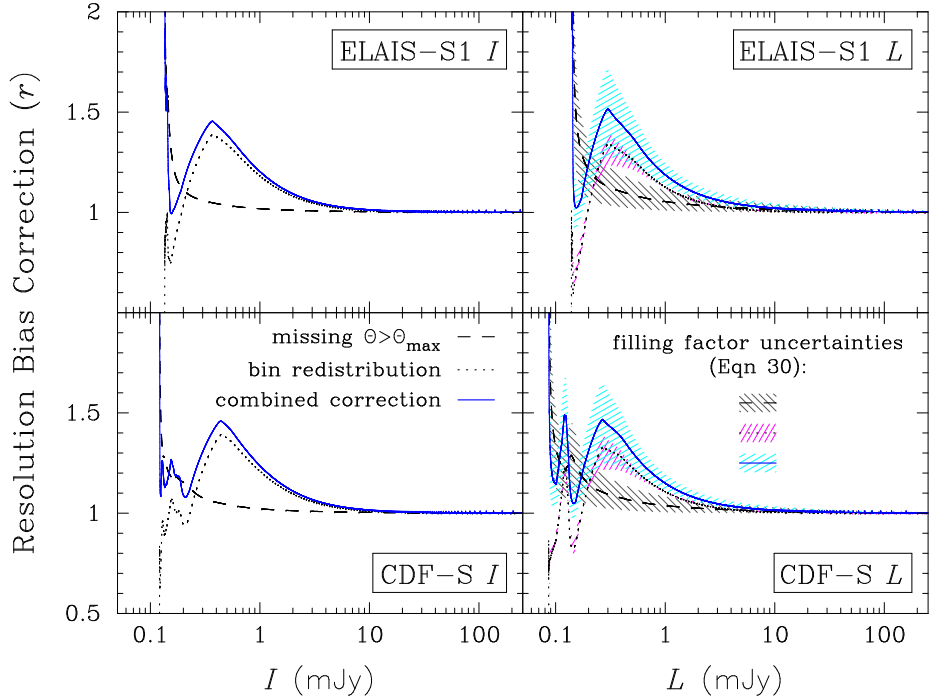} 
 \begin{minipage}{140mm}
 \caption{	Resolution bias corrections for incompleteness to resolved components
 		[Equation~(\ref{ch4:eqn:RB1}); dashed curves], flux density undervaluation for
		unresolved components [Equation~(\ref{ch4:eqn:RB2}); dotted curves], and
		the overall combined corrections [Equation~(\ref{ch4:eqn:RB3}); blue curves].
		The panel layout and details are similar to Fig.~\ref{ch4:fig:resbiasCnts}.
}
 \label{ch4:fig:resbias}
 \end{minipage}
\end{figure*}

We predicted the differential number-counts for detectable components [those with angular sizes
$\le\Theta_\trm{\scriptsize max}(S)$] by evaluating Equation~(\ref{ch4:eqn:windT}) with
Equation~(\ref{ch4:eqn:maxang}) for each total intensity and linear polarization ATLAS field, namely
\begin{equation}\label{ch4:eqn:dashed}
	\frac{dN_{\tnm{\scriptsize detectable}}}{dS}\!\left(S\right) =
	\frac{dN_{\tnm{\scriptsize true}}}{dS}\!\left(S\right) \,
	\left\{ 1-h\!\left[>\Theta_\trm{\scriptsize max}(S),S\right] \right\} \, .
\end{equation}
Equation~(\ref{ch4:eqn:dashed}) is displayed in Fig.~\ref{ch4:fig:resbiasCnts}
(dashed curves), assuming true differential component counts (solid green curves) modelled
in total intensity by the H03 distribution from Equation~(\ref{ch4:eqn:H03}) and in
linear polarization by the $\tnm{H03}*f_{\ms \Pi}(\Pi)$ distribution from \S~\ref{ch4:SecExtSubDB}.
The correction to account for the first form of resolution bias, regarding
incompleteness to resolved components, is then
\begin{equation}\label{ch4:eqn:RB1}
	r_{\tnm{\scriptsize effect-1}}(S) =
	\frac{dN_{\tnm{\scriptsize detectable}}}{dS}\!\left(S\right) \,\div\,
	\frac{dN_{\tnm{\scriptsize true}}}{dS}\!\left(S\right) \, .
\end{equation}
Equation~(\ref{ch4:eqn:RB1}) is displayed in Fig.~\ref{ch4:fig:resbias} for each ATLAS field.

\subsubsection{Effect 2: Flux Density Undervaluation for Unresolved Components}\label{ch4:SecCNCSubRB2}

We accounted for the second form of resolution bias, regarding the undervaluation
of flux densities for components classified as unresolved, as follows.

First, we estimated the minimum intrinsic angular size required for a component to
be classified as resolved, $\Theta_\trm{\scriptsize min}(S)$, following a similar
formalism to that described earlier for $\Theta_\trm{\scriptsize max}(S)$. By relating
Equation~(\ref{ch4:eqn:resunres}) with Equation~(\ref{ch4:eqn:RURlocus}), we obtained
\begin{eqnarray}
	\left[\Theta_\trm{\scriptsize min}(S)\right]^{2} &=&
	\Bigg\{	\int_{0}^{S/A_\tnm{\scriptsize S}}
	B_\trm{\scriptsize maj} \, B_\trm{\scriptsize min} \,
	a^{-b/\left( S/z \right)^{\ms c}} \,
	f_{\widetilde{\sigma}}\!\left(z\right) dz \, \times \nonumber \\
	& & \left[\int_{0}^{S/A_\tnm{\scriptsize S}}
	f_{\widetilde{\sigma}}\!\left(z^{\ms\prime}\right)
	dz^{\ms\prime}\right]^{-1} \Bigg\}^2 -
	B_\trm{\scriptsize maj} \, B_\trm{\scriptsize min} \, . \label{ch4:eqn:minang}
\end{eqnarray}
Similar to the relationships between Equation~(\ref{ch4:eqn:maxang}) and
Equations~(\ref{ch4:eqn:maxangLow}) and (\ref{ch4:eqn:maxangHigh}), the limiting
behaviours of Equation~(\ref{ch4:eqn:minang}) at low and high flux densities
are given by
\begin{equation}\label{ch4:eqn:minangLow}
	\left[\Theta_{\trm{\scriptsize min}^\prime}(S)\right]^{2} =
	B_\trm{\scriptsize maj} \, B_\trm{\scriptsize min} \,
	a^{-b/\left( S/\widetilde{\sigma}_\tnm{\scriptsize min} \right)^{\ms c}} -
	B_\trm{\scriptsize maj} \, B_\trm{\scriptsize min} \, ,
\end{equation}
and
\begin{eqnarray}
	\left[\Theta_{\trm{\scriptsize min}^{\prime\prime}}(S)\right]^{2} &=&
	\left[	\int_{0}^{\infty}
	B_\trm{\scriptsize maj} \, B_\trm{\scriptsize min} \,
	a^{-b/\left( S/z \right)^{\ms c}} \,
	f_{\widetilde{\sigma}}\!\left(z\right) dz \right]^2 - \nonumber \\
	& & B_\trm{\scriptsize maj} \, B_\trm{\scriptsize min} \, , \label{ch4:eqn:minangHigh}
\end{eqnarray}
respectively. The locus defined by Equation~(\ref{ch4:eqn:minang}) is indicated
by a dashed curve for each ATLAS field in Fig.~\ref{ch4:fig:AngSize}. These
curves are bounded by Equations~(\ref{ch4:eqn:minangLow}) and (\ref{ch4:eqn:minangHigh}),
as indicated by the relevant dotted curves.

Next, we predicted the differential number-counts for detectable components
classified as resolved by evaluating
\begin{eqnarray}
	\frac{dN_{\tnm{\scriptsize resolved}}}{dS}\!\left(S\right) &=&
	\frac{dN_{\tnm{\scriptsize detectable}}}{dS}\!\left(S\right) \times \nonumber \\
	& & ( h\!\left\{>\min\left[\Theta_\trm{\scriptsize min}(S),
	\!\Theta_\trm{\scriptsize max}(S)\right],S\right\} - \nonumber \\
	& & h\!\left[>\Theta_\trm{\scriptsize max}(S),S\right] ) \, \div \nonumber \\
	& & \left( 1-h\!\left[>\Theta_\trm{\scriptsize max}(S),S\right] \right) \;. \label{ch4:eqn:dotdash}
\end{eqnarray}
We will assume that observed flux densities for resolved components are equal to
their true flux densities. The predicted counts for detectable components
classified as unresolved are
\begin{equation}\label{ch4:eqn:dotted}
	\frac{dN_{\tnm{\scriptsize unresolved}}}{dS}\!\left(S\right) =
	\frac{dN_{\tnm{\scriptsize detectable}}}{dS}\!\left(S\right) - 
	\frac{dN_{\tnm{\scriptsize resolved}}}{dS}\!\left(S\right) \, .
\end{equation}
Equations~(\ref{ch4:eqn:dotdash}) and (\ref{ch4:eqn:dotted}) are displayed in
Fig.~\ref{ch4:fig:resbiasCnts} (dot-dashed and dotted curves, respectively).

If measurement errors were zero, flux densities for unresolved
components could be obtained from observation of their integrated
surface brightnesses. The observed differential counts for these unresolved
components would then match the $dN_{\tnm{\scriptsize unresolved}}/dS$ curves.
In reality, however, their flux densities
are set by their peak surface brightnesses assuming zero intrinsic angular size,
resulting in the redistribution of component counts from any given flux density bin
to fainter bins because components always have physical non-zero angular sizes. To
model this effect and predict the observed distribution of components classified
as unresolved, we convolved Equation~(\ref{ch4:eqn:dotted}) by a suitably
renormalised version of Equation~(\ref{ch4:eqn:windTpdf}), namely
\begin{eqnarray}
	\frac{dN_{\tnm{\scriptsize unresolved-obs}}}{dS}\!\left(S\right) &=&
	\int_{S}^{\infty}
	\frac{dN_{\tnm{\scriptsize unresolved}}}{dS^\prime}\!\left(
	S^\prime\right) \, H\!\left({\widetilde{\Theta}}-\Theta^\prime\right) \times \nonumber \\
	& & \frac{f_{\ms \Theta}\!\left(\Theta^\prime,S^\prime\right)}
	{ \int_{0}^{\widetilde{\Theta}}
	f_{\ms \Theta}\!\left(\Theta^{\prime\prime},S^\prime\right)
	d\Theta^{\prime\prime} } \, dS^\prime \;, \label{ch4:eqn:resconv}
\end{eqnarray}
where
\begin{eqnarray}
	\Theta^\prime &=& \sqrt{B_\trm{\scriptsize maj} \, B_\trm{\scriptsize min}
	\left(\frac{S^\prime}{S}-1\right)} \,,\\
	\widetilde{\Theta} &=&
	\min\left[\Theta_\trm{\scriptsize min}\left(S^\prime\right),
	\Theta_\trm{\scriptsize max}\left(S^\prime\right)\right] \;,
\end{eqnarray}
and where $H(x)$ is a unit step function with value unity for argument $x\ge0$ and
zero otherwise. Equation~(\ref{ch4:eqn:resconv}) is displayed by the red curve
for each ATLAS field in Fig.~\ref{ch4:fig:resbiasCnts}. The wiggles in these curves at
faint flux densities are real (i.e. not due to numerical instabilities); they are caused by
the behaviour of $\Theta_\trm{\scriptsize max}(S)$ at faint flux densities, in turn
influenced by the spatial distribution of rms noise and bandwidth smearing in
the ATLAS mosaics.

The correction to account for the second form of resolution bias was then calculated as
\begin{eqnarray}
	r_{\tnm{\scriptsize effect-2}}(S) &=&
	\frac{dN_{\tnm{\scriptsize detectable}}}{dS}\!\left(S\right) \,\div \nonumber \\
	& & \left[ \frac{dN_{\tnm{\scriptsize resolved}}}{dS}\!\left(S\right) +
	\frac{dN_{\tnm{\scriptsize unresolved-obs}}}{dS}\!\left(S\right) \right] \, . \label{ch4:eqn:RB2}
\end{eqnarray}
This correction is shown as the dotted curve for each ATLAS field in Fig.~\ref{ch4:fig:resbias}.

We note that the form of resolution bias investigated in this section is only
relevant for measurement schemes that use peak surface brightness as a proxy
for unresolved component flux density (mostly relevant at low SNR; e.g. see
Fig.~\ref{ch4:fig:IPR}; see also \citeauthor{2010ApJS..188..384S}
\citeyear{2010ApJS..188..384S}). The alternative is to use integrated surface
brightness measurements for both unresolved and resolved components. However,
such schemes will exhibit new and more significant biases in recovered flux
densities at low and even moderate SNRs due to both increased statistical
errors from the larger number of free parameters required to obtain an integrated
measurement compared to a peak measurement (relevant to both 2D elliptical
Gaussian and flood-fill fits; see Fig.~6 in \citeauthor{2012MNRAS.425..979H}
\citeyear{2012MNRAS.425..979H}), and increased systematic flux density
errors (particularly relevant for Gaussian fits; see Fig. 6 in
\citeauthor{2012MNRAS.425..979H} \citeyear{2012MNRAS.425..979H}).

\subsubsection{Combined Correction}

The overall resolution bias correction factors for ATLAS DR2 were calculated by
multiplying Equations~(\ref{ch4:eqn:RB1}) and (\ref{ch4:eqn:RB2}) together,
\begin{equation}\label{ch4:eqn:RB3}
	r(S) = r_{\tnm{\scriptsize effect-1}}(S) \, r_{\tnm{\scriptsize effect-2}}(S) \, .
\end{equation}
Equation~(\ref{ch4:eqn:RB3}) is displayed by the blue curve for each ATLAS field in Fig.~\ref{ch4:fig:resbias}.
For decreasing flux density, the correction factors for each field rise to a peak due to
increasing incompleteness, fall due to the redistribution of components classified as unresolved,
and then rise again at the faintest levels as incompleteness again dominates the correction
(the correction rises to infinity at levels below the faintest flux density bin because the
number of detectable components drops to zero). The blue
curves in Fig.~\ref{ch4:fig:resbias} are consistent with the results from Monte Carlo
simulations presented by \citet{2008ApJ...681.1129B}, who found that resolution bias
correction factors were not maximised for the faintest flux density bin, but
rather for a higher flux density bin due to combination of the two effects
described above.

\citet{2001A&A...365..392P} and \citet{2005AJ....130.1373H}
have derived resolution bias correction factors exhibiting similar rise-fall behaviour
to that presented by \citet{2008ApJ...681.1129B} and this work. Their solutions were
obtained by only considering the first form of resolution bias considered in this work,
regarding reduced sensitivity to resolved components. To obtain their solutions,
\citet{2001A&A...365..392P} and \citet{2005AJ....130.1373H} described the
use of Equations~(\ref{ch4:eqn:minangLow}) or (\ref{ch4:eqn:minangHigh}) [i.e. equations
representing $\Theta_\trm{\scriptsize min}(S)$], respectively, in characterising
$\Theta_\trm{\scriptsize max}(S)$ at the faintest flux densities probed by their
data. Given that $\Theta_\trm{\scriptsize min}(S)$ rises with decreasing flux density,
eventually becoming larger than $\Theta_\trm{\scriptsize max}(S)$, their resolution
bias corrections were found to rise and then fall with decreasing flux density.
However, their procedure is not suitable; $\Theta_\trm{\scriptsize max}(S)$ represents a
strict limit to the angular size of detectable components, regardless of the size of
$\Theta_\trm{\scriptsize min}(S)$ which dictates whether a {\it detected} component
will be classified as unresolved or resolved.

\subsection{Eddington Bias}\label{ch4:SecCNCSubEB}

As described in \S~\ref{ch4:SecExtSubDB}, random measurement errors in the presence of
a non-uniformly distributed component population will redistribute components between
number-count flux density bins, resulting in Eddington bias \citep{1913MNRAS..73..359E,
1938MNRAS..98..190J,1940MNRAS.100..354E}. We accounted for Eddington bias by considering
two alternative correction schemes.

For the first method, we computed Equation~(\ref{ch4:eqn:Neff}) using the deboosted flux
densities from \S~\ref{ch4:SecExtSubDB}, with $e_{\ms i}$ set to unity. The deboosting
equations presented in \S~\ref{ch4:SecExtSubDB} offer a simple approach for mitigating
Eddington bias prior to the construction of differential component counts.
However, these equations do not account for spatial variations in rms noise or bandwidth
smearing, nor do they properly account for variations in number-count slope ($\gamma$).
To account for such specifics we considered an alternative correction scheme similar to
that proposed by \citet{1913MNRAS..73..359E}, focusing on correction factors $e_{\ms i}$
for bin counts involving raw component flux densities rather than deboosted values.
We now describe this second method.

Given an observed noise distribution and an assumed underlying true component count
distribution, the observed counts can be predicted \cite[e.g.][]{2006MNRAS.372..741S}.
The ratio between the predicted and true distributions gives the Eddington bias;
the correction factors $e_{\ms i}$ are therefore given by the reciprocal of this ratio.
We modelled the predicted (i.e. biased) counts in total intensity, which we denote by
$dN_{\trm{\tiny Edd}}/dS$, by assuming that the underlying counts were distributed
according to the H03 model. We assess the suitability of this assumption
in \S~2.3 of Paper II. By accounting for the proportion of components
with true flux density $S+\epsilon$ that may be observed with flux density $S$ due
to Gaussian measurement error $-\epsilon$, we have
\begin{eqnarray}
	\frac{dN_{\trm{\tiny Edd}}}{dS}
	\left(S\right) &=& \int_{-\infty}^{\infty}
	\int_{0}^{z^{\prime\prime}}
	\frac{1}{\sqrt{2\pi}}\exp\left(\frac{-\xi^2}{2}\right)
	\times \nonumber \\
	&& \frac{dN_{\trm{\tiny H03}}}{dS}\left(S+\xi z\right)
	\frac{f_{\widetilde{\sigma}}\!\left(z\right)}
	{ \int_{0}^{z^{\prime\prime}}
	f_{\widetilde{\sigma}}\!\left(z^\prime\right) dz^\prime} \,
	dz \, d\xi \;, \label{ch4:eqn:eddI}
\end{eqnarray}
where
\begin{equation}\label{ch4:eqn:eddIzpp}
	z^{\prime\prime} = \left\{
	\begin{array}{l l}
		-S/\xi & \quad \textrm{if $\xi<0$}\\
		\infty & \quad \textrm{if $\xi\ge0$}\,,\\
	\end{array} \right.
\end{equation}
and $f_{\widetilde{\sigma}}$ is the effective noise distribution for each ATLAS field
taking into account bandwidth smearing (as introduced in \S~\ref{ch4:SecCNCSubRB}).
The parameter $z^{\prime\prime}$ prevents the argument to $dN_{\trm{\tiny H03}}/dS$
from becoming negative (i.e. unphysical). To obtain a similar relationship for
the predicted counts in linear polarization, which we denote $dN_{\trm{\tiny Edd}}/dL$,
we replaced the Gaussian error distribution from Equation~(\ref{ch4:eqn:eddI}) by
the distribution for $L_{\ms RM}$ given by Equation~(28) from \citet{2012MNRAS.424.2160H}.
Assuming an underlying $dN/dL$ distribution given by the $\tnm{H03}*f_{\ms \Pi}(\Pi)$
model introduced in \S~\ref{ch4:SecExtSubDB}, we obtained
\begin{eqnarray}
	\frac{dN_{\trm{\tiny Edd}}}{dL}
	\left(S\right) \!\!\! &=& \!\!\! \int_{-\infty}^{\infty}
	\int_{0}^{z^{\prime\prime}} \nonumber \\
	&& \!\!\! f\!\left(L_\tnm{\scriptsize RM}=S
	\;|\;M=28,L_{\ms 0}=S+\xi z,
	\sigma_\tnm{\scriptsize RM}=z\right) \times \nonumber\\
	&& \!\!\! \frac{dN_{\tnm{\scriptsize H03}*f_{\ms \Pi}(\Pi)}}{dL}
	\left(S+\xi z\right) \times \nonumber \\
	&& \!\!\! \frac{f_{\widetilde{\sigma}}\!\left(z\right)}
	{ \int_{0}^{z^{\prime\prime}}
	f_{\widetilde{\sigma}}\!\left(z^\prime\right) dz^\prime} \,
	dz \, d\xi \;,\label{ch4:eqn:eddL}
\end{eqnarray}
where
\begin{equation}\label{ch4:eqn:eddLzpp}
	z^{\prime\prime} = \left\{
	\begin{array}{l l}
		-S/\xi & \quad \textrm{if $\xi<0$}\\
		\infty & \quad \textrm{if $\xi\ge0$}\,,\\
	\end{array} \right.
\end{equation}
and $f\!\left(L_\tnm{\scriptsize RM}\,|\,M,L_{\ms 0},\sigma_\tnm{\scriptsize RM}\right)$
is the PDF for $L_{\ms RM}$ given $M$, $L_{\ms 0}$, and $\sigma_\tnm{\scriptsize RM}$.

\begin{figure*}
\centering
 \includegraphics[trim=10mm 30mm 10mm 30mm, height=90mm]{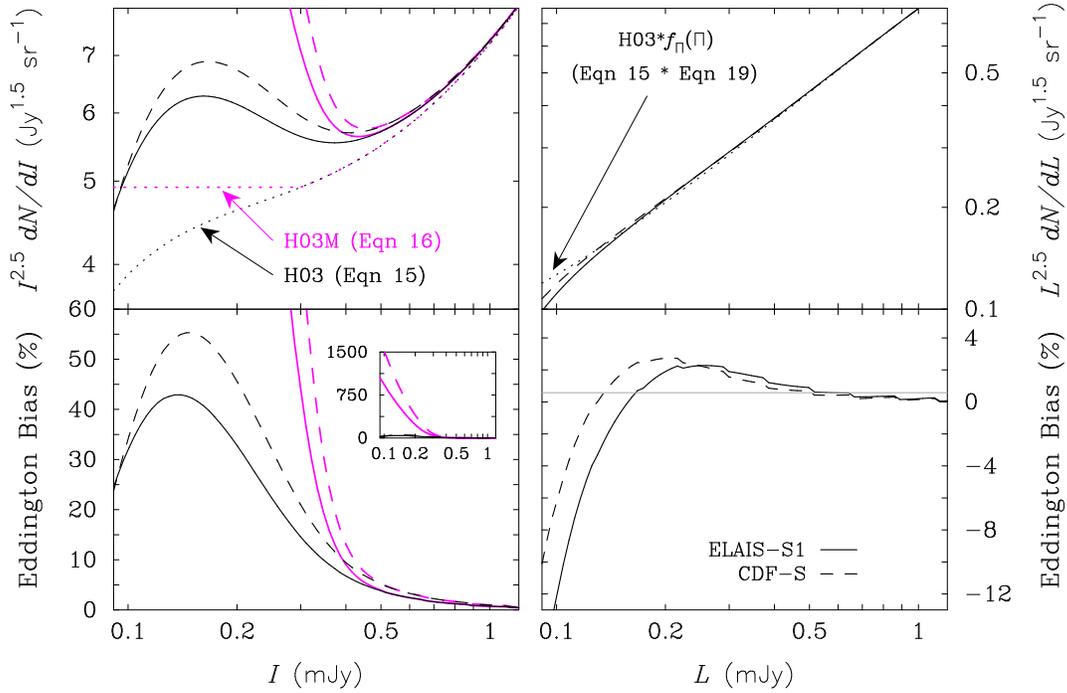}
 \begin{minipage}{140mm}
 \caption{	Modelled effects of Eddington bias on differential component
 		counts in total intensity (left column) and linear polarization
		(right column). Upper panels display predicted counts for each
		ATLAS field (solid or dashed curves) assuming true underlying count
		distributions given by the H03 (dotted black) or H03M (dotted magenta)
		models. Polarization results utilising the H03M model are not shown,
		as they are identical to the H03 results over the flux density range
		shown. Curves in the lower panels indicate the percentages
		by which the predicted distributions overestimate the underlying
		distributions.
		}
 \label{ch4:fig:EB}
 \end{minipage}
\end{figure*}

In Fig.~\ref{ch4:fig:EB} we display Equations~(\ref{ch4:eqn:eddI})
and (\ref{ch4:eqn:eddL}), their assumed underlying count distributions, and the resulting
Eddington biases for each ATLAS field; these are indicated by the black curves.
As expected, we find that Eddington bias becomes stronger with decreasing flux
density in both total intensity and linear polarization. Deviations from this
general trend are observed for the faintest flux density bins, predominantly due
to the changing slope of the counts in total intensity, and to the
positive-semidefinite and non-Gaussian nature of noise fluctuations in polarization.
The positive nature of polarization measurements largely prevents components with
low-SNR from having their flux densities underestimated \citep[e.g. see the
effective noise distribution presented in the lower panel of Fig.~1 in][]{2012MNRAS.424.2160H},
in turn causing the faintest polarized counts to be underestimated.

For the purpose of illustration, following \S~\ref{ch4:SecExtSubDB} we also predicted
the Eddington bias that would be exhibited if the true underlying counts were
described by the H03M model. The results in total intensity are displayed
as magenta curves in Fig.~\ref{ch4:fig:EB}; polarization results are not show
as they are identical to the H03 results over the flux density range displayed.
Our approach of modelling the underlying counts in order to
compare with the observed counts is similar to the forward modelling described
by \citet{2012ApJ...750..139M}. However, unlike their work, our Eddington bias
calculations take into account a more suitable statistical form to describe
$L_{\ms RM}$ \citep[see discussion in][]{2012MNRAS.424.2160H}.

We now make some remarks about our Eddington bias calculations.
Like the deboosting relationship for linear polarization presented in
Equation~(\ref{ch4:eqn:EBcorrCmpL2}), Equation~(\ref{ch4:eqn:eddL}) performs implicit
polarization debiasing. However, their treatments differ:
Equation~(\ref{ch4:eqn:EBcorrCmpL2}) assumes Ricean statistics for simplicity, whereas
Equation~(\ref{ch4:eqn:eddL}) incorporates the full PDF for $L_{\ms RM}$.

Ideally, the Eddington bias corrections presented in Equations~(\ref{ch4:eqn:eddI})
and (\ref{ch4:eqn:eddL}) require underlying count distributions that represent
unresolved components, and not all components as implemented here (note that
this distinction was made for
our deboosting corrections in \S~\ref{ch4:SecExtSubDB}). Our solutions above
implicitly assume that all components are unresolved, such that their true
peak surface brightnesses may be perturbed by noise fluctuations, in turn
directly perturbing their observed flux densities. However, this
simplification in our analysis is unlikely to result in any significant
systematics, because the flux density range over which significant Eddington
bias is observed in Fig.~\ref{ch4:fig:EB} consists overwhelmingly of unresolved
components (see Fig.~\ref{ch4:fig:resbiasCnts}). In the future, a potential
refinement may be to combine both the resolution and Eddington bias corrections,
rather than splitting them as presented in this work.

Finally, we note that Equation~(\ref{ch4:eqn:eddL}) and thus our Eddington bias predictions
may be inaccurate for two reasons. First, our treatment of the correlation between the
error distribution and the underlying signal may not include all nuisance parameters.
Second, the Eddington bias predictions for our data are relatively
small, which is perhaps unusual given the non-Gaussian PDF for $L_{\ms RM}$. However,
as will be demonstrated in Paper II, differences between the two independent Eddington
bias correction schemes described above are largely negligible when applied to both the
total intensity and linear polarization number-counts, providing confidence in both approaches.

\section{Results}\label{ch5:SecRes}

\subsection{Mosaics}\label{ch5:SecResSubMos}

Figs.~\ref{ch5:fig:mosCI}--\ref{ch5:fig:mosEL} display the total intensity and linear
polarization mosaicked images of the CDF-S and ELAIS-S1 ATLAS fields. Residual sidelobes
are observed around strong sources in the total intensity images, giving rise to minor residual
polarization leakage about these sources in the polarization images. Residual
sidelobes remaining after cleaning are observed in the south-east quadrant of the total intensity
ELAIS-S1 field (Fig.~\ref{ch5:fig:mosEI}), which originate from the 3.8~Jy source
PKS~B0039$-$445 outside the field of view.
The random pattern of pixels with zero intensity in the polarization images is an artefact
of our data processing (difficult to see in printed images; look to field edges where
intensity contrast is greatest). This pattern was caused by an error recognition scheme we
implemented during the RM cleaning stage, in which we automatically set
Equation~(\ref{ch4:eqn:LrmDEF}) to zero if the maximum polarized intensity was located
at $\pm\phi_{\trm{\scriptsize max}}$. Thus $\sim$1 in 800 pixels was artificially
set to zero. We note that this scheme did not affect subsequent data analysis.

\subsection{Component and Source Catalogues}\label{ch5:SecResSubCmpCat}

The ATLAS 1.4~GHz DR2 component catalogue is presented in Appendix~A.
This catalogue lists a total of 2588 components in total intensity and linear polarization;
no components were detected in circular polarization.

The ATLAS 1.4~GHz DR2 source catalogue is presented in Appendix~B.
This catalogue lists a total of 2221 sources as identified through the cross-identification
and classification schemes presented in \S~\ref{ch4:SecClass}.

\section{Conclusion}\label{ch4:SecConcl}

We have presented data reduction and analysis procedures
for the second data release of the Australia Telescope Large
Area Survey. We produced and analysed sensitive 1.4~GHz images of
the CDF-S and ELAIS-S1 regions across a combined area of 6.392~deg$^2$ in
total intensity ($I$), linear polarization ($L$), and circular polarization ($V$).
The data for $L$ were processed using RM synthesis and RM clean. Typical
sensitivities across each of the mosaicked multi-pointing images are
$\sim30$~$\mu$Jy~beam$^{-1}$, falling to $<25$~$\mu$Jy~beam$^{-1}$ within
smaller areas. The typical spatial resolutions are $12\arcsec\times6\arcsec$.

We performed component detection and extraction independently
in $I$, $L$, and $V$ using a combination of {\tt BLOBCAT} and {\tt IMFIT},
accounting for spatial variations in image sensitivity, bandwidth smearing
and instrumental polarization leakage. Corrections for clean bias were
not required, due to our implemented cleaning strategy.
ATLAS DR2 is the first survey to have been analysed using {\tt BLOBCAT}. We catalogued
a total of 2416, 172, and 0 components in $I$, $L$, and $V$, respectively,
and determined flux densities for each of these components by considering
their angular sizes. We catalogued 2221 sources by matching single or
multiple $I$ components with SWIRE mid-infrared sources, and by matching
$L$ components to their $I$ counterparts. We classified these sources as
AGNs, SFGs, or stars according to four diagnostic criteria. Our source
catalogue is slightly biased against the detection of multi-component sources
due to our nearest-neighbour cross-identification method,
and toward the classification of AGNs due to lack of optical spectroscopy for
the majority of sources.

We presented a comprehensive prescription for handling
multi-pointing data consistently in both total intensity and linear polarization.
We described our data reduction and analysis procedures in detail in order
to inform future surveys and to highlight our novel extensions of processing
techniques from total intensity to linear polarization. We developed new analytic
techniques to account for bandwidth smearing with a non-circular beam, and resolution
bias in differential number-counts. We extended the analytic framework for Eddington
bias corrections from total intensity to linear polarization.

In Paper II we present the ATLAS DR2 cross-identification and number-count results,
and discuss statistics of the faint polarized 1.4~GHz sky.

\section*{Acknowledgments}

We thank the following for helpful discussions and support: Julie Banfield,
Mark Calabretta, Chris Carilli, Tim Cornwell, Ron Ekers, Juliana Kwan, Tim Robishaw,
Bob Sault, Nick Seymour, Chris Simpson, Rogier Windhorst, and Peter Zinn.
We thank the anonymous referee for their comprehensive review of the
manuscript and helpful comments.
C.~A.~H. acknowledges the support of an Australian Postgraduate Award,
a CSIRO OCE Scholarship, and a Jansky Fellowship from the National Radio
Astronomy Observatory. B.~M.~G. and R.~P.~N. acknowledge the support
of the Australian Research Council Centre of Excellence for All-sky
Astrophysics (CAASTRO), through project number CE110001020.
The Australia Telescope Compact Array is part of the Australia Telescope
National Facility which is funded by the Commonwealth of Australia for
operation as a National Facility managed by CSIRO. This paper includes
archived data obtained through the Australia Telescope Online Archive
({\tt http://atoa.atnf.csiro.au}).

%
\begin{landscape}
\setlength{\headheight}{170pt}
\begin{figure*}
\includegraphics[trim = 9mm 17mm 9mm 15mm, clip, angle=-90, totalheight=\textwidth]{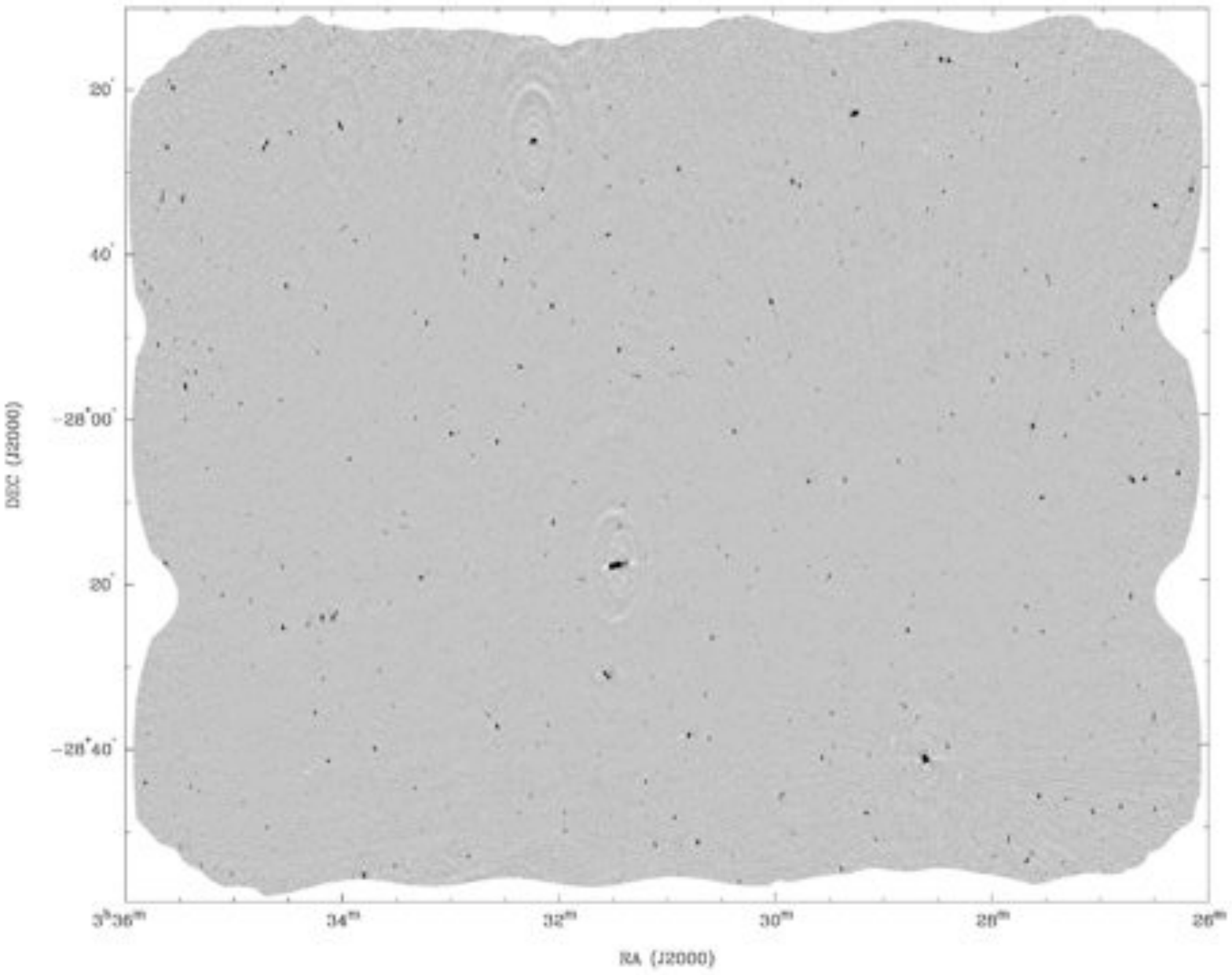}
\caption{
	1.4~GHz total intensity mosaic ($I_{\trm{\tiny MFS}}$) of the CDF-S ATLAS field
	in zenithal equal-area projection at a resolution of $13\farcsecd0\times6\farcsecd0$.
	The peak surface brightness is 0.89~Jy~beam$^{-1}$ and the typical rms noise is 40~$\mu$Jy~beam$^{-1}$.
	The intensity scale is linear, saturating white below -0.3~mJy~beam$^{-1}$ and
	black above 1~mJy~beam$^{-1}$.
	}
\label{ch5:fig:mosCI}
\end{figure*}
\end{landscape}
%
%
\begin{landscape}
\setlength{\headheight}{170pt}
\begin{figure*}
\includegraphics[bb = 100 100 500 500, trim = 9mm 17mm 9mm 15mm, clip, angle=-90, totalheight=\textwidth]{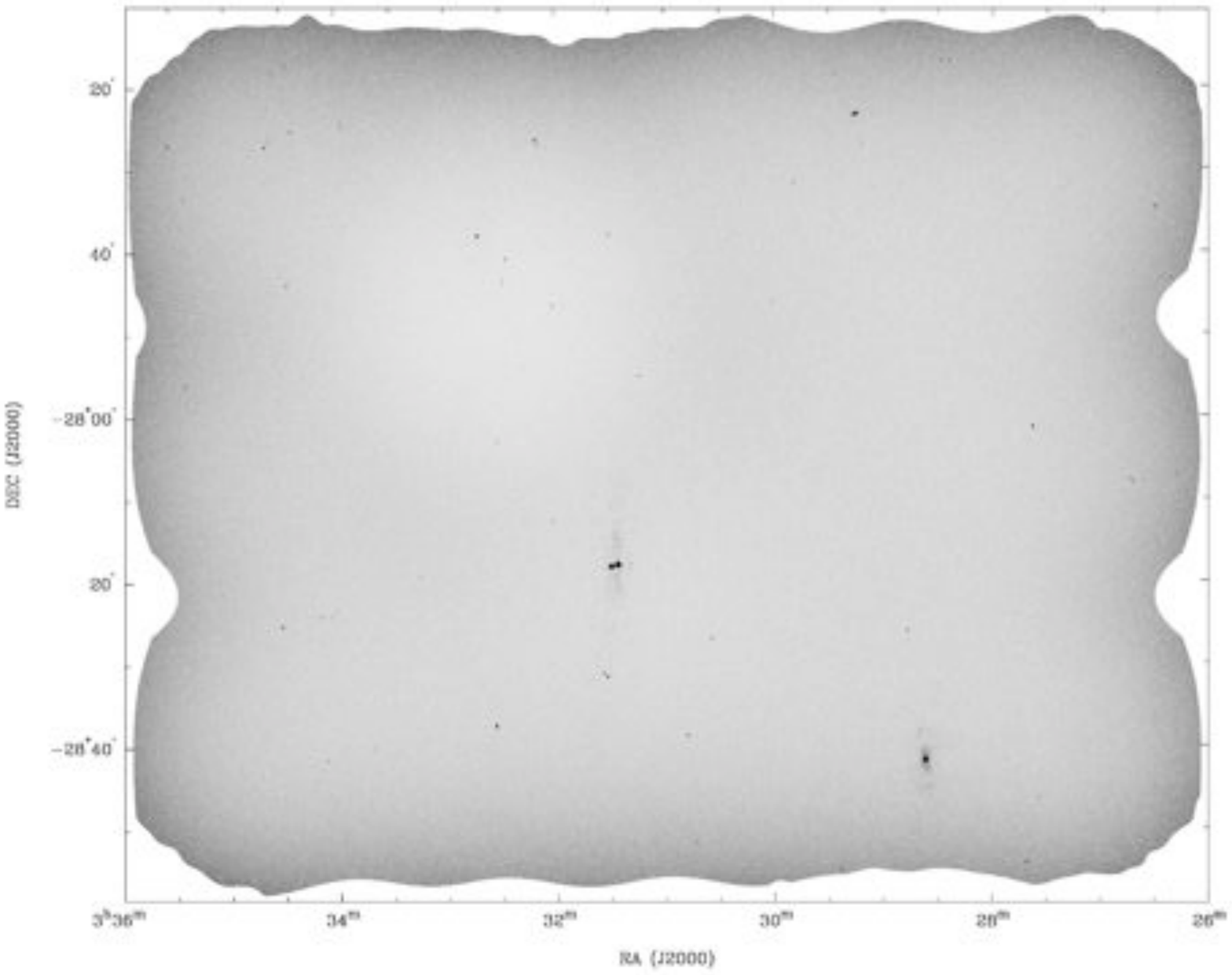}
\caption{
	1.4~GHz leakage-corrected linear polarization mosaic
	($L_{\trm{\tiny RM}}^{\trm{\tiny CORR}}$) of the CDF-S ATLAS field
	in zenithal equal-area projection at a resolution of $14\farcsecd6\times5\farcsecd4$.
	The peak surface brightness is 42~mJy~beam$^{-1}$ and the typical rms noise is 25~$\mu$Jy~beam$^{-1}$.
	The intensity scale is linear, saturating white at 0~mJy~beam$^{-1}$ and black
	above 0.4~mJy~beam$^{-1}$.
	}
\label{ch5:fig:mosCL}
\end{figure*}
\end{landscape}
\begin{figure*}
\centering
\includegraphics[bb = 100 100 500 500, trim = 7mm 35mm 10mm 39mm, clip, angle=-90, width=\textwidth]{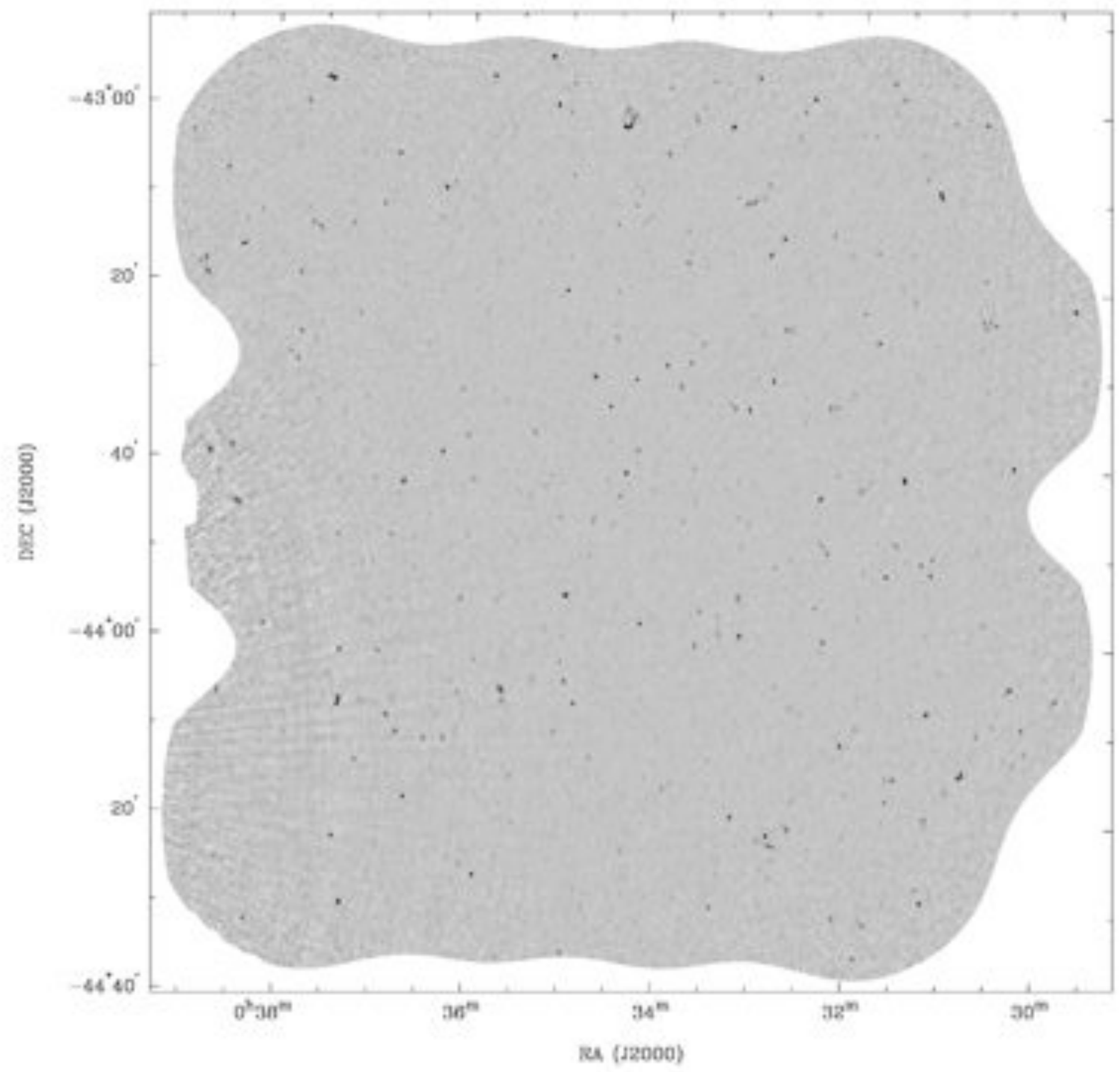}
\caption{
	1.4~GHz total intensity mosaic ($I_{\trm{\tiny MFS}}$) of the ELAIS-S1 ATLAS
	field in zenithal equal-area projection at a resolution of $9\farcsecd6\times7\farcsecd6$.
	The peak surface brightness is 0.16~Jy~beam$^{-1}$ and the typical rms noise is 40~$\mu$Jy~beam$^{-1}$.
	The intensity scale is linear, saturating white below -0.3~mJy~beam$^{-1}$
	and black above 1~mJy~beam$^{-1}$.
	}
\label{ch5:fig:mosEI}
\end{figure*}
\begin{figure*}
\centering
\includegraphics[bb = 100 100 500 500, trim = 7mm 35mm 10mm 39mm, clip, angle=-90, width=\textwidth]{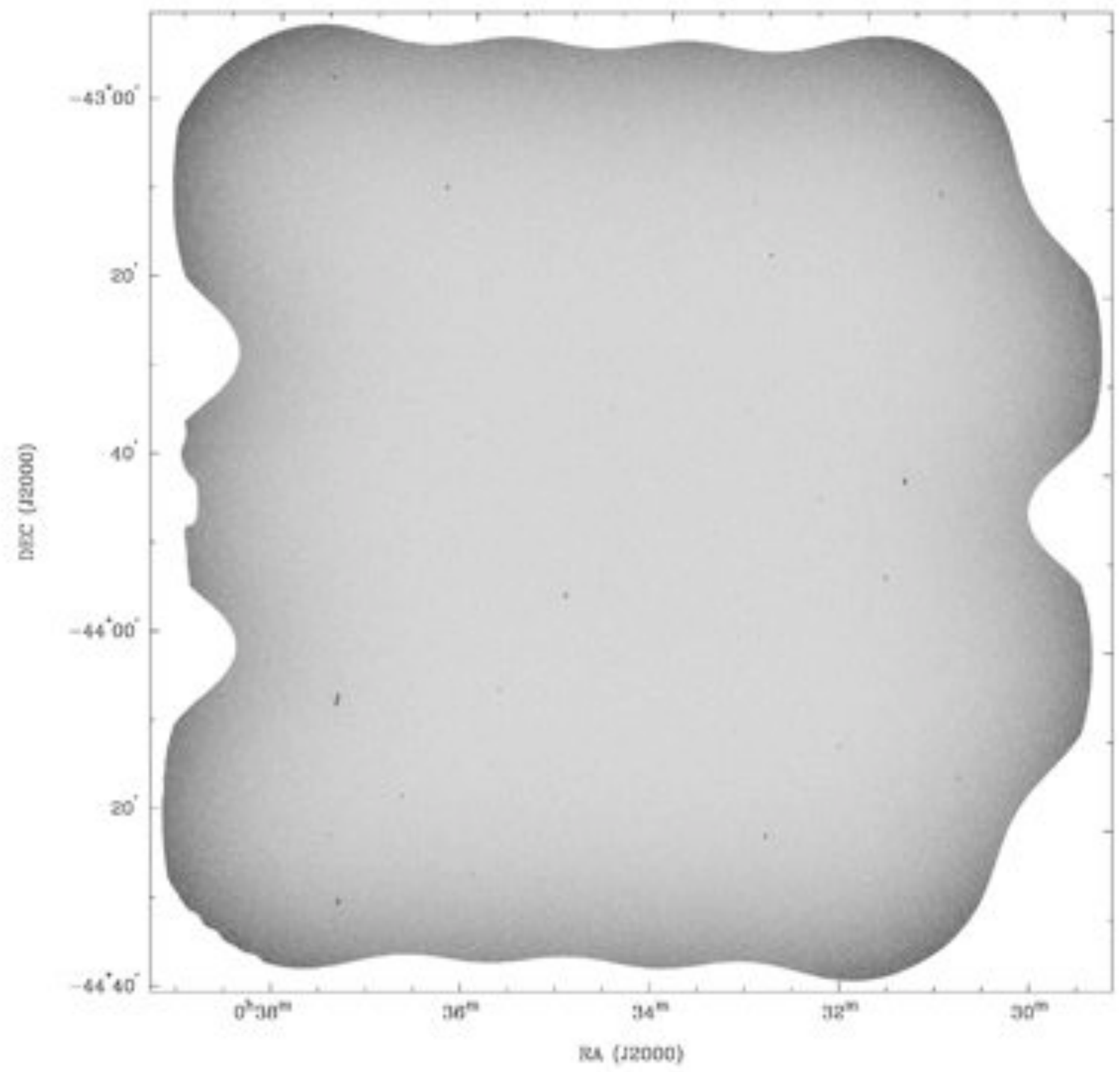}
\caption{
	1.4~GHz leakage-corrected linear polarization mosaic
	($L_{\trm{\tiny RM}}^{\trm{\tiny CORR}}$) of the ELAIS-S1 ATLAS field
	in zenithal equal-area projection at a resolution of $10\farcsecd6\times6\farcsecd2$.
	The peak surface brightness is 3.8~mJy~beam$^{-1}$ and the typical rms noise is
	25~$\mu$Jy~beam$^{-1}$. The intensity scale is linear, saturating white at
	0~mJy~beam$^{-1}$ and black above 0.4~mJy~beam$^{-1}$.
	}
\label{ch5:fig:mosEL}
\end{figure*}

\appendix

\section{Component Catalogue}

This appendix presents the ATLAS 1.4~GHz DR2 component catalogue, a portion of which
is displayed in Table~\ref{tbl:cmpcat} for guidance regarding its form and content.
The catalogue lists a total of 2588 components in total intensity and linear polarization;
no components were discovered in circular polarization. For ease of visual inspection,
components have been grouped according to their nature of detection, arranged in the order
total intensity then linear polarization. Components within each of these two groups
are ordered by increasing right ascension. The columns of Table~\ref{tbl:cmpcat} are:
\begin{description}

\item {\it Column (1)}.---Component identification number. This gives the internal
designation of the component used within our data processing. The form is a composite
of three descriptors plus an optional fourth. The first is a single character that
represents the ATLAS field, given by {\tt C} for CDF-S or {\tt E} for ELAIS-S1. The
second descriptor is a single character that represents the nature of detection,
given by {\tt T} for total intensity or {\tt L} for linear polarization. The third
descriptor is an integer that gives the blob identification number assigned by
{\tt BLOBCAT}. The fourth descriptor is only suffixed for those components that
were obtained through refitting all or part of the original blob using {\tt IMFIT},
denoted by {\tt C}$j$, or {\tt BLOBCAT}, denoted by {\tt F}$j$, for the $j$'th
extracted component from a given blob.

\item {\it Column (2)}.---Full ATLAS DR2 component name. This has been provided
in a form acceptable for International Astronomical Union (IAU) designation
\citep{1994A&AS..107..193L}. The
form is {\tt ATLAS2\_JHHMMSS.SS$+$DDMMSS.ST}
where {\tt ATLAS2} is the survey acronym, {\tt J} specifies the J2000.0 coordinate
equinox, HHMMSS.SS are the hours, minutes and truncated (not rounded) seconds of
right ascension, $+$ or $-$ is the sign of declination, DDMMSS.S are the degrees,
minutes and truncated seconds of declination, and the single character specifier
in parentheses indicates the nature of detection as {\tt T} or {\tt L}. The
position derives from Columns (3) and (4) below.

\item {\it Columns (3) and (4)}.---Right ascension and declination (J2000.0) at
intensity-weighted centroid.

\item {\it Columns (5) and (6)}.---Absolute astrometric uncertainties in right
ascension and declination. Minimum and maximum errors are 0\farcsecd11 and 1\farcsecd1
in right ascension, and 0\farcsecd17 and 2\farcsecd1 in declination, respectively.
\item {\it Columns(7)}.---SNR of raw detection, $A_\tnm{\tiny S}$.

\item {\it Column (8)}.---Local rms noise value, $\sigma_\tnm{\tiny S}$.

\item {\it Column (9)}.---Local bandwidth smearing value, $\varpi$.

\item {\it Columns (10) and (11)}.---Peak surface brightness corrected for bandwidth
smearing, $S_\tnm{\tiny peak}$, and rms error, $\sigma_{\ms S_\tnm{\tiny peak}}$.

\item {\it Columns (12) and (13)}.---Integrated surface brightness, $S_\tnm{\tiny int}$, and
rms error, $\sigma_{\ms S_\tnm{\tiny int}}$.

\item {\it Column (14)}.---Visibility area, $V^{\trm{\tiny AREA}}$.

\item {\it Columns (15) and (16)}.---Estimated deconvolved angular size or
upper bound, $\Theta$, and rms error, $\sigma_{\ms \Theta}$. If $\sigma_{\ms \Theta}>0$
then the component is resolved with flux density given by $S_\tnm{\tiny int}$. If
$\sigma_{\ms \Theta}=0$, the component is unresolved with $\Theta$ representing
an upper bound to the deconvolved angular size, and with flux density given
by $S_\tnm{\tiny peak}$.

\item {\it Column (17)}.---Deboosted flux density, $S_\tnm{\tiny ML}$.

\end{description}
See \S~\ref{ch4:SecExt} and \citet{2012MNRAS.425..979H} for details regarding the
parameters presented above.


\begin{landscape}
\setlength{\headheight}{180pt}
\thispagestyle{empty}
\begin{table*}
 \caption{
	ATLAS 1.4~GHz DR2 Component Catalogue. This table has been truncated and is
	available in its entirety in the online version of this paper.
 }\label{tbl:cmpcat}
 \begin{tabular}{@{}lcrrrrrrrrrrrrrrr@{}}
 \hline
	\multicolumn{1}{c}{ID} &
	\multicolumn{1}{c}{Name} &
	\multicolumn{1}{c}{R.A.} &
	\multicolumn{1}{c}{Decl.} &
	\multicolumn{1}{c}{$\sigma_{\trm{\tiny R.A.}}$} &
	\multicolumn{1}{c}{$\sigma_{\trm{\tiny Decl.}}$} &
	\multicolumn{1}{c}{$A_{\trm{\tiny S}}$} &
	\multicolumn{1}{c}{$\sigma_{\trm{\tiny S}}$} &
	\multicolumn{1}{c}{$\varpi$} &
	\multicolumn{1}{c}{$S_{\trm{\tiny peak}}$} &
	\multicolumn{1}{c}{$\sigma_{S_\trm{\tiny peak}}$} &
	\multicolumn{1}{c}{$S_{\trm{\tiny int}}$} &
	\multicolumn{1}{c}{$\sigma_{S_\trm{\tiny int}}$} &
	\multicolumn{1}{c}{$V^{\trm{\tiny AREA}}$} &
	\multicolumn{1}{c}{$\Theta$} &
	\multicolumn{1}{c}{$\sigma_{\Theta}$} &
	\multicolumn{1}{c}{$S_{\trm{\tiny ML}}$} \\
	& \multicolumn{1}{c}{(Prefix: ATLAS2\_J)} &
	\multicolumn{1}{c}{(\degree, J2000.0)} &
	\multicolumn{1}{c}{(\degree, J2000.0)} &
	\multicolumn{1}{c}{($^{\prime\prime}$)} &
	\multicolumn{1}{c}{($^{\prime\prime}$)} & &
	\multicolumn{1}{c}{(mJy~b$^{-1}$)} & &
	\multicolumn{1}{c}{(mJy~b$^{-1}$)} &
	\multicolumn{1}{c}{(mJy~b$^{-1}$)} &
	\multicolumn{1}{c}{(mJy)} &
	\multicolumn{1}{c}{(mJy)} & &
	\multicolumn{1}{c}{($^{\prime\prime}$)} &
	\multicolumn{1}{c}{($^{\prime\prime}$)} &
	\multicolumn{1}{c}{(mJy)} \\
	\multicolumn{1}{c}{(1)} &
	\multicolumn{1}{c}{(2)} &
	\multicolumn{1}{c}{(3)} &
	\multicolumn{1}{c}{(4)} &
	\multicolumn{1}{c}{(5)} &
	\multicolumn{1}{c}{(6)} &
	\multicolumn{1}{c}{(7)} &
	\multicolumn{1}{c}{(8)} &
	\multicolumn{1}{c}{(9)} &
	\multicolumn{1}{c}{(10)} &
	\multicolumn{1}{c}{(11)} &
	\multicolumn{1}{c}{(12)} &
	\multicolumn{1}{c}{(13)} &
	\multicolumn{1}{c}{(14)} &
	\multicolumn{1}{c}{(15)} &
	\multicolumn{1}{c}{(16)} &
	\multicolumn{1}{c}{(17)}\\
 \hline
ET383      & 002938.16$-$432946.7T    &      7.409023 &    -43.496308 &     0.45 &     0.64 &     11.06 &     0.069 &   0.84 &     0.913 &     0.095 &     0.999 &     0.086 &   1.000 &    7.08 &    0.00 &     0.897 \\
ET691      & 002940.15$-$440309.3T    &      7.417319 &    -44.052597 &     0.75 &     1.07 &      6.51 &     0.066 &   0.85 &     0.509 &     0.083 &     0.401 &     0.069 &   0.990 &   10.64 &    0.00 &     0.481 \\
ET780      & 002943.13$-$440812.1T    &      7.429743 &    -44.136712 &     0.82 &     1.18 &      5.89 &     0.066 &   0.84 &     0.462 &     0.082 &     0.309 &     0.068 &   0.977 &   11.59 &    0.00 &     0.430 \\
ET727      & 002944.33$-$433629.7T    &      7.434732 &    -43.608272 &     0.78 &     1.12 &      6.21 &     0.069 &   0.83 &     0.516 &     0.087 &     0.527 &     0.074 &   0.991 &   11.06 &    0.00 &     0.484 \\
ET71C2     & 002945.34$-$432148.4T    &      7.438933 &    -43.363466 &     0.16 &     0.20 &     73.51 &     0.071 &   0.85 &     6.141 &     0.324 &     8.495 &     0.431 &   1.000 &    5.29 &    0.50 &     8.495 \\
ET71C1     & 002946.18$-$432149.2T    &      7.442417 &    -43.363667 &     0.15 &     0.20 &     79.93 &     0.070 &   0.85 &     6.630 &     0.348 &     9.107 &     0.461 &   1.000 &    5.22 &    0.51 &     9.107 \\
ET125      & 002949.92$-$440541.3T    &      7.458010 &    -44.094817 &     0.17 &     0.23 &     46.01 &     0.059 &   0.86 &     3.164 &     0.175 &     3.461 &     0.183 &   1.000 &    2.90 &    0.00 &     3.161 \\
ET760      & 002951.14$-$432354.3T    &      7.463110 &    -43.398433 &     0.81 &     1.16 &      5.99 &     0.060 &   0.87 &     0.413 &     0.072 &     0.316 &     0.062 &   0.953 &   11.43 &    0.00 &     0.385 \\
ET197      & 002951.21$-$440556.5T    &      7.463388 &    -44.099045 &     0.23 &     0.32 &     25.40 &     0.059 &   0.86 &     1.725 &     0.111 &     1.811 &     0.108 &   1.000 &    4.12 &    0.00 &     1.721 \\
ET986      & 002953.48$-$440618.7T    &      7.472854 &    -44.105196 &     0.96 &     1.37 &      5.06 &     0.058 &   0.87 &     0.336 &     0.069 &     0.315 &     0.060 &   0.883 &   13.34 &    0.00 &     0.303 \\
ET158      & 003001.25$-$435046.6T    &      7.505240 &    -43.846287 &     0.20 &     0.26 &     33.96 &     0.058 &   0.85 &     2.300 &     0.136 &     2.589 &     0.142 &   1.000 &    3.46 &    0.00 &     2.297 \\
ET481C1    & 003003.18$-$435950.9T    &      7.513271 &    -43.997472 &     0.60 &     0.86 &      9.11 &     0.051 &   0.89 &     0.520 &     0.063 &     0.579 &     0.059 &   0.992 &    8.14 &    0.00 &     0.505 \\
ET956      & 003003.67$-$441238.4T    &      7.515330 &    -44.210670 &     0.93 &     1.33 &      5.18 &     0.060 &   0.85 &     0.363 &     0.072 &     0.370 &     0.063 &   0.915 &   13.03 &    0.00 &     0.329 \\
ET929      & 003007.66$-$441329.8T    &      7.531941 &    -44.224947 &     0.92 &     1.31 &      5.26 &     0.058 &   0.85 &     0.359 &     0.071 &     0.354 &     0.061 &   0.910 &   12.85 &    0.00 &     0.326 \\
ET997      & 003007.91$-$432725.6T    &      7.532983 &    -43.457133 &     0.97 &     1.38 &      5.01 &     0.048 &   0.90 &     0.268 &     0.055 &     0.201 &     0.049 &   0.762 &   13.48 &    0.00 &     0.240 \\
ET375C1    & 003008.06$-$441148.5T    &      7.533613 &    -44.196822 &     0.48 &     0.68 &     12.08 &     0.058 &   0.87 &     0.807 &     0.078 &     0.933 &     0.075 &   1.000 &    6.65 &    0.00 &     0.796 \\
ET590      & 003008.82$-$433320.5T    &      7.536756 &    -43.555700 &     0.67 &     0.95 &      7.35 &     0.046 &   0.89 &     0.377 &     0.055 &     0.362 &     0.049 &   0.928 &    9.62 &    0.00 &     0.361 \\
ET375C2    & 003008.89$-$441144.7T    &      7.537071 &    -44.195772 &     0.85 &     1.21 &      6.59 &     0.058 &   0.87 &     0.436 &     0.070 &     0.505 &     0.063 &   0.965 &   10.53 &    0.00 &     0.413 \\
ET62       & 003010.81$-$440907.7T    &      7.545050 &    -44.152158 &     0.14 &     0.18 &    108.20 &     0.052 &   0.88 &     6.392 &     0.331 &     7.220 &     0.365 &   1.000 &    3.07 &    0.86 &     7.220 \\
ET445      & 003012.74$-$433245.8T    &      7.553084 &    -43.546061 &     0.52 &     0.74 &      9.54 &     0.045 &   0.90 &     0.476 &     0.055 &     0.387 &     0.049 &   0.982 &    7.87 &    0.00 &     0.464 \\
 \hline
 \end{tabular}
\end{table*}
\end{landscape}

\section{Source Catalogue}

This appendix presents the ATLAS 1.4~GHz DR2 source catalogue, a portion of which
is displayed in Table~\ref{tbl:srccat} for guidance regarding its form and content.
The catalogue lists a total of 2221 sources as identified through the cross-identification
and classification schemes presented in \S~\ref{ch4:SecClass}. Sources
are ordered by increasing right ascension. The columns of the source catalogue are:
\begin{description}

\item {\it Column (1)}.---Source identification number. This gives the internal
designation of the source used within our data processing. The form is a composite
of two descriptors. The first is a single character that represents the ATLAS field,
given by {\tt C} for CDF-S or {\tt E} for ELAIS-S1. The second descriptor is an
integer that reflects the ordering of sources within each field, as described above.

\item {\it Column (2)}.---Full ATLAS DR2 source name. This has been provided
in a form appropriate for future IAU designation. The form is\linebreak
{\tt ATLAS2\_JHHMMSS.SS$+$DDMMSS.S} where {\tt ATLAS2} is the survey
acronym, {\tt J} specifies the J2000.0 coordinate equinox, HHMMSS.SS are the hours,
minutes and truncated seconds of right ascension, $+$ or $-$ is the sign of
declination, and DDMMSS.S are the degrees, minutes and truncated seconds of
declination. If a SWIRE cross-identification was available, the position was
specified by that of the infrared source. Otherwise, the position was calculated
as the unweighted centroid of all total intensity components comprising the source.

\item {\it Columns (3)--(8)}.---Component identification numbers for all total
intensity components belonging to the source, corresponding to column~(1) of
Table~\ref{tbl:cmpcat}. Up to 6 components may be provided; blanks are
indicated by $0$.

\item {\it Columns (9)--(11)}.---Component identification numbers for all linearly
polarized components belonging to the source, corresponding to column~(1) of
Table~\ref{tbl:cmpcat}. Up to 3 components may be provided; blanks are indicated
by $0$.

\item {\it Columns (12)--(14)}.---Component groupings and their Type classifications
according to the linear polarization--total intensity association scheme presented
in \S~\ref{ch4:SecClassSubPA}. Up to 3 groups may be provided, labelled A, B, and C.
Each group is specified by a 9-digit string, where from left to right each digit
corresponds to the respective components given in columns~(3)-(11). If a component
is not included in a particular group, then its respective digit is set to 9. All
components belonging to a given group have their respective digit set to the Type
classification for that group. Group A always represents the collection of all
components comprising the source. Groups B and C represent subsets of components.

\item {\it Columns (15) and (16)}.---Sum of deboosted flux densities for all total
intensity components in Group A, and associated rms error. The error is given by
the quadrature sum of the uncertainties from column~(11) or (13) of
Table~\ref{tbl:cmpcat}, for unresolved or resolved components respectively.

\item {\it Columns (17) and (18)}.---Same as Columns (15) and (16), but for
deboosted flux densities of linearly polarized components in Group A.

\item {\it Columns (19) and (20)}.---Same as Columns (15) and (16), but for Group B.

\item {\it Columns (21) and (22)}.---Same as Columns (17) and (18), but for Group B.

\item {\it Columns (23) and (24)}.---Same as Columns (15) and (16), but for Group C.

\item {\it Columns (25) and (26)}.---Same as Columns (17) and (18), but for Group C.

\item {\it Column (27)}.---Name of SDR3 counterpart. Listed as {\tt none} for
sources without an infrared cross-identification.

\item {\it Columns (28) and (29)}.---Right ascension and declination (J2000.0),
following the position definition provided for column~(2).

\item {\it Columns (30) and (31)}.---Sum of deboosted flux densities for all total
intensity components belonging to the source, and associated rms error. The error
is given by the quadrature sum of the uncertainties from column~(11) or (13) of
Table~\ref{tbl:cmpcat}, for unresolved or resolved components respectively.

\item {\it Columns (32) and (33)}.---Same as columns~(30) and (31), but for linearly
polarized components. For unpolarized sources, column~(32) specifies the weakest
polarization upper limit for any of the source's total intensity components,
calculated using Equation~(\ref{ch4:eqn:Lupper}), while column~(33) is set to zero.

\item {\it Columns (34)--(38)}.---SWIRE infrared flux densities for the
3.6, 4.5, 5.8, 8.0, and 24.0~$\mu$m bands. Following \citet{2006AJ....132.2409N}
we selected aperture extractions for unresolved infrared sources and extended
(Kron) extractions for resolved sources. Entries specified as zero indicate that
the infrared source was undetected.

\item {\it Column (39)}.---Classification based on the criteria presented in
\S~\ref{ch4:SecClassSubRIOCI}. The categories are $0=\tnm{AGN}$, $1=\tnm{SFG}$,
$2=\tnm{star}$, and $9=\tnm{unknown}$.

\end{description}
To illustrate use of Table~\ref{tbl:srccat} we interpret the data for source E26.
The total intensity and linearly polarized flux densities for this source are
given in Columns~(30) and (32), respectively. Column~(39) indicates that the source
was classified as an AGN. Source E26 has a SWIRE cross-match given in Column~(27)
with infrared flux densities given in Columns~(34)-(38). Source E26 comprises 2
total intensity components, ET71C1 and ET71C2, and a single linearly polarized
component, EL26. Details for each of these components are given in Table~\ref{tbl:cmpcat}.
Continuing with Table~\ref{tbl:srccat}, the polarization properties of source E26 are
detailed in Columns~(12)-(26). Column~(12) indicates that the total intensity components
from Columns~(3) and (4), and the polarization component from Column~(9), form a
group (Group~A) which has a Type 3 linear polarization--total intensity classification.
This means that source E26 comprises a linearly polarized component situated midway
between two total intensity components. The lack of classification information
in Columns~(13) and (14) indicates that smaller sub-groupings cannot be formed
for source E26; it is not possible to form an unambiguous one-to-one cross-match
between component EL26 and either of ET71C1 or ET71C2. The flux density information
for Group~A is given in Columns~(15)-(18).

\begin{table*}
 \centering
 \caption{ATLAS 1.4~GHz DR2 Source Catalogue -- Part I of III. This table has been truncated and is
 available as part of a single master table in the online version of this paper.
 }\label{tbl:srccat}
 \begin{tabular}{@{}lclllllllllccc@{}}
 \hline
	\multicolumn{1}{c}{ID} &
	\multicolumn{1}{|c}{Name} &
	\multicolumn{6}{|c}{Total Intensity Component ID} &
	\multicolumn{3}{|c}{Linear Pol. Component ID} &
	\multicolumn{3}{|c}{Group} \\
	& \multicolumn{1}{|c}{(Prefix: ATLAS2\_J)} &
	\multicolumn{1}{|c}{I1} &
	\multicolumn{1}{c}{I2} &
	\multicolumn{1}{c}{I3} &
	\multicolumn{1}{c}{I4} &
	\multicolumn{1}{c}{I5} &
	\multicolumn{1}{c}{I6} &
	\multicolumn{1}{|c}{L1} &
	\multicolumn{1}{c}{L2} &
	\multicolumn{1}{c}{L3} &
	\multicolumn{1}{|c}{A} &
	\multicolumn{1}{c}{B} &
	\multicolumn{1}{c}{C} \\
	\multicolumn{1}{c}{(1)} &
	\multicolumn{1}{|c}{(2)} &
	\multicolumn{1}{|c}{(3)} &
	\multicolumn{1}{c}{(4)} &
	\multicolumn{1}{c}{(5)} &
	\multicolumn{1}{c}{(6)} &
	\multicolumn{1}{c}{(7)} &
	\multicolumn{1}{c}{(8)} &
	\multicolumn{1}{|c}{(9)} &
	\multicolumn{1}{c}{(10)} &
	\multicolumn{1}{c}{(11)} &
	\multicolumn{1}{c}{(12)} &
	\multicolumn{1}{c}{(13)} &
	\multicolumn{1}{c}{(14)}\\
 \hline
E232  & 002938.07$-$432947.9 & ET383    & 0        & 0        & 0        & 0        & 0        & 0        & 0        & 0        & 999999999 & 999999999 & 999999999 \\
E386  & 002940.19$-$440309.6 & ET691    & 0        & 0        & 0        & 0        & 0        & 0        & 0        & 0        & 999999999 & 999999999 & 999999999 \\
E420  & 002943.15$-$440813.6 & ET780    & 0        & 0        & 0        & 0        & 0        & 0        & 0        & 0        & 999999999 & 999999999 & 999999999 \\
E385  & 002944.36$-$433630.2 & ET727    & 0        & 0        & 0        & 0        & 0        & 0        & 0        & 0        & 999999999 & 999999999 & 999999999 \\
E26   & 002945.64$-$432149.3 & ET71C1   & ET71C2   & 0        & 0        & 0        & 0        & EL26     & 0        & 0        & 339999399 & 999999999 & 999999999 \\
E106  & 002949.92$-$440541.3 & ET125    & 0        & 0        & 0        & 0        & 0        & 0        & 0        & 0        & 999999999 & 999999999 & 999999999 \\
E468  & 002951.14$-$432355.3 & ET760    & 0        & 0        & 0        & 0        & 0        & 0        & 0        & 0        & 999999999 & 999999999 & 999999999 \\
E160  & 002951.26$-$440556.4 & ET197    & 0        & 0        & 0        & 0        & 0        & 0        & 0        & 0        & 999999999 & 999999999 & 999999999 \\
E577  & 002953.51$-$440617.8 & ET986    & 0        & 0        & 0        & 0        & 0        & 0        & 0        & 0        & 999999999 & 999999999 & 999999999 \\
E129  & 003001.30$-$435046.2 & ET158    & 0        & 0        & 0        & 0        & 0        & 0        & 0        & 0        & 999999999 & 999999999 & 999999999 \\
E367  & 003003.17$-$435951.4 & ET481C1  & 0        & 0        & 0        & 0        & 0        & 0        & 0        & 0        & 999999999 & 999999999 & 999999999 \\
E540  & 003003.73$-$441236.7 & ET956    & 0        & 0        & 0        & 0        & 0        & 0        & 0        & 0        & 999999999 & 999999999 & 999999999 \\
E543  & 003007.66$-$441329.8 & ET929    & 0        & 0        & 0        & 0        & 0        & 0        & 0        & 0        & 999999999 & 999999999 & 999999999 \\
E691  & 003007.95$-$432727.2 & ET997    & 0        & 0        & 0        & 0        & 0        & 0        & 0        & 0        & 999999999 & 999999999 & 999999999 \\
E487  & 003008.77$-$433321.6 & ET590    & 0        & 0        & 0        & 0        & 0        & 0        & 0        & 0        & 999999999 & 999999999 & 999999999 \\
E194  & 003008.87$-$441144.9 & ET375C1  & ET375C2  & 0        & 0        & 0        & 0        & 0        & 0        & 0        & 999999999 & 999999999 & 999999999 \\
E56   & 003010.84$-$440907.1 & ET62     & 0        & 0        & 0        & 0        & 0        & 0        & 0        & 0        & 999999999 & 999999999 & 999999999 \\
E399  & 003012.78$-$433246.4 & ET445    & 0        & 0        & 0        & 0        & 0        & 0        & 0        & 0        & 999999999 & 999999999 & 999999999 \\
E426  & 003015.46$-$431201.1 & ET680    & 0        & 0        & 0        & 0        & 0        & 0        & 0        & 0        & 999999999 & 999999999 & 999999999 \\
E244  & 003015.62$-$441311.6 & ET295C1  & 0        & 0        & 0        & 0        & 0        & 0        & 0        & 0        & 999999999 & 999999999 & 999999999 \\
 \hline
 \end{tabular}
\end{table*}

\begin{table*}
 \centering
 \contcaption{ATLAS 1.4~GHz DR2 Source Catalogue -- Part II of III. This table has been truncated and is
 available as part of a single master table in the online version of this paper.}
 \begin{tabular}{@{}rrrrrrrrrrrr@{}}
 \hline
	\multicolumn{4}{|c}{Group A} &
	\multicolumn{4}{|c}{Group B} &
	\multicolumn{4}{|c}{Group C} \\
	\multicolumn{1}{|c}{$I_{20\tnm{\tiny cm}}$} &
	\multicolumn{1}{c}{$\sigma_{I_{\tnm{\tiny 20cm}}}$} &
	\multicolumn{1}{c}{$L_{20\tnm{\tiny cm}}$} &
	\multicolumn{1}{c}{$\sigma_{L_{\tnm{\tiny 20cm}}}$} &
	\multicolumn{1}{|c}{$I_{20\tnm{\tiny cm}}$} &
	\multicolumn{1}{c}{$\sigma_{I_{\tnm{\tiny 20cm}}}$} &
	\multicolumn{1}{c}{$L_{20\tnm{\tiny cm}}$} &
	\multicolumn{1}{c}{$\sigma_{L_{\tnm{\tiny 20cm}}}$} &
	\multicolumn{1}{|c}{$I_{20\tnm{\tiny cm}}$} &
	\multicolumn{1}{c}{$\sigma_{I_{\tnm{\tiny 20cm}}}$} &
	\multicolumn{1}{c}{$L_{20\tnm{\tiny cm}}$} &
	\multicolumn{1}{c}{$\sigma_{L_{\tnm{\tiny 20cm}}}$} \\
	\multicolumn{1}{|c}{(mJy)} &
	\multicolumn{1}{c}{(mJy)} &
	\multicolumn{1}{c}{(mJy)} &
	\multicolumn{1}{c}{(mJy)} &
	\multicolumn{1}{|c}{(mJy)} &
	\multicolumn{1}{c}{(mJy)} &
	\multicolumn{1}{c}{(mJy)} &
	\multicolumn{1}{c}{(mJy)} &
	\multicolumn{1}{|c}{(mJy)} &
	\multicolumn{1}{c}{(mJy)} &
	\multicolumn{1}{c}{(mJy)} &
	\multicolumn{1}{c}{(mJy)} \\
	\multicolumn{1}{|c}{(15)} &
	\multicolumn{1}{c}{(16)} &
	\multicolumn{1}{c}{(17)} &
	\multicolumn{1}{c}{(18)} &
	\multicolumn{1}{|c}{(19)} &
	\multicolumn{1}{c}{(20)} &
	\multicolumn{1}{c}{(21)} &
	\multicolumn{1}{c}{(22)} &
	\multicolumn{1}{|c}{(23)} &
	\multicolumn{1}{c}{(24)} &
	\multicolumn{1}{c}{(25)} &
	\multicolumn{1}{c}{(26)} \\
 \hline
 0.000 &     0.000 &     0.000 &     0.000 &     0.000 &     0.000 &     0.000 &     0.000 &     0.000 &     0.000 &     0.000 &     0.000 \\
 0.000 &     0.000 &     0.000 &     0.000 &     0.000 &     0.000 &     0.000 &     0.000 &     0.000 &     0.000 &     0.000 &     0.000 \\
 0.000 &     0.000 &     0.000 &     0.000 &     0.000 &     0.000 &     0.000 &     0.000 &     0.000 &     0.000 &     0.000 &     0.000 \\
 0.000 &     0.000 &     0.000 &     0.000 &     0.000 &     0.000 &     0.000 &     0.000 &     0.000 &     0.000 &     0.000 &     0.000 \\
17.601 &     0.631 &     0.671 &     0.075 &     0.000 &     0.000 &     0.000 &     0.000 &     0.000 &     0.000 &     0.000 &     0.000 \\
 0.000 &     0.000 &     0.000 &     0.000 &     0.000 &     0.000 &     0.000 &     0.000 &     0.000 &     0.000 &     0.000 &     0.000 \\
 0.000 &     0.000 &     0.000 &     0.000 &     0.000 &     0.000 &     0.000 &     0.000 &     0.000 &     0.000 &     0.000 &     0.000 \\
 0.000 &     0.000 &     0.000 &     0.000 &     0.000 &     0.000 &     0.000 &     0.000 &     0.000 &     0.000 &     0.000 &     0.000 \\
 0.000 &     0.000 &     0.000 &     0.000 &     0.000 &     0.000 &     0.000 &     0.000 &     0.000 &     0.000 &     0.000 &     0.000 \\
 0.000 &     0.000 &     0.000 &     0.000 &     0.000 &     0.000 &     0.000 &     0.000 &     0.000 &     0.000 &     0.000 &     0.000 \\
 0.000 &     0.000 &     0.000 &     0.000 &     0.000 &     0.000 &     0.000 &     0.000 &     0.000 &     0.000 &     0.000 &     0.000 \\
 0.000 &     0.000 &     0.000 &     0.000 &     0.000 &     0.000 &     0.000 &     0.000 &     0.000 &     0.000 &     0.000 &     0.000 \\
 0.000 &     0.000 &     0.000 &     0.000 &     0.000 &     0.000 &     0.000 &     0.000 &     0.000 &     0.000 &     0.000 &     0.000 \\
 0.000 &     0.000 &     0.000 &     0.000 &     0.000 &     0.000 &     0.000 &     0.000 &     0.000 &     0.000 &     0.000 &     0.000 \\
 0.000 &     0.000 &     0.000 &     0.000 &     0.000 &     0.000 &     0.000 &     0.000 &     0.000 &     0.000 &     0.000 &     0.000 \\
 0.000 &     0.000 &     0.000 &     0.000 &     0.000 &     0.000 &     0.000 &     0.000 &     0.000 &     0.000 &     0.000 &     0.000 \\
 0.000 &     0.000 &     0.000 &     0.000 &     0.000 &     0.000 &     0.000 &     0.000 &     0.000 &     0.000 &     0.000 &     0.000 \\
 0.000 &     0.000 &     0.000 &     0.000 &     0.000 &     0.000 &     0.000 &     0.000 &     0.000 &     0.000 &     0.000 &     0.000 \\
 0.000 &     0.000 &     0.000 &     0.000 &     0.000 &     0.000 &     0.000 &     0.000 &     0.000 &     0.000 &     0.000 &     0.000 \\
 0.000 &     0.000 &     0.000 &     0.000 &     0.000 &     0.000 &     0.000 &     0.000 &     0.000 &     0.000 &     0.000 &     0.000 \\
 \hline
\end{tabular}
\end{table*}

\begin{table*}
 \centering
 \contcaption{ATLAS 1.4~GHz DR2 Source Catalogue -- Part III of III. This table has been truncated and is
 available as part of a single master table in the online version of this paper.}
 \begin{tabular}{@{}rrrrrrrrrrrrcc@{}}
 \hline
	\multicolumn{1}{c}{SWIRE Name} &
	\multicolumn{1}{c}{R.A.} &
	\multicolumn{1}{c}{Decl.} &
	\multicolumn{1}{c}{$I_{20\tnm{\tiny cm}}$} &
	\multicolumn{1}{c}{$\sigma_{I_{\tnm{\tiny 20cm}}}$} &
	\multicolumn{1}{c}{$L_{20\tnm{\tiny cm}}$} &
	\multicolumn{1}{c}{$\sigma_{L_{\tnm{\tiny 20cm}}}$} &
	\multicolumn{1}{c}{$I_{3.6\mu\trm{m}}$} &
	\multicolumn{1}{c}{$I_{4.5\mu\trm{m}}$} &
	\multicolumn{1}{c}{$I_{5.8\mu\trm{m}}$} &
	\multicolumn{1}{c}{$I_{8.0\mu\trm{m}}$} &
	\multicolumn{1}{c}{$I_{24.0\mu\trm{m}}$} &
	\multicolumn{1}{c}{Class} \\
	\multicolumn{1}{c}{(Prefix: SWIRE3\_J)} &
	\multicolumn{1}{c}{(\degree, J2000.0)} &
	\multicolumn{1}{c}{(\degree, J2000.0)} &
	\multicolumn{1}{c}{(mJy)} &
	\multicolumn{1}{c}{(mJy)} &
	\multicolumn{1}{c}{(mJy)} &
	\multicolumn{1}{c}{(mJy)} &
	\multicolumn{1}{c}{($\mu$Jy)} &
	\multicolumn{1}{c}{($\mu$Jy)} &
	\multicolumn{1}{c}{($\mu$Jy)} &
	\multicolumn{1}{c}{($\mu$Jy)} &
	\multicolumn{1}{c}{($\mu$Jy)} & \\
	\multicolumn{1}{c}{(27)} &
	\multicolumn{1}{c}{(28)} &
	\multicolumn{1}{c}{(29)} &
	\multicolumn{1}{c}{(30)} &
	\multicolumn{1}{c}{(31)} &
	\multicolumn{1}{c}{(32)} &
	\multicolumn{1}{c}{(33)} &
	\multicolumn{1}{c}{(34)} &
	\multicolumn{1}{c}{(35)} &
	\multicolumn{1}{c}{(36)} &
	\multicolumn{1}{c}{(37)} &
	\multicolumn{1}{c}{(38)} &
	\multicolumn{1}{c}{(39)} \\
 \hline
002938.07$-$432947.9 &      7.408625 &    -43.496639 &     0.897 &     0.095 &     0.524 &     0.000 &        25.22 &        26.13 &         0.00 &         0.00 &         0.00 & 0 \\
002940.19$-$440309.6 &      7.417458 &    -44.052667 &     0.481 &     0.083 &     0.487 &     0.000 &        96.35 &        90.47 &       137.53 &       268.84 &      2278.89 & 0 \\
002943.15$-$440813.6 &      7.429792 &    -44.137111 &     0.430 &     0.082 &     0.504 &     0.000 &        15.87 &        20.46 &         0.00 &        44.49 &         0.00 & 0 \\
002944.36$-$433630.2 &      7.434833 &    -43.608389 &     0.484 &     0.087 &     0.535 &     0.000 &        75.55 &        80.69 &        74.30 &         0.00 &         0.00 & 0 \\
002945.64$-$432149.3 &      7.440167 &    -43.363694 &    17.602 &     0.631 &     0.671 &     0.075 &        85.60 &       109.44 &       147.30 &       177.31 &         0.00 & 0 \\
                none &      7.458010 &    -44.094817 &     3.161 &     0.175 &     0.429 &     0.000 &         0.00 &         0.00 &         0.00 &         0.00 &         0.00 & 0 \\
002951.14$-$432355.3 &      7.463083 &    -43.398694 &     0.385 &     0.072 &     0.431 &     0.000 &        86.60 &       121.34 &       209.92 &       321.67 &       943.66 & 0 \\
002951.26$-$440556.4 &      7.463583 &    -44.099000 &     1.721 &     0.111 &     0.421 &     0.000 &        12.33 &         7.72 &         0.00 &         0.00 &         0.00 & 0 \\
002953.51$-$440617.8 &      7.472958 &    -44.104944 &     0.303 &     0.069 &     0.408 &     0.000 &        77.57 &        65.96 &        96.58 &       320.22 &      3744.82 & 9 \\
003001.30$-$435046.2 &      7.505417 &    -43.846167 &     2.297 &     0.136 &     0.409 &     0.000 &        48.71 &        61.04 &        63.24 &         0.00 &         0.00 & 0 \\
003003.17$-$435951.4 &      7.513208 &    -43.997611 &     0.505 &     0.063 &     0.339 &     0.000 &        32.75 &        37.27 &        50.42 &         0.00 &         0.00 & 0 \\
003003.73$-$441236.7 &      7.515542 &    -44.210194 &     0.329 &     0.072 &     0.419 &     0.000 &        45.79 &        53.39 &        57.07 &        48.67 &         0.00 & 9 \\
                none &      7.531941 &    -44.224947 &     0.326 &     0.071 &     0.410 &     0.000 &         0.00 &         0.00 &         0.00 &         0.00 &         0.00 & 9 \\
003007.95$-$432727.2 &      7.533125 &    -43.457556 &     0.240 &     0.055 &     0.324 &     0.000 &       249.18 &       227.67 &       192.58 &       777.43 &      2843.39 & 1 \\
003008.77$-$433321.6 &      7.536542 &    -43.556000 &     0.361 &     0.055 &     0.342 &     0.000 &       237.07 &       202.42 &       152.05 &       242.06 &       906.83 & 0 \\
003008.87$-$441144.9 &      7.536958 &    -44.195806 &     1.208 &     0.105 &     0.381 &     0.000 &       126.14 &        95.56 &        67.32 &         0.00 &         0.00 & 0 \\
003010.84$-$440907.1 &      7.545167 &    -44.151972 &     7.220 &     0.365 &     0.338 &     0.000 &        52.07 &        48.28 &        51.68 &        78.78 &       424.37 & 0 \\
003012.78$-$433246.4 &      7.553250 &    -43.546222 &     0.464 &     0.055 &     0.319 &     0.000 &        77.30 &        62.01 &        49.32 &        59.59 &         0.00 & 0 \\
003015.46$-$431201.1 &      7.564417 &    -43.200306 &     0.424 &     0.072 &     0.445 &     0.000 &        20.86 &        23.14 &         0.00 &         0.00 &         0.00 & 0 \\
003015.62$-$441311.6 &      7.565083 &    -44.219889 &     0.842 &     0.075 &     0.362 &     0.000 &       441.01 &       341.78 &       183.83 &       163.94 &         0.00 & 0 \\
 \hline
 \end{tabular}
\end{table*}

\bsp

\label{lastpage}

\end{document}